\newcommand{\ifLong}[1]{#1}
\newcommand{\ifShort}[1]{}
\newcommand{\ifShortVspace}[1]{}
\renewcommand{\ifLong}[1]{}
\renewcommand{\ifShort}[1]{#1}
\renewcommand{\ifShortVspace}[1]{\vspace{#1}}

\ifLong{
  \documentclass[10pt]{article}

}
\ifShort{
  \documentclass[11pt]{article}

}

\usepackage{geometry}
\usepackage{setspace}
\usepackage{caption} 
\usepackage{graphicx}
\usepackage[dvipsnames]{xcolor}

\PassOptionsToPackage{hyphens}{url}\usepackage[pdftex, colorlinks=true, hyperfootnotes=false, hyperindex=true, plainpages=false, pagebackref=false, pdfpagelabels=true, pdfstartview=FitH, linkcolor=blue, citecolor=blue, urlcolor=blue, bookmarks=false]{hyperref}
\usepackage{tabularx}
\usepackage{subcaption}
\usepackage{lscape}
\usepackage{wrapfig}

\usepackage{mathptmx}
\usepackage[T1]{fontenc}

\usepackage{amsthm}
\newtheorem{definition}{Definition}
\usepackage{amsmath}
\usepackage{amssymb}

\usepackage{verbatim}

\usepackage[natbibapa]{apacite}

\usepackage{natbib}
\usepackage[inline]{enumitem}

 \newcommand{\luise}[1]{\textcolor{black}{#1}} 
 \newcommand{\fabian}[1]{\textcolor{black}{#1}} 
 \newcommand{\ingo}[1]{\textcolor{black}{#1}}

\newenvironment{biseabstract}{%
\begin{quote} \bf}
{\end{quote}}

\newenvironment{bisekeywords}{%
\begin{quote} \it \textbf{Keywords -}}
{\end{quote}}

\newenvironment{biseacknowledgements}{%
    \textbf{Acknowledgements:}}
{}

\title{Automatic Resource Allocation in Business Processes:\\ A Systematic Literature Survey} 

\author
{
Luise Pufahl$^{1\ast}$, Sven Ihde$^{2}$, Fabian Stiehle$^{1}$, Mathias Weske$^{2}$, Ingo Weber$^{1}$\\
\\
\normalsize{$^{1}$School of CIT, Technical University of Munich}\\
\normalsize{$^{2}$HPI, University of Potsdam}\\
\\
\normalsize{$^\ast$Corresponding author: Luise Pufahl, E-mail: luise.pufahl@tum.de}
}

\date{}

\begin{document} 

\baselineskip24pt


\maketitle

% Place your abstract within the special {biseabstract} environment.
\begin{biseabstract}
Organizations execute various business processes to operate their business and serve 
their clients. 
During execution, upcoming process tasks must be allocated to internal resources, \fabian{such as humans or machines}.
\ingo{This} 
is a complex decision-making problem with a high impact on the effectiveness and efficiency of business processes.
\luise{A wide range of system-initiated \fabian{(largely automated)} resource allocation approaches were developed during the last decades. 
\ingo{In this study, we present a comprehensive overview of this field, by discussing the results of 61 primary studies. We identified the primary studies through a rigorous structured literature review, covering publications from 1995 to 2023.}
\fabian{We report} on which allocation capabilities and goals are supported, \fabian{the use of} process models, execution data, task and resource attributes, the type of algorithmic solution, and \fabian{evaluation methods}. 
In the review, we mainly observed approaches supporting 1-to-1 allocation, process-oriented goals, \fabian{the use of} process models, and rule-based approaches.
\fabian{Based on our results, we envision future research to explore data-driven and context-adaptable solutions and
contribute to a better understanding of approaches' performance impact.}
}

\end{biseabstract}

\begin{bisekeywords}
business process, resource management, resource allocation, optimization
\end{bisekeywords}

\begin{biseacknowledgements}
%I'd like to say thanks to... Please add the acknowledgements \textbf{only in the final version after acceptance} in case they would identify or hint towards the authors.
\end{biseacknowledgements}

% In setting up this template for *BISE* papers, we've used both
% the \section* command and the \paragraph* command for topical
% divisions. Either way, use the asterisk (*) modifier, as shown, to

\section{Introduction}

Business Process Management (BPM) is applied by organizations to run their operations effectively and efficiently. 
It is a management paradigm that positions business processes, which consist of a set of connected activities
carried out to reach a specific business goal~\citep{DBLP:books/sp/Weske19}, in the center of its efforts to
facilitate operational excellence and continuous improvement~\citep{dumas2018fundamentals}. 
For the successful execution of business processes, organizations need a rich set of internal resources, such as human resources, machines, vehicles, materials etc.~\citep{huang2011reinforcement,cabanillas2016process}.
Often, resources are valuable assets, frequently costly and limited~\citep{arias2018human}. \fabian{Thus, }
%, such that
the success of a business process is closely intertwined with the efficient use of resources.
\luise{
\emph{Resource allocation} aims at ensuring that each activity of a particular process case (i.e., a task) is executed at the right time and with the right resources~\citep{kumar2002dynamic,cabanillas2016process}.
A specific characteristic of business processes is that their focus is typically not on a single resource or \luise{activity} but on coordinating \fabian{various} activities to reach a business goal\fabian{:} constraints regarding the order of process activities need to be reflected \fabian{during} allocation~\citep{xu2013incorporating}.} 

\luise{However, in general resources are not only involved \fabian{or} required in one single process case of a single business process, but in several cases of several different business processes, which may run concurrently~\citep{zhao2016entropy}.
Consider the case of a physician who has to diagnose and treat different patients during the day. The physician is not only involved in the treatment processes but also in the reimbursement processes of the hospital or clinic, in which patients or their respective health insurers get invoiced.
By not only taking into account the resource needs for one individual task but for the set of tasks to be executed, \emph{resource allocation} is a multi-objective decision-making problem~\citep{zhao2017resource} that has to handle ``the allocation of limited resources among competing [tasks]''~\citep{luss2012equitable}.} 
\fabian{Additionally,} each business process has certain time, cost, and quality goals that must be considered~\citep{xu2013incorporating}.
%while allocating process tasks to resources~\citep{xu2013incorporating}.

\luise{%Due to the importance and complexity of resource allocation problems in business processes, many approaches have been developed to provide automatic support.
Resource allocation can be done manually by a human expert or by an IT system that proposes or enforces a resource allocation, which we call system-initiated resource allocation.
Traditionally, in the BPM domain, requirements for the needed resources are specified per activity as soon as a process model is planned to be executed \citep{dumas2018fundamentals}. Business Process Management Systems (BPMSs) often support the allocation of tasks to resources with allocation rules, \fabian{which were} structurally studied by \citet{russell2005workflow}.
Additionally, operations research offers a broad range of optimization approaches for resource allocation in different application areas~\citep{kamrani2012framework}. Numerous studies have emerged that have built on those, adapting them to the needs of business processes and integrating them into BPMS\fabian{s}~\citep{xie2016dynamic,arias2018all}. Additionally, \fabian{increased} computational power allows the \fabian{use} of meta-heuristics, \fabian{like} genetic algorithms (e.g.,~\citep{huang2012task}) and \fabian{machine learning approaches, leveraging increasing amounts of process execution data. Among others, these developments have led to a broad spectrum of studies exploring system-initiated (largely automated) resource allocation approaches for business processes.} These approaches might not necessarily use an executable process model as an artifact.}

\fabian{The variety of approaches and techniques for resource allocation in business processes renders a systematic survey valuable for researchers and practitioners alike.} 
Existing literature studies in the area of resource allocations in business processes focus on resource management more generally and have studied fewer works~\citep{cabanillas2016process} or did not discuss in detail the targeted resource allocation problems and solution approaches~\citep{arias2018human}.

\fabian{In this study, we aim to fill this gap. We present a comprehensive overview of the field, discussing 61 identified studies and reporting which allocation capabilities and goals are supported. We discuss their use of process models, execution data, task and resource attributes, the type of algorithmic solution, and evaluation methods. Our various classifications can help to gain insights into the design of such approaches and help to identify trends and research gaps. For practitioners, they can guide the design of future practical implementations. %The focus of our study lies on approaches that have been developed in addition to the traditional BPMS resource allocation rules~\citep{russell2005workflow}.
}

\fabian{We have followed a methodology for rigorous and replicable systematic literature reviews (SLRs) \citep{kitchenham2004procedures,okoli2010guide}. }
%By searching four databases and with the help of a backward and forward search, we have identified \luise{61} relevant studies in this review. Resulting from various countries, they were mainly published within the last ten years. 
\fabian{We found that most of the identified resource allocation approaches assign one task to precisely one resource; approaches supporting many-to-1 and 1-to-many were \fabian{less prevalent}. The identified approaches often have process-oriented resource allocation goals, such as minimizing process costs. The SLR offers insights into the different applied solution techniques and finds that mainly rules, heuristics and machine learning approaches are used. Lastly, the review reveals that many studies lack replicability.}

\fabian{In Section~\ref{sec:background}, we first establish the background on resource allocation in business processes. Then, }
%In the remainder of this paper, first, a background on resource allocation in business processes is given in Section~\ref{sec:background}, and 
the related work is presented in Section~\ref{sec:relatedwork}.
\ingo{Subsequently, }
%Then, 
the applied research method---the SLR protocol---is explained in \fabian{detail in} Section~\ref{sec:method}.
The \fabian{identified} studies of the SLR are \ingo{summarized, analyzed, and discussed} in Section~\ref{sec:result}.
\fabian{In Section~\ref{sec:dis}}, \ingo{the implications of our findings} are discussed, along with open research fields and limitations of the research. Finally, \fabian{Section~\ref{sec:conclusion} concludes the paper.}

%%%%%%%%%%%%%%%%%%%%%%%%%%%%%%%%%%%%%%%%%%%%%%%%%%%%%%%%%%%%%%%%%%%%%%%%%%%%%%%%%%%%%%%%%%%%%%%
\section{Background}
\label{sec:background}

To achieve its business goals, an organization runs a set of business processes that are executed with the help of different internal resources, such as humans, IT services, or machines. The following section presents the essential concepts of business processes, resources, and their allocation.

\subsection{Business Process}
\label{subsec:process_def}

\begin{figure}[!ht]
	\centering
	\includegraphics[width=0.9\textwidth]{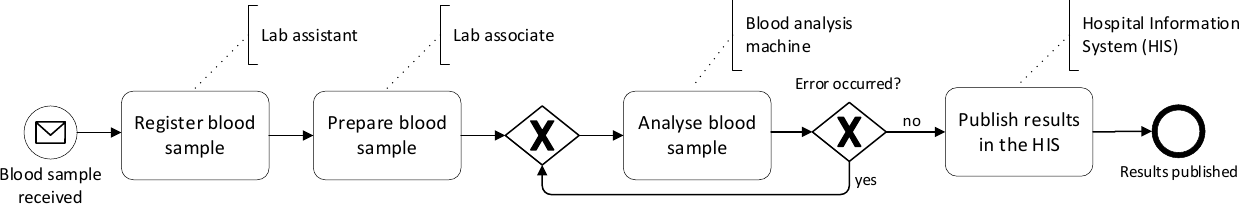}
	\caption{Laboratory process of a hospital given as BPMN process diagram, with resource information in activity annotations.} 
	\label{fig:BPMNdiagram}
\end{figure}

\paragraph{Process model.}``A business process consists of a set of activities that are executed in coordination in an organizational and technical environment to reach a specific business goal''~\citep[Chapt. 1]{DBLP:books/sp/Weske19}. For managing the documentation, redesign, execution, monitoring, and analysis of a business process, often process models are used as a formal representation of a business process. An example of a business process model from healthcare is shown in Figure~\ref{fig:BPMNdiagram}, represented in the industry standard Business Process Model and Notation \fabian{(BPMN)}~\citep{BPMN}, specifically as a ``process diagram''. The shown laboratory process describes handling a blood sample: once received, the sample is registered, prepared, and then analyzed by the blood analysis machine. The results are published in the hospital information system if no errors occur.
\ingo{Apart from BPMN, many process modeling languages have been proposed over time. Prominent examples include} \fabian{Event-driven Process Chains (EPCs)~\citep{scheer2005process}, Petri net~\citep{van1998application}---as a rather formal specification language---or Declare~\citep{van2009declarative} as a declarative modeling language.}

\ifLong{Once a blood sample has been received, the process starts, which is represented by a message start event in the process model. Next, a lab assistant registers the blood sample, followed by a lab associate preparing it for analysis. The blood analysis machine now analyses the blood sample. If no error occurs, the result is uploaded to the hospital information system, and the process ends.
In case of an error, the analysis has to be repeated. An XOR split gateway represents this. Only one of its outgoing paths is triggered (i.e., the path with the condition "no" or "yes"). 
The XOR join gateway between the activities \emph{Prepare blood sample} and \emph{Analyse blood sample} awaits one incoming flow, and as soon it is received, it triggers the next activity. BPMN provides additional gateways to realize different control-flow behavior, such as a parallel gateway for the concurrent execution of activities or an OR gateway that can trigger a subset of paths if certain conditions are fulfilled.}

\begin{comment}
Formally, a process model is defined as follows:
\begin{definition}[\textbf{Business Process Model}]
A business process model $m = (N,CF)$ consists of a finite non-empty set $N = A \cup E \cup G $ of control flow nodes being activities A, events E, and gateways G (A, E, and G are pairwise disjoint) and a control flow relation $CF \subseteq N \times N$ such that $(N, CF)$ is a connected graph.
\end{definition}
\label{def:BP}
\end{comment}

%For specifying a process model, different process modeling languages are available, such as the BPMN standard~\citep{BPMN} specified by the Object Management Group, Event-driven Process Chains (EPCs)~\citep{scheer2005process}, Petri net~\citep{van1998application} as a rather formal specification language or Declare~\citep{van2009declarative} as a declarative modeling language.

\paragraph{Process and activity instances.} A business process model acts as a blueprint for a set of business process instances. 
A process instance represents the concrete execution of a business process case.
For a concrete process instance, activity instances are initiated during runtime, which need time and resource(s) to be executed.
In this paper, we call them tasks\footnote{Please note that this definition of a \emph{task} is different from the BPMN standard~\citep{BPMN}, in which it is defined as a single unit of work in a process diagram. In contrast, as does part of the literature, we use \emph{task} as a shorthand for \textit{activity instance}. % of a process instance.
}.
Each task follows a specific \emph{lifecycle}~\citep[Chapt. 3]{DBLP:books/sp/Weske19}\footnote{\luise{Please be aware that this is an abstracted version of the lifecycle of a task. A more detailed lifecycle is provided by \citet{russell2005workflow}, where it is also considered that tasks can be \emph{offered} to resources, who then can decide on the allocation and start on their own. We will discuss later in Sect.~\ref{subsec:resources} how this is realized.}}: After its initialization, it is ready for execution as soon as its incoming control flow edges have been triggered. Then, it is allocated to one or more resources. At a certain point, the allocated resource(s) start(s) it, and after the work has been done, the task is completed. Furthermore, each task has specific characteristics, so-called task data (e.g., due date), based on which the resource allocation decision can be made.

\paragraph{Process performance.} 
Reaching its business goals is typically essential for an organization, and as per the above definition, business processes are the activities that the organization undertakes to achieve its goals. Thus, organizational success depends strongly on effective processes.
%to reach the business goals. 
% Reaching a business goal with its business processes is typically the main objective for an organization. i.e., the processes should be effective in reaching their goals. 
In addition, a process should be efficient in its execution. According to~\citet[Chapt. 2]{dumas2018fundamentals}, four main process performance dimensions can be distinguished\fabian{: }time, cost, quality, and flexibility.
% NOT CLEAR: -- that refined into specific process performance measures. 
In the following paragraph, we sketch sample measures for each dimension and refer the interested reader to~\citep[Chapt. 2,7,8]{dumas2018fundamentals} for more details on process performance measures.

A relevant measure for the \emph{time} dimension is the cycle time, which measures the time between the start and end of a process instance (i.e., processing times plus waiting times between activities). \ifLong{Additionally, the processing time for individual activities can be measured, and the waiting time between activities.}
Regarding the \emph{cost} dimension, the goal is usually to minimize the costs. The operational process costs can be calculated by measuring the time used to execute a task and multiplying it by the cost per time unit of the resource.
Resource utilization is the time a resource needs to execute process tasks divided by its available working time. This measure can be used to evaluate whether process costs can be further decreased by increasing resource utilization.
Process \emph{quality} can be identified by calculating, for instance, the rate of successfully executed process instances.
Process \emph{flexibility} is the capability to react to changes in the process execution, e.g., the capability to react to missing resources.

\begin{figure}[tb]
	\centering
	\includegraphics[width=0.6\textwidth]{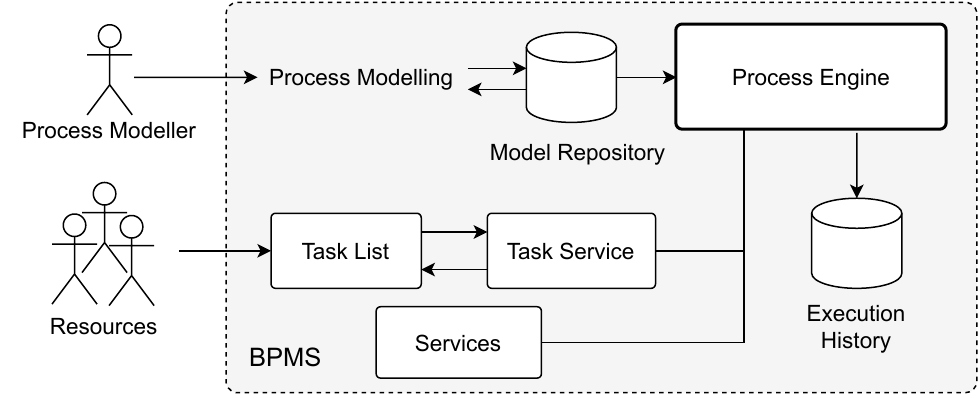}
	\caption{Architecture of a BPMS \luise{(based on \citep[Chapt.8]{DBLP:books/sp/Weske19})}, 
 including its task service and task list of resources.} 
	\label{fig:BPMS}
\end{figure}

\begin{table}[b]
\caption{Illustration of the event log structure}	
\center

\begin{tabular}{c c c c c}
	\textbf{case id} & \textbf{timestamp} & \textbf{task} & \textbf{lifecycle state} & \textbf{resource} \\
	\hline
	 ... & ... & ... & ... & ... \\
	 9845 & 14/01/2021 11:22 & Register blood sample & start & Lab assistant \\
	 9845 & 14/01/2021 11:25 & Register blood sample & complete & Lab assistant \\
	 9852 & 14/01/2021 11:26 & Register blood sample & start & Lab assistant \\
	 9852 & 14/01/2021 11:30 & Register blood sample & complete & Lab assistant \\
	 9845 & 14/01/2021 11:26 & Prepare blood sample & start & Lab associate \\
	 ... & ... & ... & ... & ... \\
\end{tabular}
\label{table:el_struc}

\end{table}

\paragraph{Process execution and data.} In the BPM domain, Business Process Management Systems (BPMSs) have been designed to ``control the execution of business processes instances, according to the logic defined in the respective process model.''~\citep{duran2019rewriting}.
As shown in the architectural model in Figure~\ref{fig:BPMS}, a BPMS consists of a process modeling and a process engine component connected to several services~\citep{DBLP:books/sp/Weske19,dumas2018fundamentals}. Business processes can be modeled with the process modeling component and stored in the process model repository. Process models can be deployed on the process engine. Then, process instances can be started in the process engine and are executed according to the behavior specified in the deployed process model.
When business processes are supported in their execution by an information system or a BPMS, then process execution data is stored, from which event logs can be generated~\citep{remy2020event}.
Event logs are useful for evidence-based process analysis with the help of process mining techniques, such as the automated discovery of process models~\citep{van_der_aalst_process_2016}.
An event log consists of a series of events for different process instances reflecting, e.g., the start or completion of a task. 
An excerpt of the event log from the laboratory process shown in  Figure~\ref{fig:BPMNdiagram} is represented in Table~\ref{table:el_struc}.
\\
Each event has a case id (e.g., \texttt{9845}) as a reference to a process instance, a timestamp (e.g., \texttt{14/01/2021 11:22}) to denote when the event has occurred, the executed activity (e.g., \texttt{Register blood sample}) and its lifecycle state (e.g., \texttt{start}). Additional other data can be given, such as resource information (e.g., \texttt{Lab assistant}) \luise{that can be useful also for the analysis of resource behavior}.
\ifLong{Event logs can be, for example, imported into the process mining toolkit \emph{ProM}, which offers a broad range of process mining techniques for discovery, conformance, and process analysis~\citep{kalenkova2014discovering}. }

\subsection{Resources and their Allocation in Business Processes}
\label{subsec:resources}
Resources are the fundamental basis for organizations to execute the tasks necessary to reach the goals of business processes. 
In the literature, the definitions for concepts like \textit{resource} vary according to the background and goals of the research. 
\ingo{The semi-formal definitions below provide clarity regarding the meaning of these terms for the context of this paper. To start, we}
define\footnote{The definition is based on a broad range of different definitions provided~\citep{arias2018all,doerner2006enriched,rhee2010increasing,xie2019integration,xu2013incorporating,yaghoibi2017cycle,zhao2016entropy,zhou2008project,bellaaj2017obstacle,xu2009resource,bussler1995policy,ihde2022framework,erasmus2018method,cabanillas2013priority}.} a resource \fabian{as} anything necessary to execute tasks (e.g., human, vehicle, software, tools) in the context of a business process:

\begin{definition}[\textbf{Resource}]
A resource $r$ is an entity required to execute a process task. It has a set of attributes describing its capabilities, capacities, and availability at a given point in time $t$;
those values might change during runtime. $R$ is the set of all resources.
\end{definition}
\label{def:Res}

\luise{\citet{ouyang2010modelling} distinguish \emph{human} and \emph{non-human} resources. Non-human resources are subdivided into \emph{application} and \emph{non-application} resources. An application resource describes anything that can execute tasks, such as a software service. In contrast, non-application resources, e.g., a transport vehicle, can be used by human or application resources to execute a task. Non-application resources can be further sub-divided into \emph{consumable} (e.g., machine oil) or \emph{durable} (e.g., surgical instruments) resources.}

\paragraph{Resource allocation.} Resource allocation\footnote{\luise{In operations research, the term \emph{resource assignment} is also used for \emph{resource allocation}~\citep{kamrani2012framework}. In contrast, in business process management, \emph{resource assignment} defines the specification of resource requirements for process activities at design time~\citep{cabanillas2016process}. In this work, we refer to this as resource allocation constraints.}} is a key part of resource management in business processes~\citep{cabanillas2016process}.
Its goal is assigning a process task to the most appropriate \ingo{resource(s)} among the available resources \luise{at runtime or planned for runtime~\citep{kamrani2012framework}}, as shown in Figure~\ref{fig:OverviewPreliminaries}.
Resource allocation\footnote{We talk about \emph{(task) scheduling} \citep{slack2018operations} when the solution of a resource allocation is a concrete schedule, including the information on which task is worked by which resource at which time.
Usually, the resource allocation \luise{happens before the} task execution, \luise{see the above discussion of the lifecycle of a task}. If the resource allocation is some time ahead (e.g., a day before the task execution), it is often called \emph{planning}. A resource allocation plan might require an adaptation later in contrast to a real-time resource allocation.} in processes is a complex decision problem that \ingo{should} consider the following aspects:
\begin{itemize}
    \item Multiple business processes and process cases \ingo{may} run concurrently, such that conflicting requests for the same resource
    have to be resolved and imbalances between the workload of resources \ingo{should be} handled~\citep{zhao2016entropy}.
    \item For a rational resource allocation, the performance goals of business processes, such as minimizing time or costs, \ingo{should} be considered in the resource allocation~\citep{zhao2016entropy}. 
    % \luise{The goal can also be unspecified.} 
    \item Often, process activities require the allocated resources to possess certain competencies, capabilities, or rights, which \ingo{should} be considered as constraints during resource allocation~\citep{arias2018human} \luise{(in many cases referred to as resource assignment~\citep{cabanillas2016process} in the BPM domain)}. 
    % \luise{The constraints can also be unspecified.}
    \item In addition, resource allocation in business processes is not only about optimizing a single activity but should take a set of activities and the process into account~\citep{xu2013incorporating}.
\end{itemize}

\begin{figure}[tb]
	\centering
	\includegraphics[width=0.65\textwidth]{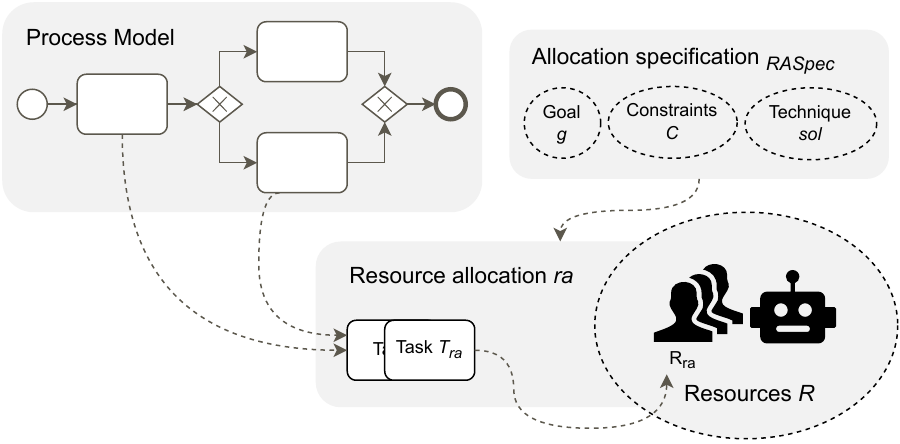}
	\caption{An organization with its business processes and resources where tasks resulting from process executions need to be allocated to one or more resources. \ingo{Note that goals, constraints, etc.\ may be left unspecified in a given setting.}} 
	\label{fig:OverviewPreliminaries}
\end{figure}

\begin{definition}[\textbf{Resource Allocation}]
\label{def:RA}
\fabian{More formally, let} $R$ be a set of resources and let $T$ be a set of tasks that are ready for execution\ingo{, which may span different processes and instances}. Furthermore, let $R_{ra} \in 2^R$ be the set of available resources and $t \in T$. Then, resource allocation is a function
$$ra: 2^R \times T \mapsto R$$
such that $ra(R_{ra},t) = r$ indicates that a resource $r$ is allocated for task $t$ in a state where resources in $R_{ra}$ are available.
To allocate several resources to a given task, the function $ra'$ is defined as follows:
$$ra': 2^R \times T \mapsto \mathcal{P}(R)$$
such that $ra'(R_{ra},t) = R_a$ indicates that a set of resources $R_a \in \mathcal{P}(R)$ is allocated for task $t$ in a state where resources in $R_{ra} \in 2^R$ are available. 
To allocate a resource to a set of tasks, $ra''$ is defined as follows:
$$ra'': 2^R \times 2^T \mapsto R$$
such that $ra''(R_{ra},T_r) = r$ indicates that a resource $r$ is allocated to a set $T_r$ of tasks in the ready state. Again, resources in $R_{ra}$ are available. 
\end{definition}

\ingo{According to Definition~\ref{def:RA}, a task can be assigned to several resources, and several tasks to a resource}. Traditionally, in BPM, 1-to-1 mappings are often supported. In this literature review, we want to analyze how frequently (if at all) the more complex resource allocation capabilities are supported. %, such as one resource being allocated to multiple tasks.

A resource allocation decision has to be made in a dynamic system where new tasks arrive, resources complete tasks, and the availability of resources is changing. Thus, in addition to the process model, the system dynamics must also be considered during resource allocation. Resource allocation approaches often capture these dynamics with distributions, e.g., the inter-arrival time of tasks, the duration of the business process activities, probabilities of executing specific process paths (i.e., branching probabilities), and a model to capture the availability of resources. In this work, we want to analyze which approaches use which kind of information regarding the system dynamics in addition to the process model.

\paragraph{\luise{Realization of a resource allocation.}} \luise{Resource allocation can be a manual effort in an organization, where a human being assigns tasks to qualified resources or the staff members select tasks independently from a shared task list.
 Supported by a BPMS, its task service handles its resource allocation if an activity is ready for execution~\citep{DBLP:books/sp/Weske19}.} The task service adds the task to at least one task list of a resource from where the resource can start and complete its execution (cf., Figure~\ref{fig:BPMS})~\citep{DBLP:books/sp/Weske19}.
\luise{One simple but often used task allocation pattern in existing BPMS\fabian{s} is the \emph{role-based distribution}~\citep{russell2005workflow}}, where the task is added to the task lists of all workers with a specific role. A role is a grouping mechanism for resources with similar capabilities and responsibilities, such as \emph{lab assistant}. A resource can be assigned to one or multiple roles.
\luise{Analyzing a comprehensive range of BPMSs, ~\citet{russell2005workflow} differentiate between pull and push patterns.
%that they support. 
Pull patterns describe situations where the resource has been made aware of a task by the system but initiates the execution itself, such as the \emph{resource-initiated execution} patterns. In contrast, the push patterns describe situations where the system actively offers or allocates a task to a resource (e.g., \emph{round robin allocation}, allocating tasks to available resources in a cyclic order).}
\\
In operations research, the problem of allocating a resource to tasks has a long tradition and is known as the \emph{Assignment Problem}; it has been discussed in different versions~\citep{kamrani2012framework}, for which also several solution techniques have been proposed \emph{to provide a more optimized allocation}. In contrast to BPM, the focus is often on a single activity or workstation. The overall business process with control flow relations between the process activities is rarely considered.
\luise{Over the last decades, several approaches have been developed \fabian{in addition} to the existing resource patterns to support the task service of a BPMS or any IT system supporting the process execution and to allow a system-initiated 
resource allocation approach for business processes, \fabian{using also} insights from operations research. In this paper, we want to investigate the research studies in this area structurally.
As a system has to know the determining factors for an allocation, we introduce the resource allocation specification in the next paragraph.}
%Having discussed the basic problem of resource allocation, we focus on the quality of allocations next.

\paragraph{Resource allocation specification.} As mentioned before, resource allocation is subject to performance measures and goals\fabian{---like} minimizing cost\fabian{---}that it should optimize for. Hence, a concrete resource allocation specification \luise{can guide a system-initiated resource allocation.}

%is needed to define how the resource allocation should be optimized.
In particular, a resource allocation specification consists of parameters that influence a resource allocation by providing an optimization goal, concrete limitations that should be fulfilled, and a technique for solving the allocation.

\begin{definition}[\textbf{Resource Allocation Specification}]
\label{def:RAS}
\fabian{More formally, }a resource allocation specification $RA_{Spec} = (g,C,sol)$ consists of 
\begin{itemize}
    \item a weighted allocation goal $g = \{g_1*w_1,g_2*w_2,...,g_n*w_n\}$ with  sub-goals $g_i$ and their respective weight $w_i \geq 0$ that the resource allocation should optimize for; notice that $\sum_{i=1}^{n} w_i =1$,
    \item a set of constraints $C$ of the resources and tasks, where we can distinguish here between hard constraints that need to be upheld and soft constraints \ingo{which may be associated} with a penalty~\citep{kumar2002dynamic}, and
    \item a concrete solution technique $sol$ to \ingo{make allocation decisions; this may also be manual}.
\end{itemize}
\end{definition}

\paragraph{Solution Quality.}
Multiple feasible solutions might exist for a particular business process scenario. \ingo{We call a solution \textit{feasible} if it satisfies all constraints of the resource allocation specification.} Given a goal evaluation function $e$ and all available information at one point in time $t$, optimization methods \ingo{typically aim} to find the best solution \luise{--the \emph{global optimum}--} among the feasible ones, i.e., the one that maximizes (or minimizes) $e$. 
\luise{Relevant information is here the resource allocation specification $RA_{Spec}$, the available resources $R_{ra}$ including their attributes, and the tasks $T$ being ready for execution including their attributes, which can also be from different processes and instances.}
Naturally, finding a global optimum is always in the best interest. 
Nonetheless, in reality, a balance between the effort and time to find a resource allocation solution 
%in contrast to 
\fabian{and} the quality of the solution has to be found. If the solution quality is high, but the time to produce it exceeds the limits of the underlying business process, then the high solution quality might not be relevant anymore.
Thus, in practice, it is often the case that a limited number of solutions are checked. 
\luise{In situations where only a limited subset of solutions is explored, there is a possibility that the best solution found is not the global optimum but rather a \emph{local optimum}. A local optimum is a solution that is the best within a specific neighborhood or region of the solution space but is not necessarily the absolute best solution across the entire solution space.}
%Thus leading to the case that the global optimum might not have been considered at all. We refer to the solution as a local optimum in such a case.
\fabian{More formally, we} define the solution quality as follows:
\ingo{\begin{definition}[\textbf{Solution and Solution Quality}]
\label{def:SQ}
Let $S$ be the set of feasible solutions, where a solution $s \in S$ comprises a set of resource allocations that jointly cover all enabled tasks: $s = \{ra(R_{ra},t) ~ | ~ t \in T\}$, where $T$ is the set of enabled tasks and $R_{ra}$ the set of available resources as above.
Let $s_{max} \in S$ be an optimal solution. 
Let $e: S \rightarrow \mathbb{R}\subseteq [0,1]$ be a function that evaluates a given solution, such that $\forall s_i \in S: e(s_i) \leq e(s_{max})$.
The solution quality is a function $q: S \rightarrow \mathbb{R}\subseteq [0,1]$ that quantifies how close a specifically feasible solution $s \in S$ is to 
the optimal solution $s_{max}$, given the evaluation function $e$ for the optimization.
% the optimum solution $s_{max}$ given the goal function $g$ that was supposed to be optimized for.
$$
q(s) = \tfrac{e(s)}{e(s_{max})} = 
\begin{cases}
1, \text{~ ~ ~ if $s$ is a global optimum}\\
< 1, \text{~ otherwise}
% < 1, \text{if $s$ is a local optimum}
\end{cases}
$$
For $s' \not\in S$ we define $q(s) = 0$, in other words, infeasible solutions are assigned minimal quality.
\end{definition}}

%\begin{definition}[\textbf{Solution Quality}]
%\label{def:SQ}
%Let $RA$ be the set of feasible resource allocations, while %$ra_{max} \in RA$ is an optimal solution. Let $e: RA \rightarrow %\mathbb{R}$ be the function that evaluates a given solution, such that $\forall ra_i \in RA: e(ra_i) \leq e(ra_{max})$.
%The solution quality is a function $q: RA \rightarrow \mathbb{R}$ that quantifies how close a specifically feasible resource allocation $ra \in RA$ is to 
%the optimal solution $ra_{max}$, given the evaluation function $e$ for the optimization.
% the optimum solution $ra_{max}$ given the goal function $g$ that was supposed to be optimized for.
%$$
%q(ra) = \tfrac{e(ra)}{e(ra_{max})} = 
%\begin{cases}
%1, \text{~ ~ ~ if $ra$ is a global optimum}\\
%< 1, \text{~ otherwise}
% < 1, \text{if $ra$ is a local optimum}
%\end{cases}
%$$
%\end{definition}

%\bigskip \bigskip \noindent
%Alternative approach to solution quality formalization:
% We define a solution $s \in S$, where $S$ is the set of feasible solutions, as a set of resource allocations that cover all enabled tasks: $s = \{ra(R_{ra},t) ~ | ~ t \in T\}$, where $T$ is the set of enabled tasks and $R_{ra}$ the set of available resources as before.
% Let $e: S \rightarrow \mathbb{R} \subseteq [0,1]$

%With this quality function, it can be decided how close a given optimized resource allocation is to the optimal resource allocation $ra_{max}$ and to distinguish between a global and local optimum. This SLR wants to analyze how different approaches target this challenge.
One approach to reducing the effort in finding the optimal solution is focusing on a single task \fabian{for} the allocation decision~\citep{xu2016resource}. As an organization's current set of to-be-executed tasks carries different levels of importance, focusing on a single task can lead to a local optimum solution. If an allocation approach considers all to-be-executed tasks, then the approach tries to reach a global optimum. Naturally, this refers to the decision at that point in time $t$.
This SLR wants to analyze how different approaches target this challenge.
%This means, it is possible that new information might appear in the future of the process execution, that renders a given solution as not the best decision in hindsight. However, as it is not easy to predict the future, if at all possible given the scenario, in this SLR we want to set the focus on analyzing whether approaches follow a single- or multi-task optimization.

\section{Related Work}
\label{sec:relatedwork}

This section presents and discusses existing BPM surveys and surveys on resource allocation in business processes.
A first comprehensive study on the use and representation of (primarily human) resources in existing BPMSs was given by \citet{russell2005workflow}.
\luise{Studying different existing process modeling languages and BPMSs, the authors provide a resource meta-model, a task lifecycle, and a collection of patterns to create, pull, push, and detour tasks to resources that help to set up a task service in a BPMS.
However, the resource patterns do not focus on the optimization problem associated with resource allocation\footnote{The resource patterns do not have an inherent allocation goal association. Therefore, it falls upon the process designer to evaluate whether a chosen set of patterns aligns with the allocation goal for a given business process. Additionally, the resource patterns predominantly adopt a localized perspective, typically focusing on allocating an individual task. An exception is here the pattern \emph{system-determined work queue content} as it provides the possibility that the system sorts the tasks, for example, based on their priority in the working list.}. 
An extension of the resource patterns \fabian{for} teams collaboratively working on tasks is presented by \citet{van2001reference}. It discusses aspects that must be considered when teams work on tasks and presents an architecture to integrate groupware, such as a shared calendar, with BPMSs.}
\luise{The focus of the resource patterns is mainly on what is supported by modeling languages and BPMSs, whereas the subject of this study is the state-of-the-art in research.}
%Since the publication of resource patterns, numerous research studies have emerged that 

%\ingo{---}\fabian{the subject of this study.}} 

In a comprehensive survey, \citet{van2013business} provides typical BPM research use cases. Resources and their allocation do not play a central role in the presented use cases. Still, it is mentioned that resource information can be discovered from event logs and enhanced process models, that BPMSs have to handle resources, and that studies that deal with optimal resource allocations were observed.

\citet{cabanillas2016process} \fabian{examines research works on resource handling in process- and resource-oriented systems. They categorize the works into:} 
\begin{enumerate*}
    \item resource assignment (i.e., the definition of resource requirements for process activities at design time),
    \item resource allocation (i.e., designation of concrete resources to a specific task during runtime) and
    \item resource analysis (i.e., evaluation of process execution with a focus on resources).
\end{enumerate*}
The author did not intend to present an exhaustive literature review but a framework with representative works, so only a select set of studies \fabian{were explored}.

\citet{arias2018human} provide a systematic mapping study on human resource allocation in business process management and process mining. The focus of the study is on human resources and reporting on  \fabian{research methodology (e.g., type of study) and where and when the studies were published. It does not discuss the identified solutions in detail.}

%where and when the studies were published, but not discussing their solved problems, etc. 
\ifLong{Further, it is limited to studies that the authors could access at their university.}
A survey on automated planning and BPM \citep{marrella2019automated} shows that artificial intelligence can leverage particular challenges in business process management, such as the automatic generation of process models, allowing process adaptations, or enabling conformance checking. It gives a concrete method for building planning problems. Although resources play an indirect role, resource allocation is not discussed as an application area.

\luise{In their structured literature review, \citet{yari2022survey} examine process-aware recommender systems that utilize event logs as their input. The authors categorize these systems based on whether they recommend the next activity or the next resource. They delve into the used recommendation approaches (e.g., data mining, process minings, statistical or hybrid approaches), the implementation environments, and the evaluation method. However, the primary focus of their review centers on the recommendation systems themselves, with limited coverage of the mechanisms for resource allocation. The authors do not investigate the optimization objectives of these approaches, the specifics of their input data, or the roles of process models and the execution data.}

% our advantages: 
%Based on the related work, we can observe that a comprehensive analysis of the research works in the context of resource allocation in business processes is still missing. 
\luise{\ingo{In contrast to prior literature reviews}, we aim to deliver a comprehensive analysis of research studies enabling a system-initiated resource allocation in the business process through an exhaustive literature review. We want to provide a structured comparison of the studies' proposed approaches concerning their optimization objectives, capabilities, utilization of process models and execution data, input data sources, and applied solution techniques.} 

%the resource allocation in business processes as an optimization problem. This work aims to deliver a comprehensive analysis through an exhaustive literature review, including a structured comparison of the identified studies with regard to their optimization objectives, capabilities, utilization of process models and execution data, input data sources, and applied solution techniques.}
%in addition to related work

%So far, no systematic review exists that represents the goals, capabilities, solution techniques, and maturity with regards to applicability and evaluation of existing resource allocation approaches and compares them in a structured way.

\section{Research Method}
\label{sec:method}

A Systematic Literature Review (SLR)~\citep{kitchenham2004procedures,okoli2010guide} allows the identification and analysis of relevant and existing literature related to specific research questions while aiming to minimize bias and maximize reproducibility.
This section describes the search protocol followed in the SLR.
First, the addressed research questions are presented in Section~\ref{subsec:researchquestions}.
The data sources and the search strings for the primary search, the backward and forward search, the relevance checks, and the data extraction scheme are described in Section~\ref{subsec:primarysearch}.  
%The resulting studies were read, and relevant information was extracted according to a predefined data scheme, which is also given in Section~\ref{subsec:primarysearch}.

\subsection{Research Questions}
\label{subsec:researchquestions}
This study aims to answer the \fabian{overarching} question \textit{"What is the state-of-the-art of \luise{system-initiated resource allocation approaches for business processes}?"}. %automatic approaches supporting resource allocation in business processes
We divided this general question into four more concrete sub-questions \fabian{to support more fine-grained analysis:}
%to investigate and describe the existing research:
\begin{itemize}
    \item[RQ1] \textit{What are the targeted resource allocation goals and capabilities?}
    \luise{When an automatic approach is responsible for the resource allocation in business processes, it usually needs an allocation goal as per Definition~\ref{def:RAS}. Furthermore, approaches can support traditional 1-to-1 \fabian{allocations of tasks to} resources \fabian{but also other allocation capabilities (see Definition~\ref{def:RA}).}
    %or also more, as given in Definition~\ref{def:RA} -- the allocation capability.
    To help practitioners and researchers in identifying allocation approaches for a particular allocation problem with a specific target or an allocation capability, }an overview of the approaches' different allocation goals and their capabilities \luise{is needed}.  
    %\luise{As defined in Definition~\ref{def:RAS}}, a resource allocation usually targets an optimization goal, e.g., minimizing the cycle time, applicable in different use cases. In response to this RQ, we aimed to generate an overview of the approaches' different resource allocation goals and resource allocation capabilities \luise{(cf. Definition~\ref{def:RA})}: Which approaches support traditional 1-to-1 mappings, and what else is supported?
    
    \item[RQ2] \textit{What \ingo{are the respective roles} of process models and process execution data in the resource allocation approach?}
    Resource allocation in business processes is unique because the \ingo{ordering constraints between} process activities needs to be considered~\citep{zhao2017resource} \luise{in the resource allocation specification (cf. Definition~\ref{def:RAS}).} 
    With this question, we wanted to understand \ingo{if and how information from process models or execution data is encoded and considered in the identified approaches}.
    
    \item[RQ3] \textit{Which input data regarding resources and tasks do the different resource allocation approaches use?}
    \luise{As outlined in Definition~\ref{def:RAS}, a resource allocation specification encompasses the incorporation of constraints. In defining these constraints, one typically leverages resource and task-related information additionally to the ordering constraints of business processes (see RQ2) to identify a useful resource allocation. To understand the broad spectrum of possibilities, we want to extract the input information that is used by the approaches for constraining the resource problem. \citet{arias2017towards} derived a taxonomy of resource attributes applied by resource allocation approaches in BPM. When addressing this research question, we \fabian{make use of} this taxonomy to understand and classify the input data used in the identified studies and extend it \fabian{by} task attributes.
    } 
    
    \item[RQ4]  \textit{Which solution strategies are used?}
    \luise{Additional to the goals and constraints, a solution technique is \ingo{applied} for solving a resource allocation problem (cf. Definition~\ref{def:RAS}).} With this research question, we want to understand the types of solution approaches used for resource allocation for business processes, such as rules, heuristics, and machine learning.
    %We want to analyze whether techniques from other disciplines are used and how they have to be transformed to be applicable to business processes.
    Furthermore, we want to \luise{investigate the targeted solution quality of the approaches (cf. Definition~\ref{def:SQ}) and want to focus here on} whether approaches follow a single-task optimization \ingo{in the context of individual}  process cases or a multi-task optimization, where the \ingo{whole} set of to-be-executed process tasks is considered. %, possibly with different levels of importance.

    \item[RQ5]  \textit{How applicable are the proposed resource allocation approaches, \ingo{given the} availability of evaluations and prototypes?}
    \ingo{The applicability of an allocation approach in real-world settings depends on whether it has been evaluated and whether a corresponding prototype has been made available. A rigorous} evaluation increases the trust of researchers and practitioners that a resource allocation approach leads to the intended results. Based on \citet{zelkowitz1998experimental}, we investigated which approaches have no evaluation, an argumentation on a toy example, a case study, or a controlled experiment. A research prototype helps enable the \ingo{implementation of an approach in a real-world setting} and \ingo{allows benchmarking different approaches and comparing them with each other} for a certain business scenario. We intend to distinguish whether a prototype is available, not provided, not accessible, or only as pseudocode.
    %Depending on the maturity level of a research approach, its applicability in real-world settings can be inferred. When addressing this RQ, we distinguish between 4 levels (low, medium, high, advanced) and base this decision on the evaluation, the availability of a prototype/pseudocode, and whether it is an approach for local or global optimization.
    %Approaches that have neither been implemented nor evaluated are categorized as \emph{low}.
    %Only approaches that (i) target a global optimization in the resource allocation, (ii) have been implemented or at least specified with pseudocode, and (iii) have been evaluated are classified as \emph{high} or more; all others as \emph{medium}. 
    %The subset of approaches fulfilling the criteria for \emph{high}, which further provide an accessible research prototype, are categorized as having \emph{advanced} maturity.
\end{itemize}

\subsection{Search for studies and data extraction}
\label{subsec:primarysearch}
Our search for studies was split into two phases: a primary search and a secondary search.
Research databases were queried with a predefined set of search terms concerning the abstract, title, and keywords in the primary search.
The resulting studies were reviewed to identify a core set of relevant studies.
This core set was then used to conduct a backward-forward search. \ifLong{This search was done by investigating studies citing our core set (forward search) and the studies cited in our core set (backward search).} By doing this, we could find additional relevant studies and maximize the completeness of our search. The complete search process, with the resulting number of studies, is illustrated as a BPMN process diagram in Figure~\ref{figsearch}.

\begin{figure}[!ht]
	\centering
	\includegraphics[width=\textwidth]{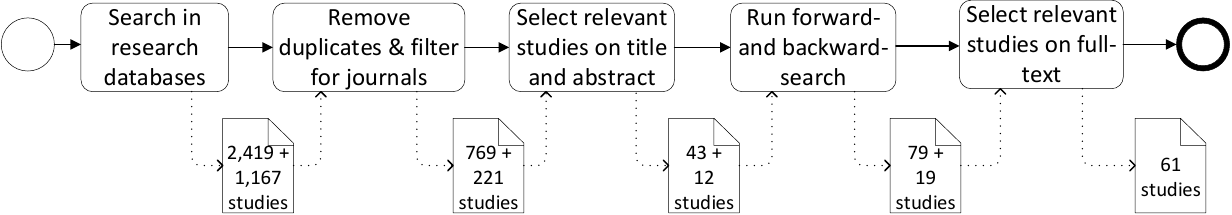}
	\caption{Search process and the number of studies as result of the difference steps\luise{, including the searches in 2019 and 2023} \fabian{indicated by the $+$.}} 
	\label{figsearch}
\end{figure}
%and described in more detail below.

In the primary search, we queried several research databases: \textit{ACM Digital Library}, \textit{IEEE Xplore}, \textit{SciVerse Scopus}, and \textit{Web of Science}, with the following general search term: (RESOURCE | STAFF | TASK) x (ALLOCATION | ASSIGNMENT | SCHEDULING | OPTIMIZATION | PLANNING) x (BUSINESS PROCESS | PROCESS MINING).
The search was conducted in June 2019 \luise{and repeated in September 2023 % for the years from 2019 until 2023
\ingo{in order to cover more recent publications}}.
We observed that the defined search led to an enormous set of resulting studies in some preceding searches. This set included a high proportion of short conference papers.
Thus, we decided to focus on journal papers, which are usually more detailed and \ingo{on average} of higher quality, and filtered for such articles in those databases (where possible).
We were aware that high-quality works are also published at conferences, which we planned to include in the backward-forward search.
The exact search queries per database, including synonyms, and the resulting number of studies are shown in Table~\ref{tab:primarysearch}.
% filtered only for journal papers

\begin{table}[ht]
  \centering
  \scriptsize
    \caption{Database, search queries, and resulting number of studies in the primary search \luise{done in 2019 and the repeated search in 2023 for 2019-2023}. TS of Web of Science can be used for searches for topic terms within a record, such as search in abstract, author keywords, etc.}
  \label{tab:primarysearch}
  \begin{tabularx}{15cm}{>{\hsize=.1\hsize}X>{\hsize=0.7\hsize}X>{\hsize=.075\hsize}X>{\hsize=.075\hsize}X}
    \hline 
\textbf{Database} & \textbf{Search queries} & \textbf{\# 2019} & \textbf{\# 2023}\\

\hline
\textbf{ACM Digital Library} & \texttt{recordAbstract:(resource staff task) + (allocation assignment scheduling optimization planning) + (``business processes'' ``process mining'')} & 535 & 27\\
\textbf{IEEEXplore} & \texttt{((task OR staff OR resource) AND (allocation OR assignment OR scheduling OR optimization OR planning) AND (``process mining'' OR ``business processes''))} & 1086 & 836\\
\textbf{Science Direct} & \texttt{(resource OR staff OR task) AND (allocation OR assignment OR scheduling OR optimization OR planning)  AND (``process mining'' OR ``business process management'')} & 61 & 36\\ %recheck with only business process
\textbf{Web of Science} & \texttt{(TS=(task OR staff OR resource) AND (allocation OR assignment OR scheduling OR optimization OR planning) AND (``process mining'' OR ``business processes'')) AND LANGUAGE: (English)} & 722 & 268\\ 
%&& total 1582, only 767 - 27 manually identified duplicates\\
%Primary search result 740 studies
\hline 
\end{tabularx}
\end{table}

The resulting 2,419 studies \ingo{from 2019} \luise{and the 1,167 additional studies from 2023} were, with the help of the literature \ingo{management} tool \textit{Citavi}, checked for duplicates, and non-journal publications were filtered out.
%Some duplicates needed to be removed manually in an additional step.
The remaining 769 studies \luise{and 221 studies from the more recent search}  were checked independently by two authors of this paper for their relevance, based on the title and abstract and with the help of the inclusion and exclusion criteria given in Table~\ref{tab:criteria}. \luise{Thereby, also some duplicates were identified manually and removed.}

\begin{table}{}
  \centering
  \scriptsize
    \caption{Inclusion and exclusion criteria}
  \label{tab:criteria}
  \begin{tabularx}{15cm}{>{\hsize=.05\hsize}X>{\hsize=0.95\hsize}X}
    \hline 
\textbf{\#} & \textbf{Criteria} \\

\hline
\textbf{IN1} & The study describes an algorithm or technique to support resource allocation in the context of business processes.\\
\textbf{EX1} & The study provides exclusively a survey on topics related to resources or business processes.\\
\textbf{EX2} & The study is not written in English.\\
\textbf{EX3} & The study is not published in a peer-reviewed journal.\\
\textbf{EX4} & The study focuses only on modeling resources or resource analysis.\\
\textbf{EX5} & The study focuses only on the planning of activities only, not their allocation to resources.\\
\textbf{EX6} & The study focuses on the design, configuration, or application of an ERP system.\\
\textbf{EX7} & The study only describes a resource allocation approach for one specific use case\luise{, such as manufacturing processes~\citep{howard1999application}, and not a generalized approach for business processes}.\\
\textbf{EX8} & The study describes an approach for allocating complete business process instances for their execution in an execution environment. The process instance is considered as an entire block and not as a set of related tasks (for instance, some studies on executing process instances in the cloud, like~\citep{wei2016proactive}).\\
    \hline
\end{tabularx}
\end{table}

All studies that both authors categorized as relevant were accepted. Studies for which a disagreement existed were discussed, and a final decision was jointly made.
The resulting core set of relevant studies, comprising 43 papers \luise{and 12 papers of the more recent search}, was then used for the backward-forward search to find additional studies.
We also considered conference or workshop papers in the backward search because we assumed that these studies referenced by a journal article have a high implication for the research field.
For this step, \emph{Web of Science} was used. Identified papers were immediately checked for their relevance.
\fabian{Any uncertainties regarding their relevance were discussed among co-authors and jointly resolved.}
%Studies for which a researcher was not sure were discussed with the co-authors. 
Relevant papers were added to the core set of studies, which resulted in 79 studies \luise{\fabian{plus} 19 of the recent search}.

%As we ran the first search in June 2019, we updated the set of relevant studies by rerunning the searches in the research databases in February 2021 and selecting relevant studies manually from the databases.
%We found four additional studies.
Next, we read the full text of the \luise{98} studies and excluded another \luise{37} because they did not fulfill the inclusion and exclusion criteria \luise{or were duplicates}.
Each exclusion decision was discussed with the group of co-authors.
Reasons were often that the focus was on process modeling or resource analysis, the presented resource allocation approach was too domain-specific, or that the work had been published in different types of papers, whereby the most extensive versions had been kept.
One study~\citep{liu2013accelerating} was added based on an expert's suggestion.
The final set of \luise{61} studies was read thoroughly, and relevant information was extracted according to the following predefined data schema
%\footnote{The \luise{included and excluded} studies, as well as the data extraction result, are collected in a spreadsheet. \luise{It can be found at \url{https://figshare.com/s/2c76210a250e1ab8ad75.}}}
:
\begin{enumerate*}
    \item year and country,
    \item type of resources,
    \item allocation capability (1-to-1, 1-to-m, m-to-1, m-to-n),
    \item allocation goal,
    \item usage of process model and execution data,
    \item task and resource attributes,
    \item solution technique,
    \item single- vs. multi-task optimization,
    \item type of evaluation (none, simulation experiments, experiments with real-world data, case study etc.), and
    \item prototype (available, only pseudocode, etc.).
\end{enumerate*}

%%%%%%%%%%%%%%%%%%%%%%%%%%%%%%%%%%%%%%%%%%%%%%%%%%%%%%%%%%%%%%%%%%%%%%%%%%%%%%%%%%%%%%%%%%%%%%%
\section{Resource Allocation Approaches in Business Processes}
\label{sec:result}

In this section, we \fabian{discuss} the results of the SLR, structured by \fabian{our} research questions. Before we do so, we \fabian{report on} basic statistics about the final set of \luise{61} primary studies.
We extracted the originating country based on the affiliation of the corresponding author. In rare cases where the corresponding author could not be identified, the author's affiliation first listed was chosen. 
% \fabian{The interested reader is referred to Figure~\ref{fig:country}.}
\ingo{Figure~\ref{fig:country} shows the results.}
%As a result, we found that around 30\% of the primary studies % analyzed in this SLR on resource allocation in business processes originate from China (\luise{16} studies) as shown in Figure~\ref{fig:country}, followed by Austria (6 studies) and Germany (4 studies). Next to Asia and Europe, global interest in this topic can be observed: works by authors from North (USA (\luise{6} studies)) and South America (Chile (1 study) \luise{Colombia (1 study)}), Africa (Tunisia (\luise{3} studies)), and Australia (\luise{3} studies) were also found.
The first work of our study collection was published in 1995, followed by \ingo{few studies published around 2002} (see Figure~\ref{fig:year}).
After 2010, more published studies on resource allocation in processes can be observed, with a peak in 2016.
Most papers studied in this review were published \luise{between 2016 and 2021}.

\luise{Additionally, we observed that most of the studies focused on human and application resources based on the classification by \citet{ouyang2010modelling}. Only one study is here an exception: \citet{doerner2006enriched} considers all kinds of resources, including consumable ones.}

\begin{figure}
     \centering
     \begin{subfigure}[b]{0.49\textwidth}
         \centering
         \includegraphics[width=\linewidth]{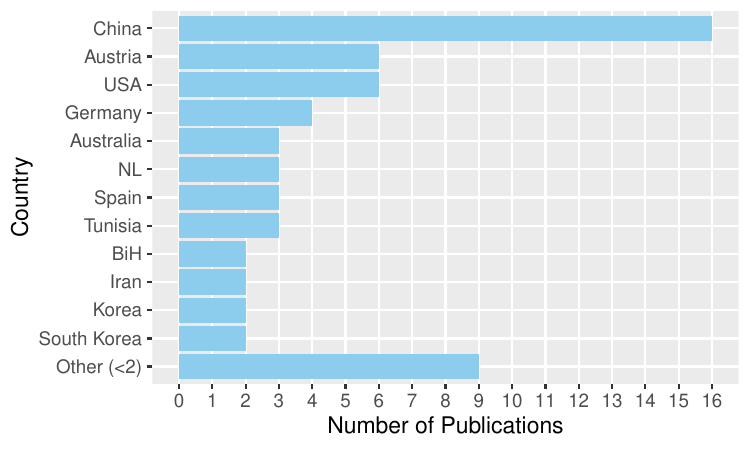}
         \caption{Originating country}
         \label{fig:country}
     \end{subfigure}
     \hfill
          \begin{subfigure}[b]{0.49\textwidth}
         \centering
         \includegraphics[width=\linewidth]{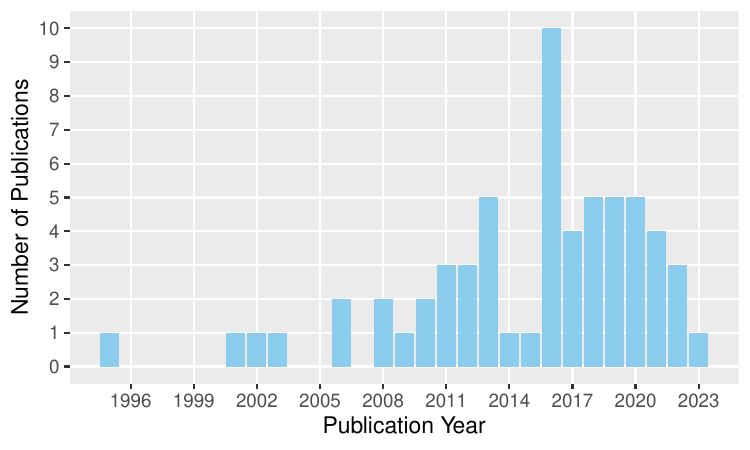}
         \caption{Publishing year}
         \label{fig:year}
     \end{subfigure}
     \caption{Resulting studies categorized by the originating country (corresponding author and the published year). \ingo{Countries with one publication each are shown as ``Other''. Numbers for 2023 are not comparable to earlier years due to the timing of the search.}}
\end{figure}

\luise{Tables~\ref{tab:studies} and ~\ref{tab:studies2} show} an overview of all studies sorted by year of publication and the categorization according to the data extraction scheme. The following subsections present details of these results.
The resource allocation types and the targeted optimization goals of the approaches are presented in Section~\ref{subsec:goals}, addressing RQ1.
Next, the role of process models and process execution data in the resource allocation approaches is discussed in Section~\ref{subsec:processmodel}, addressing RQ2.
The considered input data, that is, the attributes of tasks and resources, are presented and classified in Section~\ref{subsec:attributes}, addressing RQ3.
Solution approaches are discussed in Section~\ref{subsec:solution}, addressing RQ4.
Finally, the evaluation techniques and the usage of research prototypes are presented in Section~\ref{subsec:maturity}, addressing RQ5.

\subsection{RQ1: Resource allocation capabilities and optimization goals}
\label{subsec:goals}
In this subsection, 
\emph{RQ1: What are the targeted resource allocation goals and capabilities?} is explored. To do so, we present the results regarding the resource allocation capabilities supported in the literature, followed by targeted optimization goals.

Figure~\ref{fig:capability} shows how many studies support the different resource allocation capabilities \luise{based on Definition~\ref{def:RA}} .
Thereby, \luise{three} resource allocation capabilities can be differentiated (relations as |tasks|-to-|resources|):
\begin{itemize}
    \item 1-to-1 allocation, where one resource is assigned to exactly one task; 
    \item 1-to-many allocation, where a team of resources is assigned to one task; and
    \item many-to-1 allocation, where the capacity of a resource is greater than one, and thus, several tasks can be assigned to it.
%    \item many-to-many allocation, where a team of resources is assigned to a set of tasks.
\end{itemize}
The concrete studies supporting a certain capability are given in Table~\ref{tab:studies}.
Most studies support an assignment of one resource to one task.
For example, \citet{doerner2006enriched} represent the business process as a Petri net and resources as tokens on resource-places in the net. 
\ingo{Depending on whether there is an arc from a resource-place to a transition and/or vice versa, a firing transition (representing an activity execution) consumes and/or produces \textit{at most one} resource token. Hence, this approach offers 1-to-1 allocation.}
% At maximum, a transition representing an activity consumes one resource. 
% After the transition is completed, it returns the resource token.   

\luise{13} studies support 1-to-many allocations. Some of these studies consider the possibility of having one or more resources assigned to a task. 
An example is \citet{kamrani2012framework} that offer the possibility to define a constraint in the resource allocation specification, which encodes the number of resources needed for an activity. \ifLong{\citet{djedovic2018innovative} offer a technique to identify the number of resources for each process activity that should be available at runtime to solve their tasks. }
\luise{\citet{bessai2016business} provide an approach for assigning tasks to crowd workers. If a certain task needs multiple skills that can only be fulfilled by several crowd workers, their approach can also assign a team to a task.}

Furthermore, \luise{seven} studies consider that resources can have capacities greater than one (e.g., a nurse can transport one or several blood tubes at once) or that resources should work on similar or related tasks in sequence. In either case, these studies support many-to-1 allocations.
For instance, \citet{xu2016resource} allow resources \fabian{to have} several slots available for a particular time. Thus, the resource can process several tasks in parallel.
\citet{pflug2016application} provide an approach that regularly classifies the existing tasks into sets of similar tasks; then, a set of similar tasks is assigned to a resource that can work on them one after another.

\begin{figure}
     \centering
     \begin{subfigure}[t]{0.32\textwidth}
         \centering
         \includegraphics[width=\linewidth]{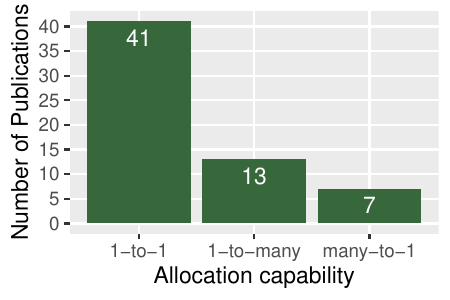}
         \caption{Supported allocation capabilities (|tasks|-to-|resources|)}
         \label{fig:capability}
     \end{subfigure}
     \hfill
     \begin{subfigure}[t]{.64\textwidth}
         \centering
         \includegraphics[width=\linewidth]{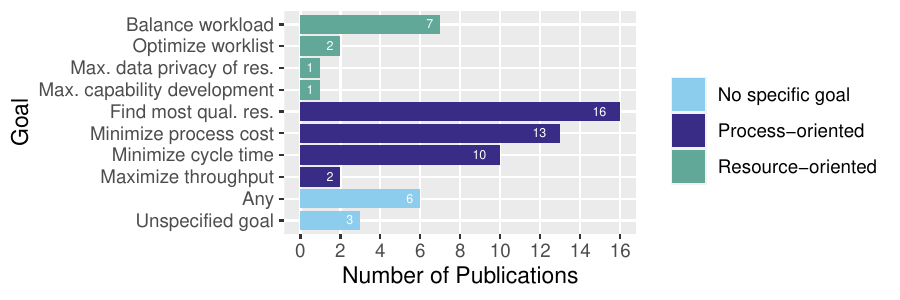}
         \caption{Optimization goals for allocation}
         \label{fig:goal}
     \end{subfigure}
     \caption{Support allocation capabilities and optimization goals of the studies.} 
\end{figure}

Many studies support a specific resource allocation goal \fabian{(c.f. Definition~\ref{def:RAS})} in their approach.
\fabian{We can group these goals based on}
\luise{the two entities involved in a resource allocation: \fabian{the business process with its tasks and the resource responsible for executing the work.}}
It can be observed that a majority of studies focus on goals concerning the improvement of the performance of the \emph{process} (e.g., minimizing the process costs or cycle time). A smaller portion of studies focuses on improving \luise{the organization of the resources' workload and a \fabian{resource orientation} in their goal definition}. Few studies have no (specific) optimization goal. An overview of the different optimization goals---and how many studies consider them---can be found in Figure~\ref{fig:goal}.
In the following, we discuss these categories in more depth.

\begin{landscape}
\begin{table}[]
\tiny
    \centering
    \begin{tabular}{p{4.5cm}|p{0.5cm}|p{0.9cm}|p{1.3cm}|p{3cm}|p{0.8cm}|p{0.5cm}|p{0.6cm}|p{2.5cm}|p{3cm}|p{2cm}}
        \hline
         \textbf{Reference} & \textbf{Year} & \textbf{Country} & \textbf{Capability} & \textbf{Optimization Goal} & \textbf{PM/PD} & \textbf{Role} & \textbf{Type} & \textbf{Solution technique} & \textbf{Evaluation} & \textbf{Prototype}\\
         \hline
\citet{bussler1995policy}&1995&Germany&many-to-1&Unspecified goal&PM&Prep.&Single&Rule&No evaluation&not acc., pseudoc.\\
\citet{van2001optimal}&2001&NL&1-to-many&Minimize cycle time&PM&Prep.&Single&Heuristic&Toy example&no implementation\\
\citet{kumar2002dynamic}&2002&USA&1-to-1&Find most qual. res.&PM&Input&Single&Rule&Toy example&no implementation\\
\citet{eder2003personal}&2003&Austria&1-to-1&Balance workload&PM&Prep.&Multiple&Heuristic&No evaluation&no implementation\\
\citet{doerner2006enriched}&2006&Austria&1-to-1&Minimize process cost&PM&Prep.&Multiple&Metaheuristic&Case study&pseudocode\\
\citet{ha2006development}&2006&Korea&1-to-1&Balance workload&PM&Prep.&Multiple&Rule&(Comp.) sim. experiments&pseudocode\\
\citep{rhee2010increasing}&2008&USA&many-to-1&Balance workload&PM&Input&Multiple&Heuristic&(Comp.) sim. experiments&not acc., pseudoc.\\
\citet{zhou2008project}&2008&China&1-to-1&Minimize process cost&PM&Prep.&Multiple&Metaheuristic&Toy example&no implementation\\
\citet{xu2009resource}&2009&Australia&1-to-many&Minimize process cost&PM&Input&Multiple&Rule&(Comp.) sim. experiments&pseudocode\\
\citet{huang2010adaptive}&2010&NL&1-to-1&Any&PM&Input&Single&Machine Learning&(Comp.) experiments&available\\
\citet{delias2010optimizing}&2010&Greece&1-to-1&Balance workload&PM&Prep.&Multiple&Exact Algorithm&(Comp.) sim. experiments&not accessible\\
\citet{huang2011mining}&2011&China&1-to-1&Unspecified goal&PD&Input&Single&Mined Rule&Experiments + Case study&not acc., pseudoc.\\
\citet{huang2011reinforcement}&2011&China&1-to-1&Minimize process cost&PM&Prep.&Multiple&Machine Learning&(Comp.) experiments&available\\
\citet{kamrani2012framework}&2011&Sweden&1-to-many&Minimize process cost&PM&Input&Multiple&Heuristic&(Comp.) sim. experiments&pseudocode\\
\citet{huang2012task}&2012&China&1-to-1&Minimize process cost&PM&Prep.&Multiple&Metaheuristic&(Comp.) sim. experiments&not acc., pseudoc.\\
\citet{huang2012resource}&2012&China&1-to-1&Find most qual. res.&PD&Prep.&Single&Rule&Case study&available\\
\citet{liu2012mining}&2012&China&1-to-1&Find most qual. res.&PD&Input&Multiple&Mined Rule&(Comp.) experiments&not acc., pseudoc.\\
\citet{kumar2013optimal}&2013&USA&1-to-1&Any&PD&Prep.&Multiple&Heuristic&Case study&not acc., pseudoc.\\
\citet{barba2013user}&2013&Spain&1-to-1&Optimize worklist&PM&Prep.&Single&Logic Programming&(Comp.) sim. experiments&not acc., pseudoc.\\
\citet{cabanillas2013priority}&2013&Austria&1-to-1&Find most qual. res.&PM&Input&Single&Heuristic&No evaluation&available\\
\citet{liu2013accelerating}&2013&USA&1-to-many&Find most qual. res.&PD&Prep.&Single&Rule&Toy example&no implementation\\
\citet{xu2013incorporating}&2013&Australia&1-to-1&Minimize cycle time&PM&Input&Multiple&Rule&(Comp.) sim. experiments&not acc., pseudoc.\\
\citet{schall2014crowdsourcing}&2014&Austria&1-to-many&Find most qual. res.&PM&Prep.&Single&Rule&(Comp.) sim. experiments&not acc., pseudoc.\\
\citet{zhao2015optimization}&2015&China&1-to-1&Minimize cycle time&PD&Prep.&Single&Rule&(Comp.) experiments&not accessible\\
\citet{djedovic2016optimization}&2016&BiH&1-to-many&Minimize process cost&PM&Prep.&Multiple&Metaheuristic&Case study&not accessible\\
\citet{havur2016resource}&2016&Austria&1-to-1&Minimize cycle time&PM&Input&Multiple&Logic Programming&(Comp.) sim. experiments&no implementation\\
\citet{bessai2016business}&2016&France&1-to-many&Minimize cycle time&PM&Input&Multiple&Heuristic&(Comp.) sim. experiments&not acc., pseudoc.\\
\citet{pflug2016application}&2016&Austria&many-to-1&Maximize throughput&PD&Input&Multiple&Machine Learning&Case study&not acc., pseudoc.\\
\citet{schonig2016framework}&2016&Germany&1-to-1&Unspecified goal&PD&Input&Multiple&Mined Rule&(Comp.) sim. experiments&not accessible\\
\citet{wibisono2016dynamic}&2016&Indonesia&1-to-1&Minimize cycle time&PM&Input&Single&Rule&(Comp.) sim. experiments&pseudocode\\
\citet{xie2016dynamic}&2016&China&1-to-many&Minimize cycle time&PM&Prep.&Multiple&Exact Algorithm&Experiments + Case study&available\\
\hline
    \end{tabular}
    \caption{Studies and their categorizations ordered by publication year, Part 1 \fabian{(PD = Process data, PM = Process model).}}
    \label{tab:studies}
\end{table}
\end{landscape}

\begin{landscape}
\begin{table}[]
\tiny
    \centering
    \begin{tabular}{p{4.5cm}|p{0.5cm}|p{0.9cm}|p{1.3cm}|p{3cm}|p{0.8cm}|p{0.5cm}|p{0.6cm}|p{2.5cm}|p{3cm}|p{2cm}}
        \hline
         \textbf{Reference} & \textbf{Year} & \textbf{Country} & \textbf{Capability} & \textbf{Optimization Goal} & \textbf{PM/PD} & \textbf{Role} & \textbf{Type} & \textbf{Solution technique} & \textbf{Evaluation} & \textbf{Prototype}\\
         \hline
\citet{xu2016resource}&2016&China&many-to-1&Balance workload&PM&Input&Multiple&Metaheuristic&(Comp.) sim. experiments&not acc., pseudoc.\\
\citet{yaghoubi2016resource}&2016&Iran&1-to-1&Optimize worklist&PD&Prep.&Multiple&Machine Learning&(Comp.) experiments&not accessible\\
\citet{zhao2016entropy}&2016&China&1-to-1&Find most qual. res.&PD&Prep.&Multiple&Mined Heuristic&(Comp.) experiments&pseudocode\\
\citet{bellaaj2017obstacle}&2017&Tunisia&1-to-1&Find most qual. res.&PD&Prep.&Single&Rule&No evaluation&not acc., pseudoc.\\
\citet{hirsch2017information}&2017&USA&1-to-1&Minimize cycle time&PM&Input&Multiple&Exact Algorithm&(Comp.) sim. experiments&not acc., pseudoc.\\
\citet{zhao2017resource}&2017&China&1-to-1&Minimize cycle time&PM&Input&Multiple&Metaheuristic&(Comp.) experiments&not accessible\\
\citet{yaghoibi2017cycle}&2017&Iran&many-to-1&Balance workload&PM&Prep.&Multiple&Rule&(Comp.) experiments&pseudocode\\
\citet{arias2018all}&2018&Chile&1-to-1&Find most qual. res.&PD&Prep.&Multiple&Exact Algorithm&Experiments + Case study&available\\
\citet{djedovic2018innovative}&2018&BiH&1-to-many&Minimize process cost&PD&Prep.&Multiple&Metaheuristic&(Comp.) sim. experiments&not accessible\\
\citet{erasmus2018method} &2018&NL&1-to-1&Find most qual. res.&PM&Input&Multiple&Rule&Case study&not acc., pseudoc.\\
\citet{abdulhameed2018resource}&2018&Egypt&1-to-1&Find most qual. res.&PD&Prep.&Single&Mined Rule&Case study&no implementation\\
\citet{si2018petri}&2018&Macau&1-to-1&Minimize process cost&PM&Input&Single&Metaheuristic&Case study&not acc., pseudoc.\\
\citet{duran2019rewriting}&2019&Colombia&1-to-many&Any&PM&Prep.&Multiple&Heuristic&(Comp.) sim. experiments&available\\
\citet{lee2019dynamic}&2019&Korea&1-to-1&Find most qual. res.&PD&Prep.&Single&Rule&Experiments + Case study&no implementation\\
\citet{Luo2019}&2019&China&1-to-1&Find most qual. res.&PD&Prep.&Single&Machine Learning&No evaluation&not acc., pseudoc.\\
\citet{soeffker2019adaptive}&2019&Germany&1-to-1&Maximize throughput&PM&Prep.&Multiple&Heuristic&(Comp.) sim. experiments&not acc., pseudoc.\\
\citet{xie2019integration}&2019&China&1-to-1&Minimize process cost&PM&Input&Multiple&Metaheuristic&Experiments + Case study&not acc., pseudoc.\\
\citet{bellaaj2020avoiding}&2020&Tunisia&1-to-1&Minimize process cost&PD&Prep.&Single&Machine Learning&(Comp.) sim. experiments&not acc., pseudoc.\\
\citet{jemel2020rpminter}&2020&Tunisia&1-to-many&Max. data privacy of res.&PM&Prep.&Multiple&Logic Programming&(Comp.) sim. experiments&not accessible\\
\citet{pereira2020new}&2020&Portugal&1-to-1&Find most qual. res.&PM&Prep.&Single&Rule&No evaluation&no implementation\\
\citet{yu2020task}&2020&China&1-to-1&Minimize cycle time&PM&Prep.&Single&Rule&Toy example&no implementation\\
\citet{zhao2020human}&2020&China&1-to-1&Minimize process cost&PD&Prep.&Single&Machine Learning&(Comp.) sim. experiments&not acc., pseudoc.\\
\citet{barba2021flexible}&2021&Spain&1-to-1&Any&PM&Prep.&Multiple&Logic Programming&Case study&not acc., pseudoc.\\
\citet{duran2021resource}&2021&Spain&1-to-many&Minimize process cost&PM&Prep.&Multiple&Rule&Experiments + Case study&not accessible\\
\citet{hou2021bottleneck}&2021&China&many-to-1&Balance workload&PM&Input&Multiple&Exact Algorithm&(Comp.) sim. experiments&not acc., pseudoc.\\
\citet{pika2021machine}&2021&Australia&1-to-1&Max. capability development&PD&Prep.&Single&Machine Learning&Experiments + Case study&available\\
\citet{ihde2022framework}&2022&Germany&many-to-1&Any&PM&Prep.&Multiple&Any&Case study&available\\
\citet{liu2022multi}&2022&USA&1-to-many&Find most qual. res.&PD&Prep.&Multiple&Rule&Experiments + Case study&not accessible\\
\citet{yeon2022experimental}&2022&South Korea&1-to-1&Find most qual. res.&PD&Prep. &Single&Mined Rule&Case study&no implementation\\
\citet{park2023optimizing}&2023&South Korea&1-to-1&Any&PD&Input&Single&Machine Learning&Experiments + Case study&available\\
\hline
    \end{tabular}
    \caption{Studies and their categorizations ordered by publication year, Part 2 \fabian{(PD = Process data, PM = Process model)}.}
    \label{tab:studies2}
\end{table}
\end{landscape}

\paragraph{Process-oriented optimization.} A majority of the identified studies aim at \emph{finding the most qualified resource}. A resource allocation tries to identify the most qualified resource for a specific process task based on the task's needs. The focus is here to optimize the process.
As shown in Figure~\ref{fig:goal}, the second-most frequent goal is to \emph{minimize the process costs}. 
\luise{Followed by the goal to \emph{minimize the cycle time} of process executions and by a smaller portion of studies that aims at \emph{maximizing the throughput}. The latter two goals are both time-oriented. But whereas \emph{minimizing the cycle time} tries to ensure that each instance has the shortest possible time between start and end, \emph{maximizing the throughput} aims at increasing the total output of a process within a given timeframe. This can also lead to longer \ingo{cycle} times for certain process instances.}

\paragraph{Resource-oriented optimization.}
Most studies focusing on a resource-oriented optimization try to \emph{balance the workload} between resources, i.e., tasks are distributed equally to the available resources. In contrast to finding the most qualified resource, tasks might also be allocated to less appropriate resources to equalize workloads. 
Two studies aim at \emph{optimizing the worklist}: one study focuses on prioritizing tasks~\citep{barba2013user}, and the other one on minimizing the entropy of worklists~\citep{yaghoubi2016resource}.
\luise{\citet{pika2021machine} want to maximize the capability development of the workers involved in business processes. \fabian{To increase workers capabilities, their approach assigns tasks \ingo{with which resources are less familiar,} based on historical information.}  
Finally, \citet{jemel2020rpminter} aim at maximizing the data privacy of resources. They are providing an allocation approach for inter-organizational business processes and want to reduce the \fabian{amount} of information shared of involved resources.}

\paragraph{No explicit optimization goal.}
\luise{Six} studies support the idea that the optimization goal can be individually defined when applying and implementing the resource allocation approach.
\citet{huang2010adaptive} maximize the allocation reward, and the user can specify the calculation of the reward. \citet{duran2019rewriting} describe that a multi-objective optimization problem has to be defined for a resource allocation, and \citet{ihde2022framework} provide a way to define the optimization goal individually for each process activity at design-time.
\luise{\citet{kumar2013optimal,barba2021flexible,park2023optimizing} \ingo{all propose approaches where the objective function can be individually defined}.}
These studies provide more generalized approaches that are usable in more application scenarios. Practitioners are enabled to define their resource allocation goals individually.
\luise{Three} studies working with resource allocation rules~\citep{bussler1995policy,huang2011mining, schonig2016framework} are not associated with a specific resource allocation goal. The process expert has to evaluate whether the rules are supporting the process goals.

\subsection{RQ2: Role of Process Models and Execution Data}
\label{subsec:processmodel}

This section provides the results to answer \emph{RQ2: \textit{What \ingo{are the respective roles} of process models and process execution data in the resource allocation approach?}} When analyzing this question, it can be observed that most of the primary studies (\luise{39 studies}) use process models for resource allocation, and fewer (\luise{22 studies}) use process execution data in the form of event logs. We summarize our results in Table~\ref{tab:model}.
We further categorized the studies into those that use process model or data to prepare specific artifacts (e.g., a process simulation model), which is then used as input for the resource allocation approach, and those that directly apply it as input.

\begin{table}[]
    \centering
    \footnotesize
    \begin{tabular}{p{2.5cm}|p{6cm}|p{6cm}}
        \hline
        \footnotesize
          & \textbf{Process Model} & \textbf{Process Execution Data} \\
          \hline
          \textbf{Used as preparation for an input artifact for the allocation technique}& \textbf{Enhanced Process Models} {\citep{bussler1995policy,eder2003personal,doerner2006enriched,zhou2008project,delias2010optimizing,huang2012task,barba2013user,schall2014crowdsourcing,djedovic2016optimization,xie2016dynamic,jemel2020rpminter,pereira2020new,yu2020task,barba2021flexible,ihde2022framework}}& \textbf{Characteristics of resources} {\citep{huang2012resource,kumar2013optimal,liu2013accelerating,zhao2015optimization,zhao2016entropy,bellaaj2017obstacle,arias2018all,abdulhameed2018resource,lee2019dynamic,Luo2019,zhao2020human,pika2021machine,liu2022multi}}\\
          &\textbf{Stochastic model} {\citep{van2001optimal,huang2011reinforcement,soeffker2019adaptive,duran2019rewriting,duran2021resource}}& \textbf{Characteristics of tasks} {\citep{yaghoubi2016resource,bellaaj2020avoiding}}\\
          & \textbf{Queue Model} {\citep{ha2006development,yaghoibi2017cycle}}& \textbf{Discovered enhanced process model and simulation model} {\citep{djedovic2018innovative,yeon2022experimental}}\\
         \hline
          \textbf{Used as direct input for the allocation technique and their usage}& \textbf{To create a heuristic or metaheuristic}  {\citep{rhee2010increasing,kamrani2012framework,cabanillas2013priority,zhao2015optimization,xu2016resource,bessai2016business,xie2019integration}}& \textbf{To mine allocation rules} {\citep{huang2011mining,liu2012mining,schonig2016framework}}\\
          & \textbf{To create rule or logical programming} {\citep{kumar2002dynamic,xu2009resource,xu2013incorporating,havur2016resource,wibisono2016dynamic,erasmus2018method}}& \textbf{To cluster tasks} {\citep{pflug2016application}}\\
          & \textbf{To create a linear programming model} {\citep{hirsch2017information,hou2021bottleneck}}& \textbf{To train a machine learning model} {\citep{park2023optimizing}}\\
           & \textbf{To create a ML model} {\citep{huang2010adaptive}}& \\
    \hline
    \end{tabular}
    \caption{Role of process models and data in resource allocation approaches.}
    \label{tab:model}
\end{table}

\ifLong{
\begin{figure}
    \begin{subfigure}{.39\textwidth}
        \includegraphics[width=\linewidth]{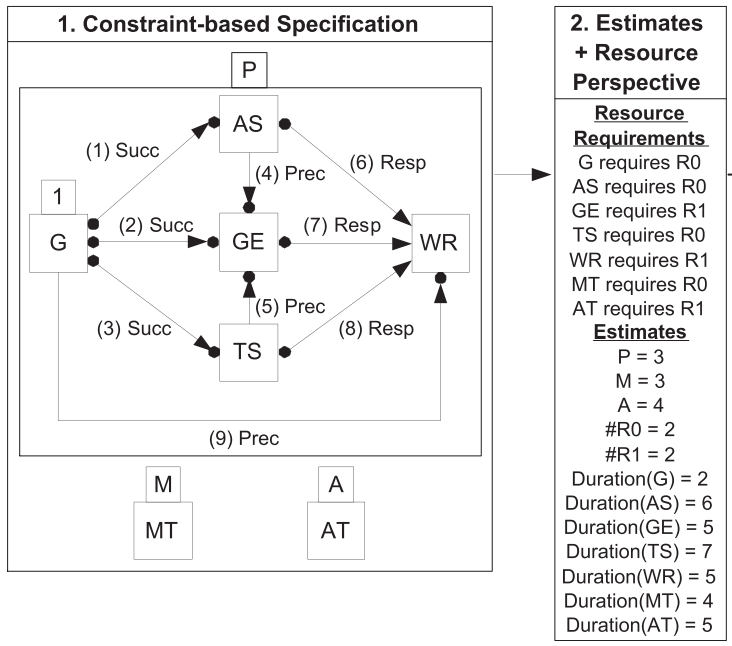}
        \caption{Enhanced process model, an enhanced declarative model~\citep{barba2013user}}
    \end{subfigure}
    \hfill
    \begin{subfigure}{.59\textwidth}
        \includegraphics[width=\linewidth]{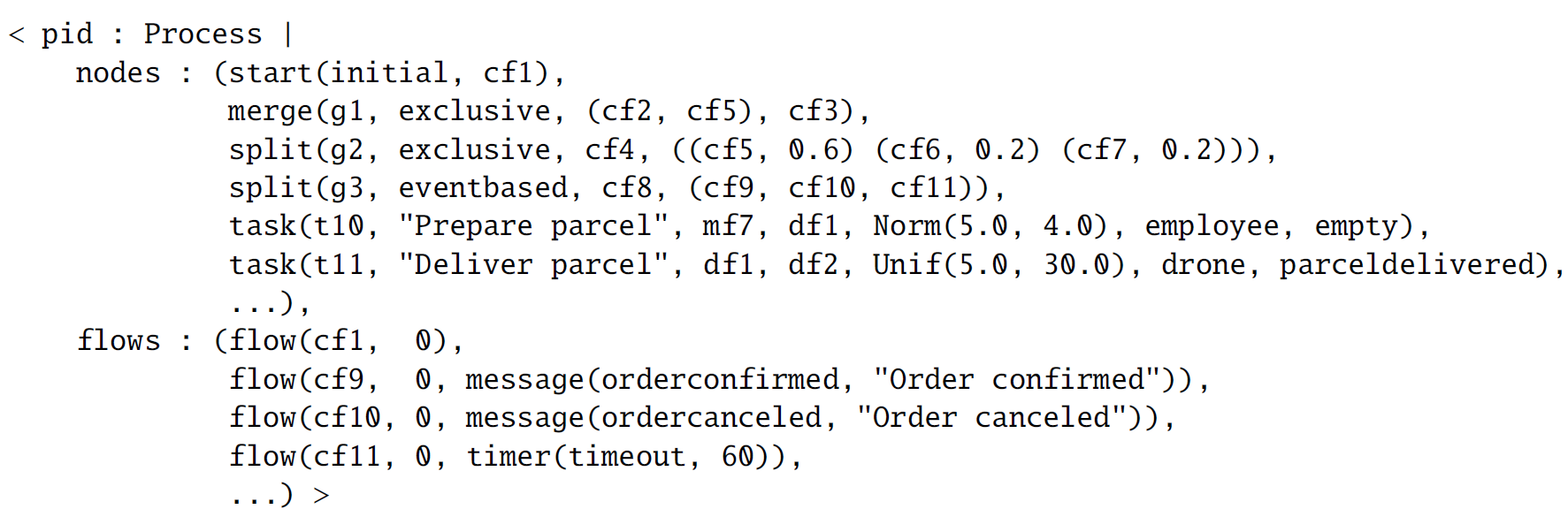}
        \caption{Stochastic model, a Maude simulation model~\citep{duran2019rewriting} }
   \end{subfigure}
   \begin{subfigure}{.48\textwidth}
        \includegraphics[width=\linewidth]{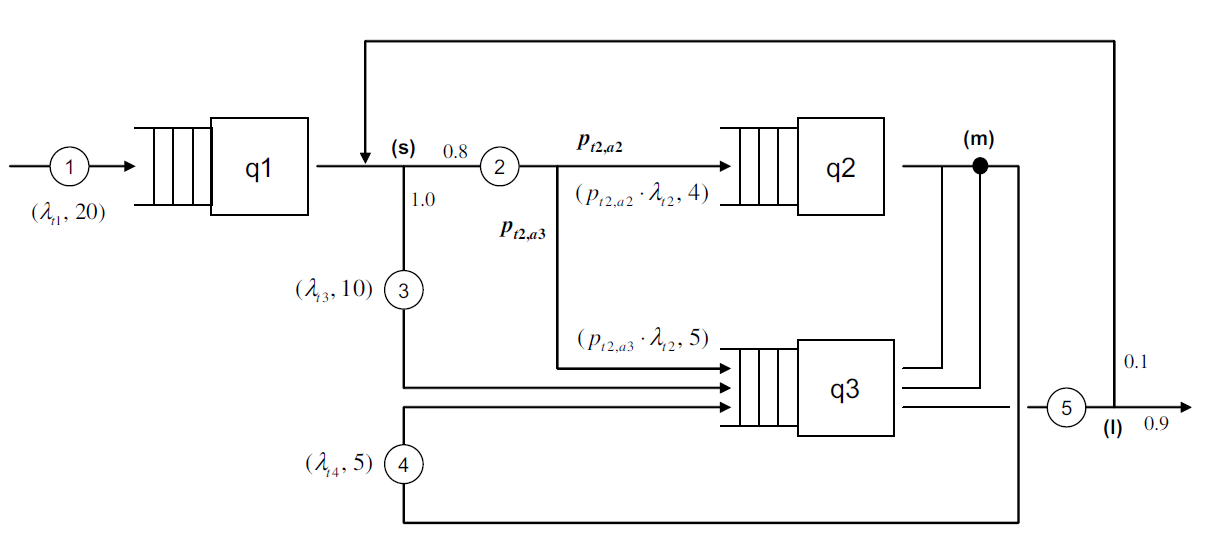}
        \caption{Queueing model, a queueing network~\citep{ha2006development} }
   \end{subfigure}
    
    \caption{Examples of the usage of process models to prepare a certain input for the resource allocation approach.}
    \label{fig:example_PM}
\end{figure}
}

%Role of process models (a) and data (b) in resource allocation approaches, categorized into approaches using the respective artifact either as direct input, or to prepare an input for their approach; and the artifacts' usages. Number of studies in a particular class are shown in parentheses.

\subsubsection{Usage of process models}  

Process models capture the activities, events, and their control flow relations. They are used to prepare a specific artifact as input (\luise{22} studies) or as direct input (\luise{17} studies) to the resource allocation technique (see Table \ref{tab:model}). For these studies, the process expert has to feed the proposed approaches by the studies with the process model and additional information, leading to considerable manual work.

A process model can capture further information as attributes of its process elements. The first category of studies (\luise{15} studies) enhances process models with additional information. 
On the one hand, requirements of the activities for the resource allocation are captured, such as required resource types, policies, time constraints, or data\ifLong{~\citep{bussler1995policy,doerner2006enriched,zhou2008project,huang2012task,schall2014crowdsourcing,pereira2020new}}.
For example, \citet{ihde2022framework} require for each process activity the definition of needed resource types, the allocation problem, and the solution technique.
On the other hand, business process dynamics are also captured, such as estimates or distributions of the activity execution times, inter-arrival times of process cases, or branching probabilities\ifLong{~\citep{eder2003personal,delias2010optimizing,barba2013user,djedovic2016optimization,xie2016dynamic}}.
\ifLong{The example given in Figure \ref{fig:example_PM}(a) shows a declarative model enhanced with required resources and estimated activity execution. 
\citet{delias2010optimizing} use the enhanced process model with the needed resource type, the activity duration, and the branching probabilities to calculate for each new process instance at runtime a task schedule with the start and finish times. The schedule and the needed resource type are then provided as input for their resource allocation approach.}

Other studies use process models to create a stochastic model (\luise{4} studies) for the resource allocation technique. A stochastic model represents the system's dynamics of the process in which resource allocation occurs.
In~\citet{huang2011reinforcement,soeffker2019adaptive}, the process model is used to create a Markov decision process capturing the process states, possible actions, and reward functions. 
\luise{\citet{van2001optimal} use a stochastic workflow net and} \citet{duran2019rewriting,duran2021resource} transform a given BPMN diagram into a Maude simulation model\ifShort{.}\ifLong{ as shown in Figure \ref{fig:example_PM}(b).}
\luise{Both models reflect} the order of the process nodes, the distribution of the activities' duration, the branching probabilities, etc.

A third category creates queueing models (2 studies) as a basis \fabian{for} their resource allocation approach. \ifLong{Figure \ref{fig:example_PM}(c) shows a queueing network of \citep{ha2006development} (also used by~\citep{yaghoibi2017cycle} )} \ifShort{\fabian{\citet{ha2006development, yaghoibi2017cycle} create queueing networks}} where each resource type used for \fabian{a} process activity presents a node with a queue. Based on the process model, the flows and the probabilities between these resource types are deduced. This network is later used to define rules for balancing the workload \fabian{of} the resources.

As direct input (\luise{17} studies), traditional process models are used to create a heuristic or metaheuristic (\luise{8} studies) for resource allocation, rules, or logic programming (\luise{6} studies). We discuss the concrete solution techniques in Section~\ref{subsec:solution} in detail.
\luise{Three} studies use the model as input for linear programming. Finally, one study uses the process model as input to create a machine learning (ML) model. 

%Organizations that own data on their business processes can also apply the just described studies applying process models. With the help of process discovery~\citep{van_der_aalst_process_2016}, process models can be derived from event logs.

\subsubsection{Usage of process execution data} Process execution data \fabian{contains} information about the past process executions. \fabian{This data can be leveraged to \ingo{extract} insights and apply these insights to} resource allocations in the future. Process execution data is used for resource allocation in \luise{23} studies. Most studies use event logs (cf. Section~\ref{subsec:process_def}), also including resource information. Only \citet{Luo2019} use a simplified log, an activity-employee log where the executed activities, the employee, and needed duration are included. 

\luise{17} primary studies use data to prepare an artifact for the resource allocation technique.
\luise{A majority of these} studies \luise{(13 studies)} employ an event log to identify some insights on resources of the process (i.e., characteristics of resources), e.g., their previous performance\ifLong{~\citep{arias2018all,huang2012resource,lee2019dynamic,zhao2015optimization}}, their expertise\ifLong{~\citep{liu2013accelerating,arias2018all,Luo2019,bellaaj2017obstacle}}, their workload\ifLong{~\citep{arias2018all,bellaaj2017obstacle,zhao2015optimization}}, their team compatibility\ifLong{~\citep{kumar2013optimal}}, or their social context\ifLong{~\citep{liu2013accelerating,lee2019dynamic}}.
Other studies identify characteristics of tasks during preparation for resource allocation, such as the similarity between tasks~\citep{yaghoubi2016resource} or misallocations of tasks in the past~\citep{bellaaj2020avoiding}.
\luise{Finally, \citet{yeon2022experimental} discover an enhanced process model with the information on how often a performer has executed an activity from the event log, and}
\citet{djedovic2018innovative} learn the process dynamics from the event log and discover a process simulation model with the distribution of the activities' duration, etc. 

\fabian{Six studies use process data as direct input.
Of these, three studies use data to} mine rules for the allocation. Additionally, \citet{pflug2016application} use an event log to cluster similar tasks so that they can jointly be allocated and \luise{\citet{park2023optimizing} train a machine learning \ingo{models from event logs}.}

\subsection{RQ3: Resource and Task Attributes}
\label{subsec:attributes}

In this subsection, we investigate \emph{RQ3: Which input data are used for resource allocation in business processes?}
In a resource allocation, characteristics of both the tasks and resources can be considered to identify a fitting match. These characteristics are often encoded as task and resource attributes \luise{ and then used as constraints in the resource allocation specification (cf. Definition~\ref{def:RAS})}.

\subsubsection{Resource attributes}
\label{subsubsec:resource-attributes}

\begin{table}[tb]
    \centering
    \scriptsize
     \begin{tabular}{p{2.5cm}p{12cm}}
        \hline
        \textbf{Resource Attribute} & \textbf{Description} \\
         \hline 
         Previous Performance& All performance attributes that are based on the execution history of the resource. The performance of resources can be, for example, the cost, quality, or execution time of previous executions of the resource. \\
         \hline 
         Workload & Attributes that are based on the schedule of a resource. The workload includes attributes such as availability, backlog of allocated tasks, or idle level.\\
        \hline
        Role & Attributes pertaining to the role of a resource, such as authorizations, organizational position, or responsibilities. \\
        \hline
        Expertise  & Attributes encoding a resource's capabilities, skills, and knowledge. The expertise includes functional attributes associated with the resource directly, such as adaptability, and non-functional attributes, which may include environmental factors and employed aids. It also includes attributes based on work variety, i.e., the analysis of similar and dissimilar tasks in the execution history of a resource. \\
        \hline
        \fabian{Resource Amount}  & Attributes encoding the number of resources that exist. For \luise{non-application} resources, this may encode the number of resources in stock. For \luise{human and application} resources, this may encode the number of resources of a specific type. \\
        \hline
        Social Context & Attributes based on the social network of a resource. These attributes may measure the ability to collaborate or the overall compatibility of resources, but also more abstract social constructs, such as the social position or influence of a resource within its network. \\
        \hline
        \fabian{Trustworthiness} & \fabian{Degree of trust to execute a task.} \\
        \hline
        Experience & Attributes such as years of service or other quantifiable attributes based on the experience of a resource. Note the difference to expertise that assesses the actual ability of a resource. \\
        \hline
        Preference & Attributes expressing the preference for a resource executing certain types of tasks. \\
    \hline
    \end{tabular}
    \caption{Brief description of resource attributes\fabian{, based on \citet{arias2017towards}.}}
    \label{tab:attribute_desc}
\end{table}

\citet{arias2017towards} provide a taxonomy for human resource allocation criteria based on a previously conducted mapping study. 
We utilized this taxonomy
%\footnote{Contrary to ~\citep{arias2017towards}, we did not observe any study considering \emph{trustworthiness} (a level of trust a resource may have to perform a task) as a resource attribute. Accordingly, we omit this notion from our classification for resource attributes.}
to classify the resource and task attributes discussed in the identified primary studies. 
The different resource attributes are briefly described in Table~\ref{tab:attribute_desc}. 
Despite their focus on human resources, we were still able to categorize the attributes identified from our set of primary studies. \fabian{Indeed, we found that most studies (37 studies) focus on human resources exclusively; 24 studies talk about resources in a more general sense. Of these, 15 mention machines or software systems.}
Notably, only~\citet{doerner2006enriched} considers human, application, and non-application resources; all other studies do not consider non-application resources only.

\fabian{The result of our resource attribute classification is presented in  Table~\ref{tab:resatt}, showing the different categories of attributes and the corresponding studies considering them in their allocation.} 

\begin{table}[bt]
    \centering
    \scriptsize
     \begin{tabular}{p{2.5cm}p{12.5cm}}
        \hline
        \textbf{Resource Attribute} & \textbf{Corresponding Studies} \\
         \hline 
         Previous Performance \fabian{(32)} & \fabian{\citep{van2001optimal, rhee2010increasing, zhou2008project, huang2010adaptive, huang2011mining, huang2011reinforcement, huang2012task, huang2012resource, liu2012mining, kumar2013optimal, xu2013incorporating, zhao2015optimization, havur2016resource, bessai2016business, wibisono2016dynamic, yaghoubi2016resource, zhao2016entropy, bellaaj2017obstacle, zhao2017resource, arias2018all, djedovic2018innovative, abdulhameed2018resource, duran2019rewriting, lee2019dynamic, xie2019integration, bellaaj2020avoiding, zhao2020human, duran2021resource, pika2021machine, liu2022multi, yeon2022experimental, park2023optimizing}} \\
         \hline 
         Workload \fabian{(23)} & \fabian{\citep{kumar2002dynamic, eder2003personal, ha2006development, rhee2010increasing, zhou2008project, xu2009resource, huang2012task, huang2012resource, barba2013user, kumar2013optimal, zhao2015optimization, xu2016resource, zhao2016entropy, pflug2016application, yaghoubi2016resource, yaghoibi2017cycle, bellaaj2017obstacle, arias2018all, soeffker2019adaptive, duran2019rewriting, duran2021resource, barba2021flexible, hou2021bottleneck}} \\
        \hline
        Role \fabian{(17)} & \fabian{\citep{bussler1995policy, kumar2002dynamic, xu2009resource, delias2010optimizing, kumar2013optimal, xu2013incorporating, djedovic2016optimization, havur2016resource, schonig2016framework, wibisono2016dynamic, bellaaj2017obstacle, hirsch2017information, yaghoibi2017cycle, jemel2020rpminter, yu2020task, zhao2020human, hou2021bottleneck}} \\
        \hline
        Expertise \fabian{(13)} & \fabian{\citep{huang2010adaptive, kamrani2012framework, liu2013accelerating, xu2013incorporating, schall2014crowdsourcing, djedovic2016optimization, bessai2016business, zhao2016entropy, arias2018all, Luo2019, pereira2020new, yu2020task, zhao2020human}} \\
        \hline
        Social Context \fabian{(8)} & \fabian{\citep{kumar2013optimal, liu2013accelerating, schall2014crowdsourcing, abdulhameed2018resource, lee2019dynamic, zhao2020human, liu2022multi, yeon2022experimental}} \\
        \hline
        \fabian{Resource Amount (5)} & \fabian{\citep{doerner2006enriched, xie2016dynamic, si2018petri, xie2019integration, duran2021resource}} \\
        \hline
        \fabian{Trustworthiness (1)} & \fabian{\citep{jemel2020rpminter}} \\
        \hline
        Experience (1) & \citep{zhao2020human} \\
        \hline
        Preference (1) & \citep{huang2010adaptive} \\
        \hline
        Any (5) & \citep{doerner2006enriched,bussler1995policy,ihde2022framework,erasmus2018method,cabanillas2013priority} \\
    \hline
    \end{tabular}
    \caption{Classification of considered resource attributes, based on \citep{arias2017towards}.}
    \label{tab:resatt}
\end{table}

As shown in Table~\ref{tab:resatt}, we found that most studies \fabian{(32 studies)} consider attributes from the category previous performance of a resource. In this category, most studies consider cost \fabian{(14 studies)}
%and time (4 %studies)\ifLong{~\citep{xu2013incorporating,%havur2016resource,bellaaj2017obstacle,van2001optimal}} 
as performance criteria. %For example, in \cite{arias2018all} cost is considered, among other dimensions, in a linear programming approach to suggest the best-fitting allocation of resources to tasks. Another frequent use of cost is to ensure that an allocation meets a pre-defined cost constraint, as in e.g.,~\cite{zhou2008project}.
\fabian{23} studies consider workload. For workload, availability of a resource is considered most often \fabian{(13 studies)}. 17 studies consider the role attribute and 
%We found that the role is most often encoded directly as an attribute (9 studies)\ifLong{~\citep{hirsch2017information,schonig2016framework,xu2013incorporating,havur2016resource,djedovic2016optimization,bussler1995policy,bellaaj2017obstacle,kumar2002dynamic,xu2009resource}}. 
13 studies consider attributes belonging to the expertise category. 
We found that social context and resource amount are considered less often \fabian{(8 and 5 studies)}. \fabian{Trustworthiness is considered only by~\citet{jemel2020rpminter}}, experience  only by~\citet{zhao2020human}, and preference only by~\citet{huang2010adaptive}. Five studies take a more flexible approach and can consider any quantifiable attribute. This approach, for example, can be achieved as in~\citet{doerner2006enriched}, by defining a custom cost function, or, as in  ~\citet{ihde2022framework}, by defining configurable input for the selected allocation algorithms. 
\subsubsection{Task attributes}
\label{susubbsec:task-attributes}
Contrary to ~\citet{arias2017towards}, we also considered task attributes. Task attributes can encode the requirements of what kind of resource is required; thus, they become part of the assignment constraints. These requirements can be mapped onto our previously identified attributes. For example, a task may require a resource with a certain role or past performance (requiring a certain level of quality or time in which the task \fabian{must} be performed). Additionally, we identified attributes influencing allocation decisions more indirectly, i.e., they \fabian{are} do not impose requirements on the resource:
\begin{itemize}
    \item \textbf{Estimated performance} describes performance attributes, such as duration or cost, which are considered for the allocation. Contrary to the \textit{previous performance} attribute of a resource, these are estimated from a task perspective.
    \item \textbf{Priority} contains attributes assessing the priority of a task. Priority may be encoded via a deadline or a simple scale indicating the importance of a task.
\end{itemize}
\begin{table}[tb]
    \centering
    \scriptsize
     \begin{tabular}{p{2.5cm}p{12.5cm}}
        \hline
        \textbf{Task Attributes} & \textbf{Corresponding Studies} \\
        \hline
        Req. Role (\fabian{20}) & \fabian{\citep{kumar2002dynamic, ha2006development, rhee2010increasing, xu2009resource, huang2012task, cabanillas2013priority, xu2013incorporating, schall2014crowdsourcing, djedovic2016optimization, bellaaj2017obstacle, hirsch2017information, bellaaj2020avoiding, jemel2020rpminter, pereira2020new, yu2020task, zhao2020human, barba2021flexible, duran2021resource, hou2021bottleneck, liu2022multi}} \\
        \hline
        Req. Expertise (\fabian{8}) & \fabian{\citep{kamrani2012framework, kumar2013optimal, schall2014crowdsourcing, erasmus2018method, Luo2019, liu2022multi, pereira2020new, bessai2016business}} \\
        \hline
        Req. Workload (\fabian{6}) & \fabian{\citep{eder2003personal, rhee2010increasing, wibisono2016dynamic, yaghoibi2017cycle, ihde2022framework, hou2021bottleneck}} \\
        \hline
        Req. \fabian{Resource Amount} (3) & ~\citep{djedovic2018innovative,kamrani2012framework,djedovic2016optimization} \\
        \hline
        Req. Social Context (\fabian{3}) & ~\citep{yaghoubi2016resource, yeon2022experimental, abdulhameed2018resource} \\
        \hline
        Req. Performance (1) & ~\citep{schall2014crowdsourcing} \\
        \hline
        \fabian{Req. Trustworthiness (1)} & \fabian{\citet{jemel2020rpminter}} \\
        \hline
        Estimated Performance (\fabian{27}) &  \fabian{\citep{barba2013user,doerner2006enriched,ha2006development,huang2012resource,kamrani2012framework,liu2012mining,rhee2010increasing,schonig2016framework,si2018petri,wibisono2016dynamic,xie2016dynamic,xu2013incorporating,yaghoibi2017cycle,zhao2016entropy,zhou2008project,havur2016resource,delias2010optimizing,huang2011mining,van2001optimal,zhao2015optimization,kumar2013optimal,bellaaj2020avoiding,duran2019rewriting}} \\
        \hline
        Priority (5) & \citep{djedovic2018innovative,zhao2016entropy,ihde2022framework,kumar2002dynamic,zhao2017resource} \\
        \hline
        \fabian{Any (5)} & \fabian{\citep{bussler1995policy, huang2010adaptive, pflug2016application, arias2018all, ihde2022framework}} \\
    \hline
    \end{tabular}
    \caption{Classification of the considered task attributes.}
    \label{tab:taskatt}
\end{table}

Table~\ref{tab:taskatt} shows the result of our classification. In total, \fabian{34} studies encode some requirements as task attributes.
Most studies require the resource to have a specific role (\fabian{20 studies}), certain expertise (\fabian{8} studies), or a certain workload (\fabian{6} studies). \fabian{Three studies consider \fabian{resource amount} and social context.} \fabian{Only} \citet{schall2014crowdsourcing} require a performance attribute and \citet{yaghoubi2016resource} require a social context attribute. \fabian{27} studies consider the estimated performance of a task. Of these, most studies \fabian{(23) consider time-related attributes (e.g., duration, service time) of a task}. Only five studies consider the priority of a task. \citet{jemel2020rpminter} require a certain degree of trustworthiness. While we found one study that considered experience \citep{zhao2020human} and one that considered the preference \citep{huang2010adaptive} of a resource, we did not find any study in which a task would in turn require an experience or preference attribute for allocation. \fabian{Five studies provide flexible means to encode arbitrary quantifiable task attributes to be considered for allocation.}

%- Discussion: Many estimated performance, How %many evidence based? (Trained on log etc...) %\\
%- Discussion: Many Previous performance: %Trend towards process mining and machine %learning? \\
%- Discussion: Gap: Social Context, Preference %(Human Factors?)\\
%- Discussion: Surprising: Only few consider %Priority of tasks \\
%
%Not needed as you show it already beforehand
%\begin{figure}
%    \centering
%    \begin{subfigure}[b]{0.7\textwidth}
%        \includegraphics[width=\textwidth]{R_ATTRIBUTES.PNG}
%         \caption{Resource attributes}
%         \label{fig:resource_attributes}
%    \end{subfigure}
%    \hfill
    
%    \begin{subfigure}[b]{0.7\textwidth}
%        \includegraphics[width=\textwidth]{T_ATTRIBUTES.PNG}
%         \caption{Task attributes}
%         \label{fig:task_attributes}
 %   \end{subfigure}
    
%    \caption{Attributes considered during allocation.}
%    \label{fig:attributes}
%\end{figure}

\subsection{RQ4: Solution Techniques}
\label{subsec:solution}

%At the end of the day, resource allocation can be treated as a problem that can be solved.
%Depending on the modeling of the problem, the solution varies accordingly.
Resource allocation is typically viewed as an optimization problem \luise{that needs to be solved~\citep{kamrani2012framework,zhao2016entropy,park2023optimizing}}. In the following, we investigate \emph{RQ4: Which solution strategies are used?}
In the primary studies, we could identify different types of solution techniques, namely, the categories \textit{rules or logic programming, \luise{heuristics and metaheuristics, exact algorithms, and mining and machine learning}.}
\luise{A special case that does not fit the given categorization is \citet{ihde2022framework}, categorized as \emph{any}. The approach given by \citep{ihde2022framework} allows the process designer to define any solution technique and its goals for each activity individually.}

When deciding on a solution strategy, it is crucial to balance the effort and time needed to generate a fitting solution with the quality\footnote{In general, the solution quality can be measured \ingo{on the basis of a solution quality function, as captured} in Definition~\ref{def:SQ}, that calculates how close the solution is to the most optimized solution possible.} of the resulting solution~\citep{cormen2022introduction}. \luise{In this work, we want to provide an indicative categorization of solution techniques regarding their execution cost (i.e., their computational effort) and solution quality, as \fabian{typically} observed in the average case. \fabian{Figure~\ref{fig:solutiontechniquesCompare} provides an intuitive sketch of our categorization.} Exceptions might exist but are not shown here. 
Currently, a quantitative benchmark of each individual technique from the 61 studies is not feasible. This limitation arises \ingo{not only from the sheer number but also because most approaches lack publicly available prototypes; a re-implementation of the studies for comprehensive benchmarking is hampered by the replicability of many of the studies, as discussed in Section~\ref{subsec:maturity}.}}

Solution techniques that focus primarily on the quality of the result tend to have much higher execution costs (i.e., effort and time). The most prominent representatives of these techniques are exact algorithms solving linear or non-linear programs~\citep{Nickel2022}.
%, as shown in Figure~\ref{fig:solutiontechniquesCompare}. 
On the other end of the spectrum are solution techniques such as rules and logic programming, which minimize the execution cost in exchange for not finding the best solution in all cases~\citep{havur2016resource}.
Heuristics \luise{and metaheuristics} can be \fabian{positioned} in the middle of this spectrum.
%, as visualized in  Figure~\ref{fig:solutiontechniquesCompare}. 
They tend to result in a higher quality of the solution \fabian{compared} to rule-based approaches~\citep{havur2016resource,Nickel2022}, while \fabian{exhibiting} only slightly higher execution cost. \luise{Usually, metaheuristics \fabian{posses} a higher execution cost but are also often able to \fabian{provide} higher solution quality than heuristics~\citep{xu2016resource}.
Some approaches are trained based on historical data: \fabian{mined rules/heuristics and machine learning approaches.} Rules\/heuristics are mined from historical process execution data. \fabian{As a result,} solution quality tends to be higher than the non-mined version~\citep{zhao2016entropy}. \fabian{Still}, mined rules also only consider a constraint solution space. The execution costs are, on average, higher if the mining activity is also \fabian{taken into account---for example, by apportioning it as cost} to the execution costs per allocation. \fabian{Compared to mined rules, machine learning approaches tend to find, on average, higher quality solutions.} Machine learning acts problem-specific due to the knowledge deduced from history. \fabian{However}, the effort is higher due to the more \fabian{involved} training phase.}

\begin{figure}[bt]
	\centering
	\includegraphics[width=0.45\textwidth]{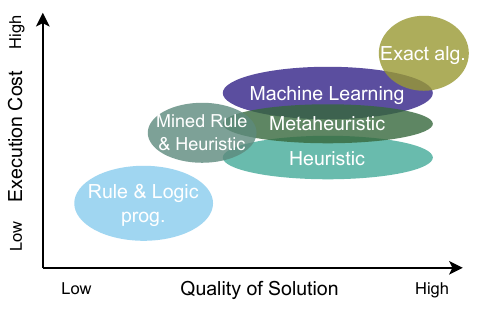}
	\caption{Identified solution strategies for resource allocation in business processes and an indicative categorization regarding their execution cost and solution quality in the average case.} 
	\label{fig:solutiontechniquesCompare}
\end{figure}

%The solution techniques can be further distinguished regarding whether a general or individualized algorithm is used. Whereas general solution techniques are defined beforehand and used in every allocation from then on, individualized algorithms try to find a solution technique that adapts to one resource allocation scenario individually. \luise{So called dynamic approaches, such as trained rules, genetic algorithms or machine learning approaches}, are used to create individualized techniques. Such approaches can deliver a high solution quality, but dependent on the training data used.

\begin{figure}[tb]
	\centering
	\includegraphics[width=0.6\textwidth]{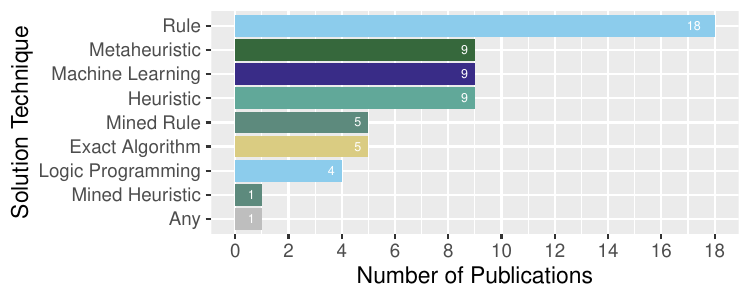}
	\caption{Found solution techniques assigned to algorithm categories.} 
	\label{fig:solutiontechniques}
\end{figure}

%\noindent In Figure~\ref{fig:solutiontechniques}, the different solution techniques of the categories mentioned above, grouped by their allocation goals, are shown. Additionally, one primary study by~\citet{ihde2022framework} provides an approach where the designer can specify any solution technique applied for a process activity.
In the following, we \fabian{discuss} the solution approaches in more detail, starting with techniques \fabian{which tend towards} lower execution cost and quality: rules and logic programming %\fabian{Followed by} 
\ingo{$\rightarrow$} heuristics \luise{and metaheuristics} \ingo{$\rightarrow$} approaches using \luise{exact algorithms}\ingo{$\rightarrow$} \luise{approaches that use historical data: mining and machine learning approaches}. 
At the end of this section, we compare the solution techniques regarding their support concerning single- or multi-task optimization.

\subsubsection{Rules \& Logic Programming}
\luise{\ingo{A large share} of the studies (22 studies) follow a rule-based solution technique. Rules minimize} the effort and time to find a suitable solution \fabian{but tend to provide lower quality results} compared to \fabian{other} types of solution techniques~\citep{cormen2022introduction}. 
There could \ingo{be} cases where a rule-based approach returns a high-quality solution. However, in the average case, this cannot be expected~\citep{havur2016resource}. \luise{These techniques are especially useful in scenarios where a solution needs to be found in a relatively short timeframe.}
\luise{For example, \citet{ha2006development} propose a dynamic rebalancing rule executed as soon as a resource is idle. It assigns a task to the resource's task list by a selected dispatching rule, such as the earliest due date.
Some studies use information about the resources previously identified from the process execution data. For instance, \citet{bellaaj2017obstacle} deduce so-called obstacle-aware resource indicators. \fabian{These indicators help assess the performance of a resource in terms of time, cost, and quality under a specific workload.} These indicators are then considered in a rule in combination with the current workload to identify an ordered list of candidate resources.}

We also decided to group logic programming approaches with rules, as logic programming defines constraints for machine execution, which are later used for BPMSs to execute processes. Therefore, logic programming approaches \ingo{strongly} resemble rule-based approaches.
\fabian{However, with only} \luise{four~\citep{barba2013user,havur2016resource,jemel2020rpminter,barba2021flexible} studies out of the 22, they represent a minority.} 

%As shown in Figure~\ref{fig:solutiontechniques}, rule-based approaches mostly concentrate on finding the most qualified resource for a task\ifLong{~\citep{huang2012resource,liu2013accelerating,schall2014crowdsourcing,erasmus2018method,lee2019dynamic,pereira2020new,kumar2002dynamic,bellaaj2017obstacle}}. 
%Furthermore, rule-based approaches support minimizing the cycle time of processes\ifLong{~\citep{zhao2015optimization,wibisono2016dynamic,havur2016resource,xu2013incorporating}}, balancing the workload\ifLong{~\citep{yaghoibi2017cycle,ha2006development}},
%minimizing the process cost\ifLong{~\citep{xu2009resource}}, or prioritizing tasks\ifLong{~\citep{barba2013user}}.
%A special case is the study by \citet{bussler1995policy} \luise{following no specific goal and by \citet{barba2021flexible} allowing to specify an objective function flexibly.} %but instead introduces a more generic approach that allows the creation of customized goals depending on the use case.

\subsubsection{Heuristics and Metaheuristics}
%The second group of solution techniques are heuristics. 
\luise{The second biggest \fabian{identified} group comprises heuristics and meta-heuristics (18 studies).
\\
\emph{Heuristics} are specialized algorithms tailored to the unique characteristics of a specific optimization problem~\citep{stork2022new}. They try to provide fast numerical solutions, but}
\ingo{aim} to minimize the chance of ending up with only a locally best solution. Therefore, they tend to generate, on average, higher-quality solutions than rule-based approaches. However, they \fabian{tend to require} more computational effort and time to produce a solution~\citep{Nickel2022}.
\luise{In sum, nine studies use a heuristic. 
For instance, \citet{kumar2013optimal} formulate an assignment problem \fabian{with the goal to maximize the resource compatibility}. Based on this, they develop a greedy heuristic where the compatibility is optimized for running cases sequentially but not globally. By comparing the heuristic with the exact solution of the optimization algorithms, the authors \fabian{find} that the solution quality of the heuristic is, on average, worse by 17--19\%.}
\\
\luise{\emph{Metaheuristics} are general-purpose optimization algorithms that are problem-independent. They are versatile and can be applied to a broad spectrum of different problems and their specific instances~\citet{stork2022new}. Six of the nine studies using a metaheuristic apply a genetic algorithm. For example, \citet{xu2016resource} use a genetic algorithm to assign the tasks of a set of running process cases to available resources, maximizing the process throughput. It is compared to a set of heuristics which it outperforms. Other works apply stochastic branch-and-bound~\citep{doerner2006enriched}, ant colony optimization~\citep{huang2012task}, and particle swarm optimization~\citep{zhao2017resource}}
%Because of their capability of balancing execution effort with solution quality, heuristics count as a best practice for most problem-solving strategies. 
%Out of the 49 studies, \luise{twelve} papers proposed a heuristics-based approach to solve the resource allocation problem.
%In contrast to rule-based approaches, the goal of heuristic approaches seems to be more balanced.
%The most common goal is to \luise{minimizing process costs}\ifLong{~\citep{kumar2013optimal,bessai2016business,xie2016dynamic,zhao2017resource}}. 
%Other process-oriented goal types are also represented as given in Table~\ref{tab:studies}. For example, minimizing cycle time\ifLong{~\citep{doerner2006enriched,kamrani2012framework}}
%, \luise{finding the most qualified resource} \ifLong{~\citep{cabanillas2013priority,zhao2016entropy}}, or maximizing the throughput of a process\ifLong{~\citep{soeffker2019adaptive}}.
%Additionally, \luise{three studies focus on resource-oriented goals} that try to balance the resource's workload across multiple processes\ifLong{~\citep{eder2003personal,rhee2010increasing}}.
%\citet{duran2019rewriting}'s approach uses a generic solution that can be redefined for any goal depending on the use case. 

\subsubsection{Exact algorithms}
\luise{Five studies have employed exact algorithms, also referred to as non-heuristic or complete algorithms~\citep{stork2022new}. These studies formulate the optimization problem as either a linear or a non-linear program. Alternatively, \citet{delias2010optimizing} formulates a  continuous optimization problem transferred into a discrete one. 
\\
\ingo{An exact} solution approach, such as the branch-and-bound algorithm, evaluates each possible solution and compares \fabian{it} to find the best one.}
This guarantees an optimal solution in contrast to the approaches presented before. The drawback is that evaluating every possible solution \fabian{requires} high computational \fabian{effort}.
\luise{An example of this group is \citep{hirsch2017information}: they formulate a nonlinear mixed-integer programming problem and produce a linearized version. The authors suggest solving it with a professional tool and, additionally, propose heuristics for the cases where no solution can be found in an acceptable time.}

%Regarding the goals, two of the studies focused on minimizing the cycle time of processes\ifLong{~\citep{xie2016dynamic,hirsch2017information}}, whereas~\citet{arias2018all} focuses on another process-oriented goal, namely searching for the most qualified resource.
%\citet{delias2010optimizing,hou2021bottleneck} propose an algorithm that balances the workload of resources, thus being resource-oriented.

\subsubsection{Mining and Machine Learning}
%So far our classification of groups only looked at the trade-off between computation time and the quality of the result. 
%The previously presented techniques have in common that they are created once and cannot be changed during runtime, thus resulting in a rather static behavior.
The last group of approaches uses solution techniques that are individually adapted to a particular business process scenario \luise{by learning from \fabian{historical data}}. Therefore, they tend to result in higher quality allocations. \luise{An additional effort is the mining or training on the historical data that could be apportioned to the execution costs per allocation. \fabian{The cost} highly depends on how often the mining or training is executed. \fabian{Further points to consider are}:} (i) the dependence on the quality of the historical data, and (ii) whenever a change occurs in the setting (i.e., a new resource gets added, processes change, etc.), the algorithm has to be retrained.
\luise{11} studies support individualized resource allocation techniques for certain business process scenarios:

\begin{itemize}
    \item \emph{Mined rules \luise{and heuristics}}: \luise{Six} papers used their adaptive approach to learn allocation rules \luise{\fabian{or} heuristics} from past process executions. \luise{Most studies use an association rule mining approach to identify relevant rules. For example, \citet{schonig2016framework} mine the resource patterns~\citep{russell2005workflow}, such as the \emph{Retain Familiar} pattern, from an event log. }
   %the goal of finding the most qualified resource was chosen in \luise{most} of the works\ifLong{~\citep{liu2012mining,abdulhameed2018resource,Luo2019,yeon2022experimental}}. Another study by~\citep{schonig2016framework} focuses on minimizing the cycle time. Lastly, ~\citet{huang2011mining} has no goal but aims to enhance existing resource allocation techniques with association rules.
    %\item \emph{Genetic algorithms}: Seven papers suggest approaches based on genetic algorithms. They belong to the so-called \emph{meta-heuristic approaches}, resulting in learned heuristic algorithms. Except for \citet{xu2016resource} focusing on optimizing the workload of resources, the remaining aim at the minimization of process costs\ifLong{~\citep{zhou2008project,djedovic2016optimization,djedovic2018innovative,xie2019integration,huang2012task,si2018petri}}.
    \item \emph{Machine learning}: \luise{The remaining nine studies use supervised machine learning approaches (\fabian{3 studies}, e.g., decision tree, bayesian neural network), unsupervised machine learning approaches (\fabian{3 studies}, e.g., clustering, generative probabilistic learning)  and reinforcement learning \fabian{(3 studies)}. An example \ingo{of the latter} is \citet{huang2011reinforcement}: The authors apply Q-learning to make allocation decisions similar to a human decision-maker as soon as a new task is enabled. Based on a predefined goal, \ingo{the new task is assigned} to a qualified %and estimated available 
    resource. The reinforcement agent continuously learns from the dynamics of the past process execution to make \ingo{better} decisions in the future. Comparing the approach to simple dispatching rules, such as FIFO, the authors show that it results in higher solution quality. \citet{yaghoubi2016resource} reuse this approach to minimize the entropy of a task list and allocate similar tasks to a resource. 
    } 
    %like reinforcement learning, decision trees, or clustering algorithms to learn heuristic solution techniques for resource allocation problems. %\luise{Most of these studies focus on minimizing the process costs, followed by time-oriented goals, such as minimizing the cycle time and throughput. Two are also focusing on resource-oriented goals, such as optimizing the worklist and maximizing the capability development of resources}. \ifLong{Two works focused on minimizing the process cost~\citep{huang2011reinforcement,bellaaj2020avoiding}. In contrast, the other papers set their goal to finding the best-fitting resource~\citep{zhao2020human}, maximizing the throughput~\citep{pflug2016application}, minimizing the cycle time~\citep{zhao2015optimization} or reducing the entropy of worklist entries~\citep{yaghoubi2016resource}.} \citet{huang2010adaptive,park2023optimizing} introduce a more generic approach that can be adapted to any goal depending on the concrete use case.
\end{itemize}

\paragraph{\fabian{Comparing} \ingo{single vs.\ multi-task} optimization.}
Generally, a resource allocation decision is made at a specific time, meaning the currently available information about tasks and resources is then used to create a solution. We categorized the approaches that focus on a narrow scope, such as finding the most qualified resource for a specific process task \fabian{as} \emph{single-task} optimization approaches (\fabian{24 studies}). Approaches that consider all available tasks of a process or an organization and their different level of importance are considered as \emph{multi-task} optimization (\fabian{37 studies}).

Resource allocation, in general, is primarily an NP-hard problem. Therefore, a common way to handle these problems is to reduce the complexity as much as possible while still \fabian{attaining} meaningful results. 
Single-task optimization achieves this by reducing the complexity by limiting the problem to select the best-matching resource for a specific task.
This might be useful in business processes where resources are not limited or the importance between tasks is the same.
Even though a single-task approach reduces the complexity of the problem, most studies further limit their approach by only allowing a 1-to-1 allocation \fabian{of tasks to resources} (except four papers~\citep{liu2013accelerating,schall2014crowdsourcing,bussler1995policy,van2001optimal,zhao2020human}).
%Another observation is that most of these studies have a process-oriented goal. Exceptions are four papers, which either had a resource-oriented goal in optimizing worklists~\citep{barba2013user} or no specific goal~\citep{bussler1995policy,huang2011mining,huang2010adaptive,park2023optimizing}.

%Nevertheless, the most expensive and limited resources are often shared between process tasks to make the allocations as efficient as possible. \fabian{In these cases,} a multi-task optimization method is \fabian{required}.
\fabian{Most studies (37)} support a multi-task optimization approach.
Despite the higher complexity of multi-task approaches, only \luise{21} papers reduce the complexity by limiting the allocation to a 1-to-1 allocation of tasks to resources.
%In contrast to single-task optimization, the global approaches support a broad range of optimization goals. %They target resource-oriented goals (\luise{nine} studies), process-oriented goals (\luise{24} studies), and no specific goals (\luise{four} studies). 

\begin{figure}[bt]
	\centering
	\includegraphics[width=.65\linewidth]{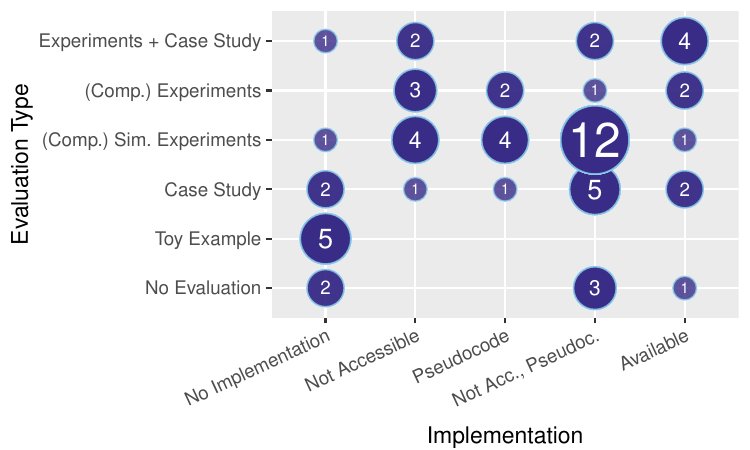}
	\caption{Evaluation methods and the associated research prototypes per method.} 
	\label{fig:eva}
\end{figure}

\subsection{RQ5: Evaluation and Research Prototypes}
\label{subsec:maturity}
In this subsection, we discuss the methods applied by the studies to evaluate their approach and the availability of research prototypes.
Thus, in this subjection, we investigate \emph{RQ5: How applicable are the proposed resource allocation approaches, \ingo{given the} availability of evaluations and prototypes?}

Based on work by~\citet{zelkowitz1998experimental}, we have categorized the approaches into those having no evaluation, argumentation on a toy example (i.e., an assertion with regards to~\citet{zelkowitz1998experimental}), case study, controlled experiments, whereby we differentiated between simulation experiments with synthetic data, and experiments with real-world data, and a combination of methods (\emph{Experiments + Case study}).
Furthermore, we distinguish between works having no implementation, a not accessible prototype, only pseudocode available, or a prototype that is not (publicly) but has accessible pseudocode, and finally, a prototype available.
Figure~\ref{fig:eva} shows the number per evaluation category and implementation type as a bubble diagram.
%In the following, we present the used evaluation methods in the order given in the previous paragraph starting from no and toy examples until the combination of methods.

Surprisingly, we can observe that \luise{five} studies provide no evaluation at all. \luise{Five} discuss their approach on a toy example, which is rather an assertion about the functionality and usefulness of their approach.
By summing up these two categories, it can be observed that 18\% of the studies have not adequately evaluated their approaches.
~\citet{cabanillas2013priority} is an interesting case in this category: the authors provide a Java implementation where the approach can be tested and used but did not include an evaluation in their paper.
Three other works with no evaluation discuss a prototype, but they are not publicly accessible; only pseudocode is provided.

\luise{Eleven} studies employ case studies to evaluate their approach. The advantages of a case study are that the implications of the resource allocation approach can be studied in detail, and interesting insights can be found. However, it is challenging to see which results are generalizable and which are not~\citep{zelkowitz1998experimental}.
Most of these approaches have a prototype in place to apply the approach to the selected case.
However, only two prototypes are publicly accessible: ~\citet{huang2012resource} provide a plugin for the process mining toolkit ProM~\citep{kalenkova2014discovering}, and~\citet{ihde2022framework} provide a stand-alone system implementation.

A majority of studies evaluate their approaches with controlled experiments (see the \emph{[Comp.] sim. experiments} and \emph{[Comp.] experiments} in Figure~\ref{fig:eva}), whereby most of the studies use synthetic data from simulations (\luise{23} studies).
Many of these simulation experiments evaluate different parameters of their approaches with regard to different settings.
\luise{Six} studies~\citep{huang2012task,wibisono2016dynamic,xu2016resource,djedovic2018innovative,soeffker2019adaptive,hou2021bottleneck} also conduct comparative simulation experiments, in which their solution is compared to other approaches.
Only \citet{duran2019rewriting} provide a prototype publicly available on a website.
For the other \luise{22}  \emph{[Comp.] sim. experiments}-studies in this category, no prototype is accessible, but most of them provide pseudocode  (\luise{16} studies).

\luise{Eight} studies have conducted experiments with real-world data. 
This type of evaluation can act on more realistic data and typically provide observations and insights with higher confidence that these would hold in practice.
Five conducted \fabian{comparative} experiments~\citep{huang2011reinforcement,liu2012mining,zhao2016entropy,zhao2017resource,yaghoibi2017cycle}.
ProM plugins as prototypes are provided by~\citet{huang2010adaptive,huang2011reinforcement}. 
The other studies of this category provide mainly pseudocode. However, several make neither a prototype nor pseudocode available.

\luise{Nine} studies use a combination of evaluation methods with controlled experiments and case studies (see the \emph{Experiments + Case studies} in Figure~\ref{fig:eva}) to strengthen the evaluation of their approaches.
\citet{arias2018all} also provide a ProM plugin and \citet{xie2016dynamic} a MathLab implementation as research prototypes. \luise{\citet{pika2021machine} and \citet{park2023optimizing} provide stand-alone implementations.}

In summary, we can observe that many studies provide no publicly available prototype, but mainly pseudocode. Pseudocode heavily depends on how detailed the concepts of an approach are described within the study. 
Some studies with no accessible prototype have linked to a stand-alone solution, which is not accessible anymore.
Of the \luise{ten} studies providing a prototype, five have used public platforms, i.e., the process mining platform ProM~\citep{huang2010adaptive,huang2011reinforcement,huang2012resource,arias2018all} or the mathematical toolbox MathLab~\citep{xie2016dynamic}, to implement their resource allocation approaches. In contrast, the other \luise{five} implemented stand-alone prototypes.

%%%%%%%%%%%%%%%%%%%%%%%%%%%%%%%%%%%%%%%%%%%%%%%%%%%%%%%%%%%%%%%%%%%%%%%%%%%%%%%%%%%%%%%%%%%%%%%
\section{Discussion}
\label{sec:dis}

\ingo{The literature survey presented above} shows a strong global research interest in automatic support for resource allocation in business processes and a variety of approaches based on different techniques. 
Based on our results, we observe several open research challenges \ingo{(see Section~\ref{subsec:open}). Furthermore, we discuss implications for practitioners in Section~\ref{subsec:practicioners} and threats to validity in Section~\ref{subsec:limit}.}

\subsection{\fabian{Observations and Future Directions}}
\label{subsec:open}

\fabian{Our various classification results can be leveraged to spot gaps and neglected research directions. In this Section, we report on our main observations and envision future research to: leverage process execution data, explore additional aspects of resource and task characteristics, increase adaptability, and conduct comprehensive performance studies.}

\subsubsection{Leverage Process Execution Data}
Most approaches use a process model with estimations of the process dynamics (e.g., activity duration, arrival rates of new process cases) as input for the resource allocation algorithm (c.f. Section~\ref{subsec:processmodel}).
%\ifLong{So far, we could not analyze what type of control-flow structures (e.g., loops) are supported by the studies because only a minority of studies report on this aspect. In the future, we suggest replicating the studies and benchmarking them with the help of different process models.}
\fabian{However, this leaves aside information on past executions of processes, which is available in historic process execution data (e.g., event logs).}
Such datasets have become more available and accessible in recent years. 
\fabian{They can be leveraged to replace or fine-tune estimates of process dynamics by investigating the past performance of processes.
Several approaches already use this data to gain insights into a resource's behavior and preferences. However, characteristics of tasks (e.g., similarity of tasks) are less investigated. Also,
only two studies use process execution data to learn about the overall dynamics of a business process. 
\citet{yeon2022experimental} discover an enhanced process model and \citet{djedovic2018innovative} create a process simulation model from execution data.
We expect that future research will increasingly make use of historic execution data. Here, we see a line for future research in making use of aspects of execution data previously not explored and using execution data as input for solution techniques.} %Generally, leveraging process execution data and learning about the process dynamics may give insights into which task and resource characteristics are particularly relevant for optimal resource allocations.

\subsubsection{\fabian{Explore Additional Aspects of Resource and Task Characteristics}}

While many approaches consider attributes like cost or availability of a resource\fabian{, only few published studies consider resources' preferences~\citep{huang2010adaptive}, their experience~\citep{zhao2020human} or trustworthiness~\citep{jemel2020rpminter}.}
%and their social context (e.g.,~\citep{schall2014crowdsourcing}) for performing an allocation additionally. These attributes seem to be especially relevant for knowledge-intensive processes, where the process is mainly managed by the knowledge of case workers~\citep{di2015knowledge}. 
Task characteristics, such as the priority of tasks, are also \fabian{considered} \ingo{less frequently}. However, these may play a similarly relevant role in finding a suitable resource allocation. 
Our classification of resource and task attributes in Section~\ref{subsec:attributes} can be leveraged to identify such gaps and investigate them in future research.
%\ifLong{
\fabian{Similarly, only \citet{doerner2006enriched} considers non-application resources. 
Such research can be important to process cases where non-application resources constitute a real bottleneck; for example, when highly specialized and costly equipment is required to complete a task. We envision a future line of research investigating such processes.}
%}

\subsubsection{\fabian{Increase Adaptability}}

Most approaches focus on human resource allocation and specific allocation goals (e.g., minimizing the process cost).
Few offer the possibility to customize the resource allocation goal, e.g.,~\citep{huang2011reinforcement}.
Many approaches published in BPM outlets follow the idea that, to reach the overall optimization goal, the resource allocation approach should be the same for all process activities. In contrast, in operations research, techniques are proposed and developed that match the specifics of a particular activity, e.g., which type of resources are needed. \citet{ihde2022framework} propose an approach to select a resource allocation technique for \ingo{specific} process activities individually. \fabian{This follows the idea of the resource patterns~\citep{russell2005workflow} in BPMSs that can be selected per activity. However, these kind of approaches give no decision support on how an allocation approach applied to a process activity influences the overall process goal.}

%Developing different versions of the solution approach might also be relevant to support different quality and time goals.
Thus, we envision future research exploring approaches adaptable to different process settings, \fabian{and providing decision support on when a given approach should take preference over another, contributing to the overall allocation goal.}

\ifLong{\subsubsection{Leveraging machine learning techniques.}
With the increased computational power, processing large amounts of data becomes more feasible. We also noticed a growing trend of approaches that use machine learning techniques to create and/or improve resource allocation algorithms.  From our comparison to other approaches, we can narrow this observation down to the following: machine learning approaches tend to create individually adapted algorithms, which often perform better than more general approaches if some assumptions hold. However, the studies' data sets included only limited resource information. They are only concerned with the past execution of processes but narrowed information about resources, like the working hours, capacity, current workload, etc., necessary for real-world resource allocation. This limitation leads to unrealistic allocations, as this information is not provided. We believe machine learning approaches show promising potential. However, whether they can be applied in real-world scenarios has to be shown on more complete data sets.}

\subsubsection{\fabian{Conduct Comprehensive Performance Studies}}\label{subsec:benchmark}
\fabian{In Section~\ref{subsec:maturity}, we found that most studies use synthetic data in simulations to demonstrate the effectiveness of their approach. 
So far, no large scale benchmark studies have been conducted comparing different resource allocation approaches, replicating and contextualizing their results. We believe this is a pressing research topic. Such research can reveal new insights and future research directions and support practitioners when transferring different approaches into real-world applications.
Such a study should go beyond runtime performance. As we discussed in Section~\ref{subsec:solution}, usually a trade-off between quality of the allocation and time to find an allocation exists. For example, rules may \ingo{be fast but lead to solutions} which might not always be optimal, whereas linear programming usually provides high-quality solutions but may \ingo{take longer to compute}.}
%The primary studies do not always discuss this aspect in detail.
%In the future, the approaches should discuss the trade-off between solution quality and time.
\fabian{However, conducting such a study will be challenging.
Only a minority of studies provide a replicable evaluation of their approach and a publicly available prototype.}
%$such that the approach is directly replicable and applicable. 
\fabian{In the future, researchers should emphasize making their approaches, prototypes, and data sets available, ideally following open science principles enabling replicability and comparability.}

\fabian{For studies utilizing comparative experiments to compare the effectiveness to other approaches we could observe that the selection of approaches for the comparison was often not done in a transparent and replicable manner. 
The results of this SLR can help researchers identify related approaches for evaluation in a more structured manner in the future.}
%Many prototypes of the discussed studies were not available, which hinders the application in practice and the comparison between approaches.
%\ifLong{We could also observe that research prototypes implemented as plugins for platforms seem to have a higher chance of remaining available and executable after some years.}

\subsection{Insights for Practitioners}
\label{subsec:practicioners}
\fabian{Classifications like ours 
can provide insights into the design process of artifacts~\citep{williamsDesignEmergingDigital2008}.
In this section, we propose a decision flow to illustrate how our various classifications can be used to guide the design of future practical implementations. 
In particular, our decision \ingo{flow---depicted in Figure~\ref{fig:practitioners}---}can be used to select relevant studies given specific requirements. This flow can be used to identify applicable studies, guide practical implementations, and use these techniques in resource allocation implementations of BPMSs.
%To support practitioners in selecting relevant studies for a business process scenario and researchers in selecting related work, we suggest a decision flow shown in Fig.~\ref{fig:practitioners}.
% The decision flow is given in Figure~\ref{fig:practitioners}.#
}
First, we suggest identifying the required allocation capability. If a 1-to-1 allocation of one task to one resource is needed, all studies can be considered because studies supporting 1-to-many or many-to-1 also support the basic variant of 1-to-1. If 1-to-many or many-to-1, the related studies should be selected (cf. Section~\ref{subsec:goals}). \ingo{Subsequently}, \fabian{it should be decided whether multiple to-be-allocated tasks of a process and their different levels of importance should be considered at any allocation}. If not, again, all studies can be considered. \fabian{Otherwise, a multi-task optimization approach is required} (cf. Section~\ref{subsec:solution}).

The presented studies support different goals in their resource allocation approach. It is helpful to select studies that support the same goal as in the intended business scenario or studies that support \emph{Any} goal (cf. Section~\ref{subsec:goals}). An alternative is to adopt and adapt an approach in a way that follows the needed optimization goal.

\begin{wrapfigure}{r}{0.35\textwidth}
	\centering
	\includegraphics[width=.34\textwidth]{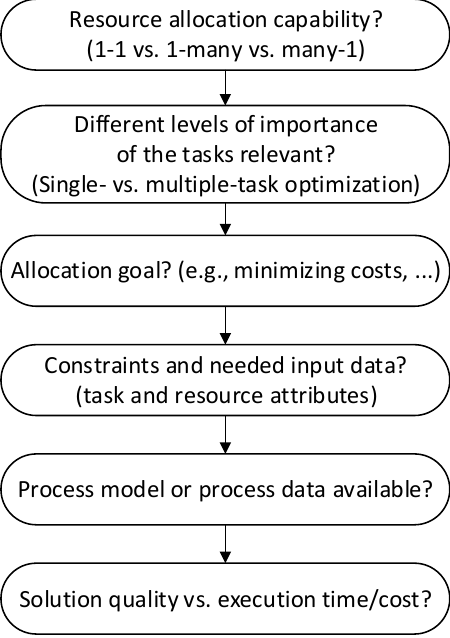}
	\caption{Decision flow for selecting relevant studies for an application process.} 
	\label{fig:practitioners}
\end{wrapfigure}
After the goal of the resource allocation has been defined; also the constraints need to be specified for which different resource and task attributes \fabian{need be considered (c.f.~\ref{subsec:attributes}).} Based on these, studies can be selected that also apply similar attributes. \fabian{Otherwise, the selected approaches could be extended by the required constraints.}

A fifth relevant selection criterion is the available process information. \fabian{If process execution data is available, a process model can be discovered with the help of process discovery techniques~\citep{dumas2018fundamentals}; then, all remaining studies can be considered.} If not, studies that support a process model as input should be selected.
\fabian{Finally, the trade-offs between solution quality an execution time/cost \ingo{have} to be considered}. As discussed in Section~\ref{subsec:solution}, \fabian{some classes of solution techniques tend to} exhibit higher solution quality but may result in higher execution time/cost, such as linear approaches. Others tend to exhibit low execution times/costs but may lead to lower solution qualities, such as rules.

%While we propose a decision flow with the aim to aid future practical implementations, we have to note that the applicability of most identified approaches is limited due to the limited replicability of conducted evaluations and limited availability of research prototypes, stressing our points made in Section~\ref{subsec:benchmark}.

\subsection{Threats to Validity}
\label{subsec:limit}
In an SLR, biases in the selection of the studies and the data extraction process can be threats to the validity of its results~\citep{cooper2015research}.
To avoid selection bias in the study search, we followed a specific search protocol described in Section~\ref{subsec:primarysearch}. We further conducted the relevance check based on defined inclusion and exclusion criteria with at least two co-authors, with discussions between them for all cases of disagreement.
The primary search needed to be limited by the search terms and was additionally limited by focusing on journal articles. 
We have observed a broad range of short conference/workshop articles on this research topic, presenting idea sketches. So, we decided to focus on more mature work expected to be found in journals.
We complemented our search by considering conference and workshop papers in the secondary search, the backward/forward search. 
The study search was initially conducted in mid-2019; to consider more recent publications, we repeated the search at the beginning of 2021 \fabian{and again at the end of 2023}.
In the full-text reading, we excluded duplicates in content. We only kept the more mature version of the papers, usually the journal article, to avoid \fabian{double counting} one approach. 
The group of co-authors discussed each exclusion.
In sum, we selected \fabian{61} primary studies which we believe represent the research field well.
Still, the risk exists that relevant studies might not have been included since they did not meet our search criteria.

Additionally, biases \fabian{can} exist in the data extraction process\fabian{, which we mitigated as follows.}
Data coding was done first individually for each paper by different co-authors.
Issues and ambiguities were discussed with the co-authors.
After having finished the data extraction from all papers, data categories were distributed among the authors, and the data extraction per category was validated and standardized.
Nevertheless, studies sometimes do not provide the information directly on a particular aspect, so the authors need to form interpretations.

%%%%%%%%%%%%%%%%%%%%%%%%%%%%%%%%%%%%%%%%%%%%%%%%%%%%%%%%%%%%%%%%%%%%%%%%%%%%%%%%%%%%%%%%%%%%%%%
\section{Conclusion}
\label{sec:conclusion}
This survey provide a structured analysis of \luise{system-initiated} approaches for resource allocation in business processes.
The complexity of the problem of assigning tasks to available resources with different capabilities has led to various approaches with different strengths and weaknesses.
The structured literature search identified \ingo{61} studies providing resource allocation approaches published mainly in the last two decades.
In this survey, for the first time, the approaches have been analyzed in terms of their goals, capabilities, input data, techniques used, and evaluation methods. 
With this SLR, we studied \ingo{five} research questions \ingo{and come to the following conclusions.}

Regarding research question RQ1 (the targeted resource allocation capabilities and goals), we found that mainly 1-to-1 allocations between tasks and resources are supported, but also studies could be identified that support \ingo{many-to-1 and 1-to-many} allocations.
Several optimization goals are \ingo{pursued}, such as minimizing process costs, whereby process-oriented goals are mainly supported.
Regarding RQ2 (the role of process models and process execution data), \ingo{we found that} process models and estimations on the process dynamics are often used as inputs to the resource allocation approaches, and process execution data taken from IT systems increasingly \ingo{plays} a role.
\fabian{To answer RQ3, we classified resource and task attributes and the relation \ingo{between} them.}
This can be used in future research to identify possible attributes that were neglected and should be considered in the future.

Allocation rules have the disadvantage of not always leading to the best solution. Nevertheless, they are utilized in many approaches, as observed \ingo{in the context of} RQ4 (solution strategies). Because \ingo{such rules} provide solutions in a short timeframe, they are still relevant for many business processes.
\luise{Additionally, the primary studies used heuristics, metaheuristics, and exact algorithms as solution techniques, and trained rules and machine-learning approaches. The latter two provide more context-sensitive solutions if the training data is of good quality and representative of future cases.}
Regarding RQ5 (\ingo{evaluation strength and prototype availability}), we showed that many studies evaluate their approaches with simulation experiments, but only a few works provide publicly available prototypical implementations.

\fabian{This survey offers researchers an overview of existing approaches, open research challenges, and the possibility of identifying structurally related approaches for comparison.} \fabian{Based on our observations, we have outlined possible future research opportunities: leverage process execution data, explore additional aspects of resource and task characteristics, increase adaptability, and conduct comprehensive performance studies.}
\fabian{Our various classifications
can be used to guide the design of future research or practical implementations. %We have given an example of how our classifications could assist practitioners.
}
%Although we observed a strong research interest in automatic support for resource allocation in business processes in this study, we also found several open research topics.
%The following \ifShort{three}\ifLong{four} main points were deduced: 
%(1) process execution data could be \ingo{utilized} more to ensure evidenced-based resource allocation decisions, 
%(2) the variability of approaches could be increased to be applicable in different use cases, 
%\ifShort{and 
%(3) evaluations and availability of research prototypes could be extended to improve the understanding of usefulness and applicability of approaches}
%\ifLong{(3) machine learning could be leveraged more to provide context-specific solution approaches, and (4) evaluations and availability of research prototypes could be extended to improve the understanding of usefulness and applicability of approaches}.
\fabian{One major limitation of many existing works is their replicability. This makes comprehensive performance studies difficult. Thus, in the future, we plan to establish a benchmarking \fabian{framework} for resource allocation approaches to allow better comparisons of the functionality and complexity of the approaches.}

% Your references go at the end of the main text. For this document we've used BibTeX, the .bib file references.bib.

\setstretch{2.0}
\bibliographystyle{apacite}
\bibliography{references}

\begin{thebibliography}{}

\bibitem [\protect \citeauthoryear {%
Abdulhameed%
, Helal%
, Awad%
\BCBL {}\ \BBA {} Ezat%
}{%
Abdulhameed%
\ \protect \BOthers {.}}{%
{\protect \APACyear {2018}}%
}]{%
abdulhameed2018resource}
\APACinsertmetastar {%
abdulhameed2018resource}%
\begin{APACrefauthors}%
Abdulhameed, N\BPBI M.%
, Helal, I\BPBI M.%
, Awad, A.%
\BCBL {}\ \BBA {} Ezat, E.%
\end{APACrefauthors}%
\unskip\
\newblock
\APACrefYearMonthDay{2018}{}{}.
\newblock
{\BBOQ}\APACrefatitle {A resource recommendation approach based on co-working
  history} {A resource recommendation approach based on co-working
  history}.{\BBCQ}
\newblock
\APACjournalVolNumPages{International Journal of Advanced Computer Science and
  Applications}{9}{7}{}.
\PrintBackRefs{\CurrentBib}

\bibitem [\protect \citeauthoryear {%
Arias%
, Munoz-Gama%
\BCBL {}\ \BBA {} Sep{\'u}lveda%
}{%
Arias%
, Munoz-Gama%
\BCBL {}\ \BBA {} Sep{\'u}lveda%
}{%
{\protect \APACyear {2018}}%
}]{%
arias2017towards}
\APACinsertmetastar {%
arias2017towards}%
\begin{APACrefauthors}%
Arias, M.%
, Munoz-Gama, J.%
\BCBL {}\ \BBA {} Sep{\'u}lveda, M.%
\end{APACrefauthors}%
\unskip\
\newblock
\APACrefYearMonthDay{2018}{}{}.
\newblock
{\BBOQ}\APACrefatitle {Towards a Taxonomy of Human Resource Allocation
  Criteria} {Towards a taxonomy of human resource allocation criteria}.{\BBCQ}
\newblock
\BIn{} \APACrefbtitle {Business Process Management Workshops} {Business process
  management workshops}\ (\BPGS\ 475--483).
\newblock
\APACaddressPublisher{}{Springer International Publishing}.
\PrintBackRefs{\CurrentBib}

\bibitem [\protect \citeauthoryear {%
Arias%
, Munoz-Gama%
, Sep{\'u}lveda%
\BCBL {}\ \BBA {} Miranda%
}{%
Arias%
, Munoz-Gama%
, Sep{\'u}lveda%
\BCBL {}\ \BBA {} Miranda%
}{%
{\protect \APACyear {2018}}%
}]{%
arias2018all}
\APACinsertmetastar {%
arias2018all}%
\begin{APACrefauthors}%
Arias, M.%
, Munoz-Gama, J.%
, Sep{\'u}lveda, M.%
\BCBL {}\ \BBA {} Miranda, J\BPBI C.%
\end{APACrefauthors}%
\unskip\
\newblock
\APACrefYearMonthDay{2018}{}{}.
\newblock
{\BBOQ}\APACrefatitle {Human resource allocation or recommendation based on
  multi-factor criteria in on-demand and batch scenarios} {Human resource
  allocation or recommendation based on multi-factor criteria in on-demand and
  batch scenarios}.{\BBCQ}
\newblock
\APACjournalVolNumPages{European Journal of Industrial
  Engineering}{12}{3}{364--404}.
\PrintBackRefs{\CurrentBib}

\bibitem [\protect \citeauthoryear {%
Arias%
, Saavedra%
, Marques%
, Munoz-Gama%
\BCBL {}\ \BBA {} Sep{\'u}lveda%
}{%
Arias%
, Saavedra%
\BCBL {}\ \protect \BOthers {.}}{%
{\protect \APACyear {2018}}%
}]{%
arias2018human}
\APACinsertmetastar {%
arias2018human}%
\begin{APACrefauthors}%
Arias, M.%
, Saavedra, R.%
, Marques, M\BPBI R.%
, Munoz-Gama, J.%
\BCBL {}\ \BBA {} Sep{\'u}lveda, M.%
\end{APACrefauthors}%
\unskip\
\newblock
\APACrefYearMonthDay{2018}{}{}.
\newblock
{\BBOQ}\APACrefatitle {Human resource allocation in business process management
  and process mining: A systematic mapping study} {Human resource allocation in
  business process management and process mining: A systematic mapping
  study}.{\BBCQ}
\newblock
\APACjournalVolNumPages{Management Decision}{56}{2}{376--405}.
\PrintBackRefs{\CurrentBib}

\bibitem [\protect \citeauthoryear {%
Barba%
, Jimenez-Ramirez%
, Reichert%
, Del~Valle%
\BCBL {}\ \BBA {} Weber%
}{%
Barba%
\ \protect \BOthers {.}}{%
{\protect \APACyear {2021}}%
}]{%
barba2021flexible}
\APACinsertmetastar {%
barba2021flexible}%
\begin{APACrefauthors}%
Barba, I.%
, Jimenez-Ramirez, A.%
, Reichert, M.%
, Del~Valle, C.%
\BCBL {}\ \BBA {} Weber, B.%
\end{APACrefauthors}%
\unskip\
\newblock
\APACrefYearMonthDay{2021}{}{}.
\newblock
{\BBOQ}\APACrefatitle {Flexible runtime support of business processes under
  rolling planning horizons} {Flexible runtime support of business processes
  under rolling planning horizons}.{\BBCQ}
\newblock
\APACjournalVolNumPages{Expert Systems with Applications}{177}{}{114857}.
\PrintBackRefs{\CurrentBib}

\bibitem [\protect \citeauthoryear {%
Barba%
, Weber%
, Del~Valle%
\BCBL {}\ \BBA {} Jim{\'e}nez-Ram{\'\i}rez%
}{%
Barba%
\ \protect \BOthers {.}}{%
{\protect \APACyear {2013}}%
}]{%
barba2013user}
\APACinsertmetastar {%
barba2013user}%
\begin{APACrefauthors}%
Barba, I.%
, Weber, B.%
, Del~Valle, C.%
\BCBL {}\ \BBA {} Jim{\'e}nez-Ram{\'\i}rez, A.%
\end{APACrefauthors}%
\unskip\
\newblock
\APACrefYearMonthDay{2013}{}{}.
\newblock
{\BBOQ}\APACrefatitle {User recommendations for the optimized execution of
  business processes} {User recommendations for the optimized execution of
  business processes}.{\BBCQ}
\newblock
\APACjournalVolNumPages{Data \& knowledge engineering}{86}{}{61--84}.
\PrintBackRefs{\CurrentBib}

\bibitem [\protect \citeauthoryear {%
Bellaaj%
, Sellami%
, Bhiri%
\BCBL {}\ \BBA {} Maamar%
}{%
Bellaaj%
\ \protect \BOthers {.}}{%
{\protect \APACyear {2017}}%
}]{%
bellaaj2017obstacle}
\APACinsertmetastar {%
bellaaj2017obstacle}%
\begin{APACrefauthors}%
Bellaaj, F.%
, Sellami, M.%
, Bhiri, S.%
\BCBL {}\ \BBA {} Maamar, Z.%
\end{APACrefauthors}%
\unskip\
\newblock
\APACrefYearMonthDay{2017}{}{}.
\newblock
{\BBOQ}\APACrefatitle {Obstacle-aware resource allocation in business
  processes} {Obstacle-aware resource allocation in business processes}.{\BBCQ}
\newblock
\BIn{} \APACrefbtitle {International Conference on Business Information
  Systems} {International conference on business information systems}\ (\BPGS\
  207--219).
\PrintBackRefs{\CurrentBib}

\bibitem [\protect \citeauthoryear {%
Bellaaj~Elloumi%
, Sellami%
\BCBL {}\ \BBA {} Bhiri%
}{%
Bellaaj~Elloumi%
\ \protect \BOthers {.}}{%
{\protect \APACyear {2020}}%
}]{%
bellaaj2020avoiding}
\APACinsertmetastar {%
bellaaj2020avoiding}%
\begin{APACrefauthors}%
Bellaaj~Elloumi, F.%
, Sellami, M.%
\BCBL {}\ \BBA {} Bhiri, S.%
\end{APACrefauthors}%
\unskip\
\newblock
\APACrefYearMonthDay{2020}{}{}.
\newblock
{\BBOQ}\APACrefatitle {Avoiding resource misallocations in business processes}
  {Avoiding resource misallocations in business processes}.{\BBCQ}
\newblock
\APACjournalVolNumPages{Concurrency and Computation: Practice and
  Experience}{32}{15}{e4888}.
\PrintBackRefs{\CurrentBib}

\bibitem [\protect \citeauthoryear {%
Bessai%
\ \BBA {} Charoy%
}{%
Bessai%
\ \BBA {} Charoy%
}{%
{\protect \APACyear {2016}}%
}]{%
bessai2016business}
\APACinsertmetastar {%
bessai2016business}%
\begin{APACrefauthors}%
Bessai, K.%
\BCBT {}\ \BBA {} Charoy, F.%
\end{APACrefauthors}%
\unskip\
\newblock
\APACrefYearMonthDay{2016}{}{}.
\newblock
{\BBOQ}\APACrefatitle {Business process tasks-assignment and resource
  allocation in crowdsourcing context} {Business process tasks-assignment and
  resource allocation in crowdsourcing context}.{\BBCQ}
\newblock
\BIn{} \APACrefbtitle {2016 IEEE 2nd International Conference on Collaboration
  and Internet Computing (CIC)} {2016 ieee 2nd international conference on
  collaboration and internet computing (cic)}\ (\BPGS\ 11--18).
\PrintBackRefs{\CurrentBib}

\bibitem [\protect \citeauthoryear {%
Bussler%
\ \BBA {} Jablonski%
}{%
Bussler%
\ \BBA {} Jablonski%
}{%
{\protect \APACyear {1995}}%
}]{%
bussler1995policy}
\APACinsertmetastar {%
bussler1995policy}%
\begin{APACrefauthors}%
Bussler, C.%
\BCBT {}\ \BBA {} Jablonski, S.%
\end{APACrefauthors}%
\unskip\
\newblock
\APACrefYearMonthDay{1995}{}{}.
\newblock
{\BBOQ}\APACrefatitle {Policy resolution for workflow management systems}
  {Policy resolution for workflow management systems}.{\BBCQ}
\newblock
\BIn{} \APACrefbtitle {Proceedings of the Twenty-Eighth Annual Hawaii
  International Conference on System Sciences} {Proceedings of the
  twenty-eighth annual hawaii international conference on system sciences}\
  (\BVOL~4, \BPGS\ 831--840).
\PrintBackRefs{\CurrentBib}

\bibitem [\protect \citeauthoryear {%
Cabanillas%
}{%
Cabanillas%
}{%
{\protect \APACyear {2016}}%
}]{%
cabanillas2016process}
\APACinsertmetastar {%
cabanillas2016process}%
\begin{APACrefauthors}%
Cabanillas, C.%
\end{APACrefauthors}%
\unskip\
\newblock
\APACrefYearMonthDay{2016}{}{}.
\newblock
{\BBOQ}\APACrefatitle {Process-and resource-aware information systems}
  {Process-and resource-aware information systems}.{\BBCQ}
\newblock
\BIn{} \APACrefbtitle {EDOC, 2016 IEEE 20th International} {Edoc, 2016 ieee
  20th international}\ (\BPGS\ 1--10).
\PrintBackRefs{\CurrentBib}

\bibitem [\protect \citeauthoryear {%
Cabanillas%
\ \protect \BOthers {.}}{%
Cabanillas%
\ \protect \BOthers {.}}{%
{\protect \APACyear {2013}}%
}]{%
cabanillas2013priority}
\APACinsertmetastar {%
cabanillas2013priority}%
\begin{APACrefauthors}%
Cabanillas, C.%
, Garc{\'\i}a, J\BPBI M.%
, Resinas, M.%
, Ruiz, D.%
, Mendling, J.%
\BCBL {}\ \BBA {} Ruiz-Cort{\'e}s, A.%
\end{APACrefauthors}%
\unskip\
\newblock
\APACrefYearMonthDay{2013}{}{}.
\newblock
{\BBOQ}\APACrefatitle {Priority-based human resource allocation in business
  processes} {Priority-based human resource allocation in business
  processes}.{\BBCQ}
\newblock
\BIn{} \APACrefbtitle {International Conference on Service-Oriented Computing}
  {International conference on service-oriented computing}\ (\BPGS\ 374--388).
\PrintBackRefs{\CurrentBib}

\bibitem [\protect \citeauthoryear {%
Cooper%
}{%
Cooper%
}{%
{\protect \APACyear {2015}}%
}]{%
cooper2015research}
\APACinsertmetastar {%
cooper2015research}%
\begin{APACrefauthors}%
Cooper, H.%
\end{APACrefauthors}%
\unskip\
\newblock
\APACrefYear{2015}.
\newblock
\APACrefbtitle {Research synthesis and meta-analysis: A step-by-step approach}
  {Research synthesis and meta-analysis: A step-by-step approach}\ (\BVOL~2).
\newblock
\APACaddressPublisher{}{Sage publications}.
\PrintBackRefs{\CurrentBib}

\bibitem [\protect \citeauthoryear {%
Cormen%
, Leiserson%
, Rivest%
\BCBL {}\ \BBA {} Stein%
}{%
Cormen%
\ \protect \BOthers {.}}{%
{\protect \APACyear {2022}}%
}]{%
cormen2022introduction}
\APACinsertmetastar {%
cormen2022introduction}%
\begin{APACrefauthors}%
Cormen, T\BPBI H.%
, Leiserson, C\BPBI E.%
, Rivest, R\BPBI L.%
\BCBL {}\ \BBA {} Stein, C.%
\end{APACrefauthors}%
\unskip\
\newblock
\APACrefYear{2022}.
\newblock
\APACrefbtitle {Introduction to algorithms} {Introduction to algorithms}.
\newblock
\APACaddressPublisher{}{MIT press}.
\PrintBackRefs{\CurrentBib}

\bibitem [\protect \citeauthoryear {%
Delias%
, Doulamis%
, Doulamis%
\BCBL {}\ \BBA {} Matsatsinis%
}{%
Delias%
\ \protect \BOthers {.}}{%
{\protect \APACyear {2010}}%
}]{%
delias2010optimizing}
\APACinsertmetastar {%
delias2010optimizing}%
\begin{APACrefauthors}%
Delias, P.%
, Doulamis, A.%
, Doulamis, N.%
\BCBL {}\ \BBA {} Matsatsinis, N.%
\end{APACrefauthors}%
\unskip\
\newblock
\APACrefYearMonthDay{2010}{}{}.
\newblock
{\BBOQ}\APACrefatitle {Optimizing resource conflicts in workflow management
  systems} {Optimizing resource conflicts in workflow management
  systems}.{\BBCQ}
\newblock
\APACjournalVolNumPages{IEEE Transactions on Knowledge and Data
  Engineering}{23}{3}{417--432}.
\PrintBackRefs{\CurrentBib}

\bibitem [\protect \citeauthoryear {%
Djedovic%
, Karabegovic%
, Avdagic%
\BCBL {}\ \BBA {} Omanovic%
}{%
Djedovic%
\ \protect \BOthers {.}}{%
{\protect \APACyear {2018}}%
}]{%
djedovic2018innovative}
\APACinsertmetastar {%
djedovic2018innovative}%
\begin{APACrefauthors}%
Djedovic, A.%
, Karabegovic, A.%
, Avdagic, Z.%
\BCBL {}\ \BBA {} Omanovic, S.%
\end{APACrefauthors}%
\unskip\
\newblock
\APACrefYearMonthDay{2018}{}{}.
\newblock
{\BBOQ}\APACrefatitle {Innovative Approach in Modeling Business Processes with
  a Focus on Improving the Allocation of Human Resources} {Innovative approach
  in modeling business processes with a focus on improving the allocation of
  human resources}.{\BBCQ}
\newblock
\APACjournalVolNumPages{Mathematical Problems in Engineering}{2018}{}{}.
\PrintBackRefs{\CurrentBib}

\bibitem [\protect \citeauthoryear {%
Djedovi{\'c}%
, {\v{Z}}uni{\'c}%
, Avdagi{\'c}%
\BCBL {}\ \BBA {} Karabegovi{\'c}%
}{%
Djedovi{\'c}%
\ \protect \BOthers {.}}{%
{\protect \APACyear {2016}}%
}]{%
djedovic2016optimization}
\APACinsertmetastar {%
djedovic2016optimization}%
\begin{APACrefauthors}%
Djedovi{\'c}, A.%
, {\v{Z}}uni{\'c}, E.%
, Avdagi{\'c}, Z.%
\BCBL {}\ \BBA {} Karabegovi{\'c}, A.%
\end{APACrefauthors}%
\unskip\
\newblock
\APACrefYearMonthDay{2016}{}{}.
\newblock
{\BBOQ}\APACrefatitle {Optimization of business processes by automatic
  reallocation of resources using the genetic algorithm} {Optimization of
  business processes by automatic reallocation of resources using the genetic
  algorithm}.{\BBCQ}
\newblock
\BIn{} \APACrefbtitle {2016 XI International Symposium on Telecommunications
  (BIHTEL)} {2016 xi international symposium on telecommunications (bihtel)}\
  (\BPGS\ 1--7).
\PrintBackRefs{\CurrentBib}

\bibitem [\protect \citeauthoryear {%
Doerner%
, Gutjahr%
, Kotsis%
, Polaschek%
\BCBL {}\ \BBA {} Strauss%
}{%
Doerner%
\ \protect \BOthers {.}}{%
{\protect \APACyear {2006}}%
}]{%
doerner2006enriched}
\APACinsertmetastar {%
doerner2006enriched}%
\begin{APACrefauthors}%
Doerner, K.%
, Gutjahr, W\BPBI J.%
, Kotsis, G.%
, Polaschek, M.%
\BCBL {}\ \BBA {} Strauss, C.%
\end{APACrefauthors}%
\unskip\
\newblock
\APACrefYearMonthDay{2006}{}{}.
\newblock
{\BBOQ}\APACrefatitle {Enriched workflow modelling and stochastic
  branch-and-bound} {Enriched workflow modelling and stochastic
  branch-and-bound}.{\BBCQ}
\newblock
\APACjournalVolNumPages{European journal of operational
  research}{175}{3}{1798--1817}.
\PrintBackRefs{\CurrentBib}

\bibitem [\protect \citeauthoryear {%
Dumas%
, La~Rosa%
, Mendling%
, Reijers%
\BCBL {}\ \protect \BOthers {.}}{%
Dumas%
\ \protect \BOthers {.}}{%
{\protect \APACyear {2018}}%
}]{%
dumas2018fundamentals}
\APACinsertmetastar {%
dumas2018fundamentals}%
\begin{APACrefauthors}%
Dumas, M.%
, La~Rosa, M.%
, Mendling, J.%
, Reijers, H\BPBI A.%
\BCBL {}\ \BOthersPeriod {.}\end{APACrefauthors}%
\unskip\
\newblock
\APACrefYear{2018}.
\newblock
\APACrefbtitle {Fundamentals of business process management} {Fundamentals of
  business process management}\ (\BVOL~2).
\newblock
\APACaddressPublisher{}{Springer}.
\PrintBackRefs{\CurrentBib}

\bibitem [\protect \citeauthoryear {%
Dur{\'a}n%
, Rocha%
\BCBL {}\ \BBA {} Sala{\"u}n%
}{%
Dur{\'a}n%
\ \protect \BOthers {.}}{%
{\protect \APACyear {2019}}%
}]{%
duran2019rewriting}
\APACinsertmetastar {%
duran2019rewriting}%
\begin{APACrefauthors}%
Dur{\'a}n, F.%
, Rocha, C.%
\BCBL {}\ \BBA {} Sala{\"u}n, G.%
\end{APACrefauthors}%
\unskip\
\newblock
\APACrefYearMonthDay{2019}{}{}.
\newblock
{\BBOQ}\APACrefatitle {A rewriting logic approach to resource allocation
  analysis in business process models} {A rewriting logic approach to resource
  allocation analysis in business process models}.{\BBCQ}
\newblock
\APACjournalVolNumPages{Science of Computer Programming}{183}{}{102303}.
\PrintBackRefs{\CurrentBib}

\bibitem [\protect \citeauthoryear {%
Dur{\'a}n%
, Rocha%
\BCBL {}\ \BBA {} Sala{\"u}n%
}{%
Dur{\'a}n%
\ \protect \BOthers {.}}{%
{\protect \APACyear {2021}}%
}]{%
duran2021resource}
\APACinsertmetastar {%
duran2021resource}%
\begin{APACrefauthors}%
Dur{\'a}n, F.%
, Rocha, C.%
\BCBL {}\ \BBA {} Sala{\"u}n, G.%
\end{APACrefauthors}%
\unskip\
\newblock
\APACrefYearMonthDay{2021}{}{}.
\newblock
{\BBOQ}\APACrefatitle {Resource provisioning strategies for BPMN processes:
  Specification and analysis using Maude} {Resource provisioning strategies for
  bpmn processes: Specification and analysis using maude}.{\BBCQ}
\newblock
\APACjournalVolNumPages{Journal of Logical and Algebraic Methods in
  Programming}{123}{}{100711}.
\PrintBackRefs{\CurrentBib}

\bibitem [\protect \citeauthoryear {%
Eder%
, Pichler%
, Gruber%
\BCBL {}\ \BBA {} Ninaus%
}{%
Eder%
\ \protect \BOthers {.}}{%
{\protect \APACyear {2003}}%
}]{%
eder2003personal}
\APACinsertmetastar {%
eder2003personal}%
\begin{APACrefauthors}%
Eder, J.%
, Pichler, H.%
, Gruber, W.%
\BCBL {}\ \BBA {} Ninaus, M.%
\end{APACrefauthors}%
\unskip\
\newblock
\APACrefYearMonthDay{2003}{}{}.
\newblock
{\BBOQ}\APACrefatitle {Personal schedules for workflow systems} {Personal
  schedules for workflow systems}.{\BBCQ}
\newblock
\BIn{} \APACrefbtitle {International Conference on Business Process Management}
  {International conference on business process management}\ (\BPGS\ 216--231).
\PrintBackRefs{\CurrentBib}

\bibitem [\protect \citeauthoryear {%
Erasmus%
\ \protect \BOthers {.}}{%
Erasmus%
\ \protect \BOthers {.}}{%
{\protect \APACyear {2018}}%
}]{%
erasmus2018method}
\APACinsertmetastar {%
erasmus2018method}%
\begin{APACrefauthors}%
Erasmus, J.%
, Vanderfeesten, I.%
, Traganos, K.%
, Jie-A-Looi, X.%
, Kleingeld, A.%
\BCBL {}\ \BBA {} Grefen, P.%
\end{APACrefauthors}%
\unskip\
\newblock
\APACrefYearMonthDay{2018}{}{}.
\newblock
{\BBOQ}\APACrefatitle {A method to enable ability-based human resource
  allocation in business process management systems} {A method to enable
  ability-based human resource allocation in business process management
  systems}.{\BBCQ}
\newblock
\BIn{} \APACrefbtitle {IFIP Working Conference on the Practice of Enterprise
  Modeling} {Ifip working conference on the practice of enterprise modeling}\
  (\BPGS\ 37--52).
\PrintBackRefs{\CurrentBib}

\bibitem [\protect \citeauthoryear {%
Ha%
, Bae%
, Park%
\BCBL {}\ \BBA {} Kang%
}{%
Ha%
\ \protect \BOthers {.}}{%
{\protect \APACyear {2006}}%
}]{%
ha2006development}
\APACinsertmetastar {%
ha2006development}%
\begin{APACrefauthors}%
Ha, B\BHBI H.%
, Bae, J.%
, Park, Y\BPBI T.%
\BCBL {}\ \BBA {} Kang, S\BHBI H.%
\end{APACrefauthors}%
\unskip\
\newblock
\APACrefYearMonthDay{2006}{}{}.
\newblock
{\BBOQ}\APACrefatitle {Development of process execution rules for workload
  balancing on agents} {Development of process execution rules for workload
  balancing on agents}.{\BBCQ}
\newblock
\APACjournalVolNumPages{Data \& Knowledge Engineering}{56}{1}{64--84}.
\PrintBackRefs{\CurrentBib}

\bibitem [\protect \citeauthoryear {%
Havur%
, Cabanillas%
, Mendling%
\BCBL {}\ \BBA {} Polleres%
}{%
Havur%
\ \protect \BOthers {.}}{%
{\protect \APACyear {2016}}%
}]{%
havur2016resource}
\APACinsertmetastar {%
havur2016resource}%
\begin{APACrefauthors}%
Havur, G.%
, Cabanillas, C.%
, Mendling, J.%
\BCBL {}\ \BBA {} Polleres, A.%
\end{APACrefauthors}%
\unskip\
\newblock
\APACrefYearMonthDay{2016}{}{}.
\newblock
{\BBOQ}\APACrefatitle {Resource allocation with dependencies in business
  process management systems} {Resource allocation with dependencies in
  business process management systems}.{\BBCQ}
\newblock
\BIn{} \APACrefbtitle {BPM} {Bpm}\ (\BPGS\ 3--19).
\PrintBackRefs{\CurrentBib}

\bibitem [\protect \citeauthoryear {%
Hirsch%
\ \BBA {} Ortiz-Pe{\~n}a%
}{%
Hirsch%
\ \BBA {} Ortiz-Pe{\~n}a%
}{%
{\protect \APACyear {2017}}%
}]{%
hirsch2017information}
\APACinsertmetastar {%
hirsch2017information}%
\begin{APACrefauthors}%
Hirsch, M\BPBI J.%
\BCBT {}\ \BBA {} Ortiz-Pe{\~n}a, H.%
\end{APACrefauthors}%
\unskip\
\newblock
\APACrefYearMonthDay{2017}{}{}.
\newblock
{\BBOQ}\APACrefatitle {Information supply chain optimization with bandwidth
  limitations} {Information supply chain optimization with bandwidth
  limitations}.{\BBCQ}
\newblock
\APACjournalVolNumPages{International Transactions in Operational
  Research}{24}{5}{993--1022}.
\PrintBackRefs{\CurrentBib}

\bibitem [\protect \citeauthoryear {%
Hou%
\ \protect \BOthers {.}}{%
Hou%
\ \protect \BOthers {.}}{%
{\protect \APACyear {2021}}%
}]{%
hou2021bottleneck}
\APACinsertmetastar {%
hou2021bottleneck}%
\begin{APACrefauthors}%
Hou, S.%
, Ni, W.%
, Wang, M.%
, Liu, X.%
, Tong, Q.%
\BCBL {}\ \BBA {} Chen, S.%
\end{APACrefauthors}%
\unskip\
\newblock
\APACrefYearMonthDay{2021}{}{}.
\newblock
{\BBOQ}\APACrefatitle {Bottleneck-Aware Resource Allocation for Service
  Processes: A New Max-Min Approach} {Bottleneck-aware resource allocation for
  service processes: A new max-min approach}.{\BBCQ}
\newblock
\APACjournalVolNumPages{International Journal of Web Services Research
  (IJWSR)}{18}{3}{1--21}.
\PrintBackRefs{\CurrentBib}

\bibitem [\protect \citeauthoryear {%
Howard%
, Kochhar%
\BCBL {}\ \BBA {} Dilworth%
}{%
Howard%
\ \protect \BOthers {.}}{%
{\protect \APACyear {1999}}%
}]{%
howard1999application}
\APACinsertmetastar {%
howard1999application}%
\begin{APACrefauthors}%
Howard, A.%
, Kochhar, A.%
\BCBL {}\ \BBA {} Dilworth, J.%
\end{APACrefauthors}%
\unskip\
\newblock
\APACrefYearMonthDay{1999}{}{}.
\newblock
{\BBOQ}\APACrefatitle {Application of a generic manufacturing planning and
  control system reference architecture to different manufacturing
  environments} {Application of a generic manufacturing planning and control
  system reference architecture to different manufacturing
  environments}.{\BBCQ}
\newblock
\APACjournalVolNumPages{Proceedings of the Institution of Mechanical Engineers,
  Part B: Journal of Engineering Manufacture}{213}{4}{381--396}.
\PrintBackRefs{\CurrentBib}

\bibitem [\protect \citeauthoryear {%
Huang%
, Lu%
\BCBL {}\ \BBA {} Duan%
}{%
Huang%
, Lu%
\BCBL {}\ \BBA {} Duan%
}{%
{\protect \APACyear {2011}}%
}]{%
huang2011mining}
\APACinsertmetastar {%
huang2011mining}%
\begin{APACrefauthors}%
Huang, Z.%
, Lu, X.%
\BCBL {}\ \BBA {} Duan, H.%
\end{APACrefauthors}%
\unskip\
\newblock
\APACrefYearMonthDay{2011}{}{}.
\newblock
{\BBOQ}\APACrefatitle {Mining association rules to support resource allocation
  in business process management} {Mining association rules to support resource
  allocation in business process management}.{\BBCQ}
\newblock
\APACjournalVolNumPages{Expert Systems with Applications}{38}{8}{9483--9490}.
\PrintBackRefs{\CurrentBib}

\bibitem [\protect \citeauthoryear {%
Huang%
, Lu%
\BCBL {}\ \BBA {} Duan%
}{%
Huang%
\ \protect \BOthers {.}}{%
{\protect \APACyear {2012}}%
{\protect \APACexlab {{\protect \BCnt {1}}}}}]{%
huang2012resource}
\APACinsertmetastar {%
huang2012resource}%
\begin{APACrefauthors}%
Huang, Z.%
, Lu, X.%
\BCBL {}\ \BBA {} Duan, H.%
\end{APACrefauthors}%
\unskip\
\newblock
\APACrefYearMonthDay{2012{\protect \BCnt {1}}}{}{}.
\newblock
{\BBOQ}\APACrefatitle {Resource behavior measure and application in business
  process management} {Resource behavior measure and application in business
  process management}.{\BBCQ}
\newblock
\APACjournalVolNumPages{Expert Systems with Applications}{39}{7}{6458--6468}.
\PrintBackRefs{\CurrentBib}

\bibitem [\protect \citeauthoryear {%
Huang%
, Lu%
\BCBL {}\ \BBA {} Duan%
}{%
Huang%
\ \protect \BOthers {.}}{%
{\protect \APACyear {2012}}%
{\protect \APACexlab {{\protect \BCnt {2}}}}}]{%
huang2012task}
\APACinsertmetastar {%
huang2012task}%
\begin{APACrefauthors}%
Huang, Z.%
, Lu, X.%
\BCBL {}\ \BBA {} Duan, H.%
\end{APACrefauthors}%
\unskip\
\newblock
\APACrefYearMonthDay{2012{\protect \BCnt {2}}}{}{}.
\newblock
{\BBOQ}\APACrefatitle {A task operation model for resource allocation
  optimization in business process management} {A task operation model for
  resource allocation optimization in business process management}.{\BBCQ}
\newblock
\APACjournalVolNumPages{IEEE Transactions on Systems, man, and cybernetics-part
  a: systems and humans}{42}{5}{1256--1270}.
\PrintBackRefs{\CurrentBib}

\bibitem [\protect \citeauthoryear {%
Huang%
, van~der Aalst%
, Lu%
\BCBL {}\ \BBA {} Duan%
}{%
Huang%
\ \protect \BOthers {.}}{%
{\protect \APACyear {2010}}%
}]{%
huang2010adaptive}
\APACinsertmetastar {%
huang2010adaptive}%
\begin{APACrefauthors}%
Huang, Z.%
, van~der Aalst, W\BPBI M.%
, Lu, X.%
\BCBL {}\ \BBA {} Duan, H.%
\end{APACrefauthors}%
\unskip\
\newblock
\APACrefYearMonthDay{2010}{}{}.
\newblock
{\BBOQ}\APACrefatitle {An adaptive work distribution mechanism based on
  reinforcement learning} {An adaptive work distribution mechanism based on
  reinforcement learning}.{\BBCQ}
\newblock
\APACjournalVolNumPages{Expert Systems with Applications}{37}{12}{7533--7541}.
\PrintBackRefs{\CurrentBib}

\bibitem [\protect \citeauthoryear {%
Huang%
, van~der Aalst%
, Lu%
\BCBL {}\ \BBA {} Duan%
}{%
Huang%
, van~der Aalst%
\BCBL {}\ \protect \BOthers {.}}{%
{\protect \APACyear {2011}}%
}]{%
huang2011reinforcement}
\APACinsertmetastar {%
huang2011reinforcement}%
\begin{APACrefauthors}%
Huang, Z.%
, van~der Aalst, W\BPBI M.%
, Lu, X.%
\BCBL {}\ \BBA {} Duan, H.%
\end{APACrefauthors}%
\unskip\
\newblock
\APACrefYearMonthDay{2011}{}{}.
\newblock
{\BBOQ}\APACrefatitle {Reinforcement learning based resource allocation in
  business process management} {Reinforcement learning based resource
  allocation in business process management}.{\BBCQ}
\newblock
\APACjournalVolNumPages{Data \& Knowledge Engineering}{70}{1}{127--145}.
\PrintBackRefs{\CurrentBib}

\bibitem [\protect \citeauthoryear {%
Ihde%
, Pufahl%
, V{\"o}lker%
, Goel%
\BCBL {}\ \BBA {} Weske%
}{%
Ihde%
\ \protect \BOthers {.}}{%
{\protect \APACyear {2022}}%
}]{%
ihde2022framework}
\APACinsertmetastar {%
ihde2022framework}%
\begin{APACrefauthors}%
Ihde, S.%
, Pufahl, L.%
, V{\"o}lker, M.%
, Goel, A.%
\BCBL {}\ \BBA {} Weske, M.%
\end{APACrefauthors}%
\unskip\
\newblock
\APACrefYearMonthDay{2022}{}{}.
\newblock
{\BBOQ}\APACrefatitle {A framework for modeling and executing task-Specific
  resource allocations in business processes} {A framework for modeling and
  executing task-specific resource allocations in business processes}.{\BBCQ}
\newblock
\APACjournalVolNumPages{Computing}{104}{11}{2405--2429}.
\PrintBackRefs{\CurrentBib}

\bibitem [\protect \citeauthoryear {%
Jemel%
, Ben~Azzouna%
\BCBL {}\ \BBA {} Ghedira%
}{%
Jemel%
\ \protect \BOthers {.}}{%
{\protect \APACyear {2020}}%
}]{%
jemel2020rpminter}
\APACinsertmetastar {%
jemel2020rpminter}%
\begin{APACrefauthors}%
Jemel, M.%
, Ben~Azzouna, N.%
\BCBL {}\ \BBA {} Ghedira, K.%
\end{APACrefauthors}%
\unskip\
\newblock
\APACrefYearMonthDay{2020}{}{}.
\newblock
{\BBOQ}\APACrefatitle {RPMInter-work: a multi-agent approach for planning the
  task-role assignments in inter-organisational workflow} {Rpminter-work: a
  multi-agent approach for planning the task-role assignments in
  inter-organisational workflow}.{\BBCQ}
\newblock
\APACjournalVolNumPages{Enterprise Information Systems}{14}{5}{611--640}.
\PrintBackRefs{\CurrentBib}

\bibitem [\protect \citeauthoryear {%
Kalenkova%
, De~Leoni%
\BCBL {}\ \BBA {} van~der Aalst%
}{%
Kalenkova%
\ \protect \BOthers {.}}{%
{\protect \APACyear {2014}}%
}]{%
kalenkova2014discovering}
\APACinsertmetastar {%
kalenkova2014discovering}%
\begin{APACrefauthors}%
Kalenkova, A\BPBI A.%
, De~Leoni, M.%
\BCBL {}\ \BBA {} van~der Aalst, W\BPBI M.%
\end{APACrefauthors}%
\unskip\
\newblock
\APACrefYearMonthDay{2014}{}{}.
\newblock
{\BBOQ}\APACrefatitle {Discovering, Analyzing and Enhancing BPMN Models Using
  ProM.} {Discovering, analyzing and enhancing bpmn models using prom.}{\BBCQ}
\newblock
\BIn{} \APACrefbtitle {BPM (Demos)} {Bpm (demos)}\ (\BPG~36).
\PrintBackRefs{\CurrentBib}

\bibitem [\protect \citeauthoryear {%
Kamrani%
, Ayani%
\BCBL {}\ \BBA {} Moradi%
}{%
Kamrani%
\ \protect \BOthers {.}}{%
{\protect \APACyear {2012}}%
}]{%
kamrani2012framework}
\APACinsertmetastar {%
kamrani2012framework}%
\begin{APACrefauthors}%
Kamrani, F.%
, Ayani, R.%
\BCBL {}\ \BBA {} Moradi, F.%
\end{APACrefauthors}%
\unskip\
\newblock
\APACrefYearMonthDay{2012}{}{}.
\newblock
{\BBOQ}\APACrefatitle {A framework for simulation-based optimization of
  business process models} {A framework for simulation-based optimization of
  business process models}.{\BBCQ}
\newblock
\APACjournalVolNumPages{Simulation}{88}{7}{852--869}.
\PrintBackRefs{\CurrentBib}

\bibitem [\protect \citeauthoryear {%
Kitchenham%
}{%
Kitchenham%
}{%
{\protect \APACyear {2004}}%
}]{%
kitchenham2004procedures}
\APACinsertmetastar {%
kitchenham2004procedures}%
\begin{APACrefauthors}%
Kitchenham, B.%
\end{APACrefauthors}%
\unskip\
\newblock
\APACrefYearMonthDay{2004}{}{}.
\newblock
{\BBOQ}\APACrefatitle {Procedures for performing systematic reviews}
  {Procedures for performing systematic reviews}.{\BBCQ}
\newblock
\APACjournalVolNumPages{Keele, UK, Keele University}{33}{2004}{1--26}.
\PrintBackRefs{\CurrentBib}

\bibitem [\protect \citeauthoryear {%
Kumar%
, Dijkman%
\BCBL {}\ \BBA {} Song%
}{%
Kumar%
\ \protect \BOthers {.}}{%
{\protect \APACyear {2013}}%
}]{%
kumar2013optimal}
\APACinsertmetastar {%
kumar2013optimal}%
\begin{APACrefauthors}%
Kumar, A.%
, Dijkman, R.%
\BCBL {}\ \BBA {} Song, M.%
\end{APACrefauthors}%
\unskip\
\newblock
\APACrefYearMonthDay{2013}{}{}.
\newblock
{\BBOQ}\APACrefatitle {Optimal resource assignment in workflows for maximizing
  cooperation} {Optimal resource assignment in workflows for maximizing
  cooperation}.{\BBCQ}
\newblock
\BIn{} \APACrefbtitle {Business Process Management} {Business process
  management}\ (\BPGS\ 235--250).
\newblock
\APACaddressPublisher{}{Springer}.
\PrintBackRefs{\CurrentBib}

\bibitem [\protect \citeauthoryear {%
Kumar%
, Van Der~Aalst%
\BCBL {}\ \BBA {} Verbeek%
}{%
Kumar%
\ \protect \BOthers {.}}{%
{\protect \APACyear {2002}}%
}]{%
kumar2002dynamic}
\APACinsertmetastar {%
kumar2002dynamic}%
\begin{APACrefauthors}%
Kumar, A.%
, Van Der~Aalst, W\BPBI M.%
\BCBL {}\ \BBA {} Verbeek, E\BPBI M.%
\end{APACrefauthors}%
\unskip\
\newblock
\APACrefYearMonthDay{2002}{}{}.
\newblock
{\BBOQ}\APACrefatitle {Dynamic work distribution in workflow management
  systems: How to balance quality and performance} {Dynamic work distribution
  in workflow management systems: How to balance quality and
  performance}.{\BBCQ}
\newblock
\APACjournalVolNumPages{Journal of Management Information
  Systems}{18}{3}{157--193}.
\PrintBackRefs{\CurrentBib}

\bibitem [\protect \citeauthoryear {%
Lee%
, Lee%
, Kim%
\BCBL {}\ \BBA {} Choi%
}{%
Lee%
\ \protect \BOthers {.}}{%
{\protect \APACyear {2019}}%
}]{%
lee2019dynamic}
\APACinsertmetastar {%
lee2019dynamic}%
\begin{APACrefauthors}%
Lee, J.%
, Lee, S.%
, Kim, J.%
\BCBL {}\ \BBA {} Choi, I.%
\end{APACrefauthors}%
\unskip\
\newblock
\APACrefYearMonthDay{2019}{}{}.
\newblock
{\BBOQ}\APACrefatitle {Dynamic human resource selection for business process
  exceptions} {Dynamic human resource selection for business process
  exceptions}.{\BBCQ}
\newblock
\APACjournalVolNumPages{Knowledge and Process Management}{26}{1}{23--31}.
\PrintBackRefs{\CurrentBib}

\bibitem [\protect \citeauthoryear {%
R.~Liu%
, Agarwal%
, Sindhgatta%
\BCBL {}\ \BBA {} Lee%
}{%
R.~Liu%
\ \protect \BOthers {.}}{%
{\protect \APACyear {2013}}%
}]{%
liu2013accelerating}
\APACinsertmetastar {%
liu2013accelerating}%
\begin{APACrefauthors}%
Liu, R.%
, Agarwal, S.%
, Sindhgatta, R\BPBI R.%
\BCBL {}\ \BBA {} Lee, J.%
\end{APACrefauthors}%
\unskip\
\newblock
\APACrefYearMonthDay{2013}{}{}.
\newblock
{\BBOQ}\APACrefatitle {Accelerating collaboration in task assignment using a
  socially enhanced resource model} {Accelerating collaboration in task
  assignment using a socially enhanced resource model}.{\BBCQ}
\newblock
\BIn{} \APACrefbtitle {Business Process Management} {Business process
  management}\ (\BPGS\ 251--258).
\newblock
\APACaddressPublisher{}{Springer}.
\PrintBackRefs{\CurrentBib}

\bibitem [\protect \citeauthoryear {%
R.~Liu%
, Kumar%
\BCBL {}\ \BBA {} Lee%
}{%
R.~Liu%
\ \protect \BOthers {.}}{%
{\protect \APACyear {2022}}%
}]{%
liu2022multi}
\APACinsertmetastar {%
liu2022multi}%
\begin{APACrefauthors}%
Liu, R.%
, Kumar, A.%
\BCBL {}\ \BBA {} Lee, J.%
\end{APACrefauthors}%
\unskip\
\newblock
\APACrefYearMonthDay{2022}{}{}.
\newblock
{\BBOQ}\APACrefatitle {Multi-level Team Assignment in Social Business
  Processes: An Algorithm and Simulation Study} {Multi-level team assignment in
  social business processes: An algorithm and simulation study}.{\BBCQ}
\newblock
\APACjournalVolNumPages{Information Systems Frontiers}{}{}{1--21}.
\PrintBackRefs{\CurrentBib}

\bibitem [\protect \citeauthoryear {%
T.~Liu%
, Cheng%
\BCBL {}\ \BBA {} Ni%
}{%
T.~Liu%
\ \protect \BOthers {.}}{%
{\protect \APACyear {2012}}%
}]{%
liu2012mining}
\APACinsertmetastar {%
liu2012mining}%
\begin{APACrefauthors}%
Liu, T.%
, Cheng, Y.%
\BCBL {}\ \BBA {} Ni, Z.%
\end{APACrefauthors}%
\unskip\
\newblock
\APACrefYearMonthDay{2012}{}{}.
\newblock
{\BBOQ}\APACrefatitle {Mining event logs to support workflow resource
  allocation} {Mining event logs to support workflow resource
  allocation}.{\BBCQ}
\newblock
\APACjournalVolNumPages{Knowledge-Based Systems}{35}{}{320--331}.
\PrintBackRefs{\CurrentBib}

\bibitem [\protect \citeauthoryear {%
{Luo}%
, {Liu}%
, {Yin}%
, {Li}%
\BCBL {}\ \BBA {} {Wu}%
}{%
{Luo}%
\ \protect \BOthers {.}}{%
{\protect \APACyear {2019}}%
}]{%
Luo2019}
\APACinsertmetastar {%
Luo2019}%
\begin{APACrefauthors}%
{Luo}, Z.%
, {Liu}, L.%
, {Yin}, J.%
, {Li}, Y.%
\BCBL {}\ \BBA {} {Wu}, Z.%
\end{APACrefauthors}%
\unskip\
\newblock
\APACrefYearMonthDay{2019}{}{}.
\newblock
{\BBOQ}\APACrefatitle {Latent Ability Model: A Generative Probabilistic
  Learning Framework for Workforce Analytics} {Latent ability model: A
  generative probabilistic learning framework for workforce analytics}.{\BBCQ}
\newblock
\APACjournalVolNumPages{IEEE Transactions on Knowledge and Data
  Engineering}{31}{5}{923-937}.
\newblock
\begin{APACrefDOI} \doi{10.1109/TKDE.2018.2848658} \end{APACrefDOI}
\PrintBackRefs{\CurrentBib}

\bibitem [\protect \citeauthoryear {%
Luss%
}{%
Luss%
}{%
{\protect \APACyear {2012}}%
}]{%
luss2012equitable}
\APACinsertmetastar {%
luss2012equitable}%
\begin{APACrefauthors}%
Luss, H.%
\end{APACrefauthors}%
\unskip\
\newblock
\APACrefYear{2012}.
\newblock
\APACrefbtitle {Equitable resource allocation: models, algorithms and
  applications} {Equitable resource allocation: models, algorithms and
  applications}.
\newblock
\APACaddressPublisher{}{John Wiley \& Sons}.
\PrintBackRefs{\CurrentBib}

\bibitem [\protect \citeauthoryear {%
Marrella%
}{%
Marrella%
}{%
{\protect \APACyear {2019}}%
}]{%
marrella2019automated}
\APACinsertmetastar {%
marrella2019automated}%
\begin{APACrefauthors}%
Marrella, A.%
\end{APACrefauthors}%
\unskip\
\newblock
\APACrefYearMonthDay{2019}{}{}.
\newblock
{\BBOQ}\APACrefatitle {Automated planning for business process management}
  {Automated planning for business process management}.{\BBCQ}
\newblock
\APACjournalVolNumPages{Journal on data semantics}{8}{2}{79--98}.
\PrintBackRefs{\CurrentBib}

\bibitem [\protect \citeauthoryear {%
Nickel%
, Rebennack%
, Stein%
\BCBL {}\ \BBA {} Waldmann%
}{%
Nickel%
\ \protect \BOthers {.}}{%
{\protect \APACyear {2022}}%
}]{%
Nickel2022}
\APACinsertmetastar {%
Nickel2022}%
\begin{APACrefauthors}%
Nickel, S.%
, Rebennack, S.%
, Stein, O.%
\BCBL {}\ \BBA {} Waldmann, K\BHBI H.%
\end{APACrefauthors}%
\unskip\
\newblock
\APACrefYear{2022}.
\newblock
\APACrefbtitle {Operations Research} {Operations research}.
\newblock
\APACaddressPublisher{Berlin, Heidelberg}{Springer Berlin Heidelberg}.
\newblock
\begin{APACrefURL} \url{https://doi.org/10.1007/978-3-662-65346-3_5}
  \end{APACrefURL}
\newblock
\begin{APACrefDOI} \doi{10.1007/978-3-662-65346-3_5} \end{APACrefDOI}
\PrintBackRefs{\CurrentBib}

\bibitem [\protect \citeauthoryear {%
Okoli%
\ \BBA {} Schabram%
}{%
Okoli%
\ \BBA {} Schabram%
}{%
{\protect \APACyear {2010}}%
}]{%
okoli2010guide}
\APACinsertmetastar {%
okoli2010guide}%
\begin{APACrefauthors}%
Okoli, C.%
\BCBT {}\ \BBA {} Schabram, K.%
\end{APACrefauthors}%
\unskip\
\newblock
\APACrefYearMonthDay{2010}{}{}.
\newblock
{\BBOQ}\APACrefatitle {A guide to conducting a systematic literature review of
  information systems research} {A guide to conducting a systematic literature
  review of information systems research}.{\BBCQ}
\newblock
\APACjournalVolNumPages{Sprouts: Working Papers on Information
  Systems}{10}{26}{http://sprouts.aisnet.org/10-26}.
\PrintBackRefs{\CurrentBib}

\bibitem [\protect \citeauthoryear {%
OMG%
}{%
OMG%
}{%
{\protect \APACyear {2011}}%
}]{%
BPMN}
\APACinsertmetastar {%
BPMN}%
\begin{APACrefauthors}%
OMG.%
\end{APACrefauthors}%
\unskip\
\newblock
\APACrefYearMonthDay{2011}{}{}.
\newblock
{\BBOQ}\APACrefatitle {Notation {BPMN} version 2.0} {Notation {BPMN} version
  2.0}.{\BBCQ}
\newblock
\APACjournalVolNumPages{OMG Specification, Object Management
  Group}{}{}{22--31}.
\PrintBackRefs{\CurrentBib}

\bibitem [\protect \citeauthoryear {%
Ouyang%
, Wynn%
, Fidge%
, ter Hofstede%
\BCBL {}\ \BBA {} Kuhr%
}{%
Ouyang%
\ \protect \BOthers {.}}{%
{\protect \APACyear {2010}}%
}]{%
ouyang2010modelling}
\APACinsertmetastar {%
ouyang2010modelling}%
\begin{APACrefauthors}%
Ouyang, C.%
, Wynn, M\BPBI T.%
, Fidge, C.%
, ter Hofstede, A\BPBI H.%
\BCBL {}\ \BBA {} Kuhr, J\BHBI C.%
\end{APACrefauthors}%
\unskip\
\newblock
\APACrefYearMonthDay{2010}{}{}.
\newblock
{\BBOQ}\APACrefatitle {Modelling complex resource requirements in business
  process management systems} {Modelling complex resource requirements in
  business process management systems}.{\BBCQ}
\newblock
\APACjournalVolNumPages{ACIS 2010 Proceedings}{}{}{}.
\PrintBackRefs{\CurrentBib}

\bibitem [\protect \citeauthoryear {%
Park%
\ \BBA {} Song%
}{%
Park%
\ \BBA {} Song%
}{%
{\protect \APACyear {2023}}%
}]{%
park2023optimizing}
\APACinsertmetastar {%
park2023optimizing}%
\begin{APACrefauthors}%
Park, G.%
\BCBT {}\ \BBA {} Song, M.%
\end{APACrefauthors}%
\unskip\
\newblock
\APACrefYearMonthDay{2023}{}{}.
\newblock
{\BBOQ}\APACrefatitle {Optimizing Resource Allocation Based on Predictive
  Process Monitoring} {Optimizing resource allocation based on predictive
  process monitoring}.{\BBCQ}
\newblock
\APACjournalVolNumPages{IEEE Access}{}{}{}.
\PrintBackRefs{\CurrentBib}

\bibitem [\protect \citeauthoryear {%
Pereira%
, Varaj{\~a}o%
\BCBL {}\ \BBA {} Uahi%
}{%
Pereira%
\ \protect \BOthers {.}}{%
{\protect \APACyear {2020}}%
}]{%
pereira2020new}
\APACinsertmetastar {%
pereira2020new}%
\begin{APACrefauthors}%
Pereira, J\BHBI L.%
, Varaj{\~a}o, J.%
\BCBL {}\ \BBA {} Uahi, R.%
\end{APACrefauthors}%
\unskip\
\newblock
\APACrefYearMonthDay{2020}{}{}.
\newblock
{\BBOQ}\APACrefatitle {A new approach for improving work distribution in
  business processes supported by BPMS} {A new approach for improving work
  distribution in business processes supported by bpms}.{\BBCQ}
\newblock
\APACjournalVolNumPages{Business Process Management Journal}{}{}{}.
\PrintBackRefs{\CurrentBib}

\bibitem [\protect \citeauthoryear {%
Pflug%
\ \BBA {} Rinderle-Ma%
}{%
Pflug%
\ \BBA {} Rinderle-Ma%
}{%
{\protect \APACyear {2016}}%
}]{%
pflug2016application}
\APACinsertmetastar {%
pflug2016application}%
\begin{APACrefauthors}%
Pflug, J.%
\BCBT {}\ \BBA {} Rinderle-Ma, S.%
\end{APACrefauthors}%
\unskip\
\newblock
\APACrefYearMonthDay{2016}{}{}.
\newblock
{\BBOQ}\APACrefatitle {Application of dynamic instance queuing to activity
  sequences in cooperative business process scenarios} {Application of dynamic
  instance queuing to activity sequences in cooperative business process
  scenarios}.{\BBCQ}
\newblock
\APACjournalVolNumPages{International Journal of Cooperative Information
  Systems}{25}{01}{1650002}.
\PrintBackRefs{\CurrentBib}

\bibitem [\protect \citeauthoryear {%
Pika%
\ \BBA {} Wynn%
}{%
Pika%
\ \BBA {} Wynn%
}{%
{\protect \APACyear {2021}}%
}]{%
pika2021machine}
\APACinsertmetastar {%
pika2021machine}%
\begin{APACrefauthors}%
Pika, A.%
\BCBT {}\ \BBA {} Wynn, M\BPBI T.%
\end{APACrefauthors}%
\unskip\
\newblock
\APACrefYearMonthDay{2021}{}{}.
\newblock
{\BBOQ}\APACrefatitle {A machine learning based approach for recommending
  unfamiliar process activities} {A machine learning based approach for
  recommending unfamiliar process activities}.{\BBCQ}
\newblock
\APACjournalVolNumPages{IEEE Access}{9}{}{104969--104979}.
\PrintBackRefs{\CurrentBib}

\bibitem [\protect \citeauthoryear {%
Remy%
, Pufahl%
, Sachs%
, B{\"o}ttinger%
\BCBL {}\ \BBA {} Weske%
}{%
Remy%
\ \protect \BOthers {.}}{%
{\protect \APACyear {2020}}%
}]{%
remy2020event}
\APACinsertmetastar {%
remy2020event}%
\begin{APACrefauthors}%
Remy, S.%
, Pufahl, L.%
, Sachs, J\BPBI P.%
, B{\"o}ttinger, E.%
\BCBL {}\ \BBA {} Weske, M.%
\end{APACrefauthors}%
\unskip\
\newblock
\APACrefYearMonthDay{2020}{}{}.
\newblock
{\BBOQ}\APACrefatitle {Event Log Generation in a Health System: A Case Study}
  {Event log generation in a health system: A case study}.{\BBCQ}
\newblock
\BIn{} \APACrefbtitle {International Conference on Business Process Management}
  {International conference on business process management}\ (\BPGS\ 505--522).
\PrintBackRefs{\CurrentBib}

\bibitem [\protect \citeauthoryear {%
Rhee%
, Cho%
\BCBL {}\ \BBA {} Bae%
}{%
Rhee%
\ \protect \BOthers {.}}{%
{\protect \APACyear {2010}}%
}]{%
rhee2010increasing}
\APACinsertmetastar {%
rhee2010increasing}%
\begin{APACrefauthors}%
Rhee, S\BHBI H.%
, Cho, N\BPBI W.%
\BCBL {}\ \BBA {} Bae, H.%
\end{APACrefauthors}%
\unskip\
\newblock
\APACrefYearMonthDay{2010}{}{}.
\newblock
{\BBOQ}\APACrefatitle {Increasing the efficiency of business processes using a
  theory of constraints} {Increasing the efficiency of business processes using
  a theory of constraints}.{\BBCQ}
\newblock
\APACjournalVolNumPages{Information Systems Frontiers}{12}{4}{443--455}.
\PrintBackRefs{\CurrentBib}

\bibitem [\protect \citeauthoryear {%
Russell%
, van~der Aalst%
, ter Hofstede%
\BCBL {}\ \BBA {} Edmond%
}{%
Russell%
\ \protect \BOthers {.}}{%
{\protect \APACyear {2005}}%
}]{%
russell2005workflow}
\APACinsertmetastar {%
russell2005workflow}%
\begin{APACrefauthors}%
Russell, N.%
, van~der Aalst, W\BPBI M\BPBI P.%
, ter Hofstede, A\BPBI H.%
\BCBL {}\ \BBA {} Edmond, D.%
\end{APACrefauthors}%
\unskip\
\newblock
\APACrefYearMonthDay{2005}{}{}.
\newblock
{\BBOQ}\APACrefatitle {Workflow resource patterns: Identification,
  representation and tool support} {Workflow resource patterns: Identification,
  representation and tool support}.{\BBCQ}
\newblock
\BIn{} \APACrefbtitle {CAiSE} {Caise}\ (\BPGS\ 216--232).
\PrintBackRefs{\CurrentBib}

\bibitem [\protect \citeauthoryear {%
Schall%
, Satzger%
\BCBL {}\ \BBA {} Psaier%
}{%
Schall%
\ \protect \BOthers {.}}{%
{\protect \APACyear {2014}}%
}]{%
schall2014crowdsourcing}
\APACinsertmetastar {%
schall2014crowdsourcing}%
\begin{APACrefauthors}%
Schall, D.%
, Satzger, B.%
\BCBL {}\ \BBA {} Psaier, H.%
\end{APACrefauthors}%
\unskip\
\newblock
\APACrefYearMonthDay{2014}{}{}.
\newblock
{\BBOQ}\APACrefatitle {Crowdsourcing tasks to social networks in BPEL4People}
  {Crowdsourcing tasks to social networks in bpel4people}.{\BBCQ}
\newblock
\APACjournalVolNumPages{World Wide Web}{17}{1}{1--32}.
\PrintBackRefs{\CurrentBib}

\bibitem [\protect \citeauthoryear {%
Scheer%
, Thomas%
\BCBL {}\ \BBA {} Adam%
}{%
Scheer%
\ \protect \BOthers {.}}{%
{\protect \APACyear {2005}}%
}]{%
scheer2005process}
\APACinsertmetastar {%
scheer2005process}%
\begin{APACrefauthors}%
Scheer, A\BHBI W.%
, Thomas, O.%
\BCBL {}\ \BBA {} Adam, O.%
\end{APACrefauthors}%
\unskip\
\newblock
\APACrefYearMonthDay{2005}{}{}.
\newblock
{\BBOQ}\APACrefatitle {Process Modeling Using Event-Driven Process Chains.}
  {Process modeling using event-driven process chains.}{\BBCQ}
\newblock
\APACjournalVolNumPages{Process-aware information systems}{119}{}{}.
\PrintBackRefs{\CurrentBib}

\bibitem [\protect \citeauthoryear {%
Sch{\"o}nig%
, Cabanillas%
, Jablonski%
\BCBL {}\ \BBA {} Mendling%
}{%
Sch{\"o}nig%
\ \protect \BOthers {.}}{%
{\protect \APACyear {2016}}%
}]{%
schonig2016framework}
\APACinsertmetastar {%
schonig2016framework}%
\begin{APACrefauthors}%
Sch{\"o}nig, S.%
, Cabanillas, C.%
, Jablonski, S.%
\BCBL {}\ \BBA {} Mendling, J.%
\end{APACrefauthors}%
\unskip\
\newblock
\APACrefYearMonthDay{2016}{}{}.
\newblock
{\BBOQ}\APACrefatitle {A framework for efficiently mining the organisational
  perspective of business processes} {A framework for efficiently mining the
  organisational perspective of business processes}.{\BBCQ}
\newblock
\APACjournalVolNumPages{Decision Support Systems}{89}{}{87--97}.
\PrintBackRefs{\CurrentBib}

\bibitem [\protect \citeauthoryear {%
Si%
, Chan%
, Dumas%
\BCBL {}\ \BBA {} Zhang%
}{%
Si%
\ \protect \BOthers {.}}{%
{\protect \APACyear {2018}}%
}]{%
si2018petri}
\APACinsertmetastar {%
si2018petri}%
\begin{APACrefauthors}%
Si, Y\BHBI W.%
, Chan, V\BHBI I.%
, Dumas, M.%
\BCBL {}\ \BBA {} Zhang, D.%
\end{APACrefauthors}%
\unskip\
\newblock
\APACrefYearMonthDay{2018}{}{}.
\newblock
{\BBOQ}\APACrefatitle {A Petri Nets based Generic Genetic Algorithm framework
  for resource optimization in business processes} {A petri nets based generic
  genetic algorithm framework for resource optimization in business
  processes}.{\BBCQ}
\newblock
\APACjournalVolNumPages{Simulation Modelling Practice and
  Theory}{86}{}{72--101}.
\PrintBackRefs{\CurrentBib}

\bibitem [\protect \citeauthoryear {%
Slack%
\ \BBA {} Brandon-Jones%
}{%
Slack%
\ \BBA {} Brandon-Jones%
}{%
{\protect \APACyear {2018}}%
}]{%
slack2018operations}
\APACinsertmetastar {%
slack2018operations}%
\begin{APACrefauthors}%
Slack, N.%
\BCBT {}\ \BBA {} Brandon-Jones, A.%
\end{APACrefauthors}%
\unskip\
\newblock
\APACrefYear{2018}.
\newblock
\APACrefbtitle {Operations and process management: principles and practice for
  strategic impact} {Operations and process management: principles and practice
  for strategic impact}.
\newblock
\APACaddressPublisher{}{Pearson UK}.
\PrintBackRefs{\CurrentBib}

\bibitem [\protect \citeauthoryear {%
Soeffker%
, Ulmer%
\BCBL {}\ \BBA {} Mattfeld%
}{%
Soeffker%
\ \protect \BOthers {.}}{%
{\protect \APACyear {2019}}%
}]{%
soeffker2019adaptive}
\APACinsertmetastar {%
soeffker2019adaptive}%
\begin{APACrefauthors}%
Soeffker, N.%
, Ulmer, M\BPBI W.%
\BCBL {}\ \BBA {} Mattfeld, D\BPBI C.%
\end{APACrefauthors}%
\unskip\
\newblock
\APACrefYearMonthDay{2019}{}{}.
\newblock
{\BBOQ}\APACrefatitle {Adaptive State Space Partitioning for Dynamic Decision
  Processes} {Adaptive state space partitioning for dynamic decision
  processes}.{\BBCQ}
\newblock
\APACjournalVolNumPages{Business \& Information Systems
  Engineering}{61}{3}{261--275}.
\PrintBackRefs{\CurrentBib}

\bibitem [\protect \citeauthoryear {%
Stork%
, Eiben%
\BCBL {}\ \BBA {} Bartz-Beielstein%
}{%
Stork%
\ \protect \BOthers {.}}{%
{\protect \APACyear {2022}}%
}]{%
stork2022new}
\APACinsertmetastar {%
stork2022new}%
\begin{APACrefauthors}%
Stork, J.%
, Eiben, A\BPBI E.%
\BCBL {}\ \BBA {} Bartz-Beielstein, T.%
\end{APACrefauthors}%
\unskip\
\newblock
\APACrefYearMonthDay{2022}{}{}.
\newblock
{\BBOQ}\APACrefatitle {A new taxonomy of global optimization algorithms} {A new
  taxonomy of global optimization algorithms}.{\BBCQ}
\newblock
\APACjournalVolNumPages{Natural Computing}{21}{2}{219--242}.
\PrintBackRefs{\CurrentBib}

\bibitem [\protect \citeauthoryear {%
Van~der Aalst%
}{%
Van~der Aalst%
}{%
{\protect \APACyear {1998}}%
}]{%
van1998application}
\APACinsertmetastar {%
van1998application}%
\begin{APACrefauthors}%
Van~der Aalst, W\BPBI M.%
\end{APACrefauthors}%
\unskip\
\newblock
\APACrefYearMonthDay{1998}{}{}.
\newblock
{\BBOQ}\APACrefatitle {The application of Petri nets to workflow management}
  {The application of petri nets to workflow management}.{\BBCQ}
\newblock
\APACjournalVolNumPages{Journal of circuits, systems, and
  computers}{8}{01}{21--66}.
\PrintBackRefs{\CurrentBib}

\bibitem [\protect \citeauthoryear {%
Van~der Aalst%
}{%
Van~der Aalst%
}{%
{\protect \APACyear {2013}}%
}]{%
van2013business}
\APACinsertmetastar {%
van2013business}%
\begin{APACrefauthors}%
Van~der Aalst, W\BPBI M.%
\end{APACrefauthors}%
\unskip\
\newblock
\APACrefYearMonthDay{2013}{}{}.
\newblock
{\BBOQ}\APACrefatitle {Business process management: a comprehensive survey}
  {Business process management: a comprehensive survey}.{\BBCQ}
\newblock
\APACjournalVolNumPages{International Scholarly Research Notices}{2013}{}{}.
\PrintBackRefs{\CurrentBib}

\bibitem [\protect \citeauthoryear {%
Van~der Aalst%
\ \BBA {} Kumar%
}{%
Van~der Aalst%
\ \BBA {} Kumar%
}{%
{\protect \APACyear {2001}}%
}]{%
van2001reference}
\APACinsertmetastar {%
van2001reference}%
\begin{APACrefauthors}%
Van~der Aalst, W\BPBI M.%
\BCBT {}\ \BBA {} Kumar, A.%
\end{APACrefauthors}%
\unskip\
\newblock
\APACrefYearMonthDay{2001}{}{}.
\newblock
{\BBOQ}\APACrefatitle {A reference model for team-enabled workflow management
  systems} {A reference model for team-enabled workflow management
  systems}.{\BBCQ}
\newblock
\APACjournalVolNumPages{Data \& Knowledge Engineering}{38}{3}{335--363}.
\PrintBackRefs{\CurrentBib}

\bibitem [\protect \citeauthoryear {%
van Der~Aalst%
, Pesic%
\BCBL {}\ \BBA {} Schonenberg%
}{%
van Der~Aalst%
\ \protect \BOthers {.}}{%
{\protect \APACyear {2009}}%
}]{%
van2009declarative}
\APACinsertmetastar {%
van2009declarative}%
\begin{APACrefauthors}%
van Der~Aalst, W\BPBI M.%
, Pesic, M.%
\BCBL {}\ \BBA {} Schonenberg, H.%
\end{APACrefauthors}%
\unskip\
\newblock
\APACrefYearMonthDay{2009}{}{}.
\newblock
{\BBOQ}\APACrefatitle {Declarative workflows: Balancing between flexibility and
  support} {Declarative workflows: Balancing between flexibility and
  support}.{\BBCQ}
\newblock
\APACjournalVolNumPages{Computer Science-Research and
  Development}{23}{2}{99--113}.
\PrintBackRefs{\CurrentBib}

\bibitem [\protect \citeauthoryear {%
van~der Aalst%
}{%
van~der Aalst%
}{%
{\protect \APACyear {2016}}%
}]{%
van_der_aalst_process_2016}
\APACinsertmetastar {%
van_der_aalst_process_2016}%
\begin{APACrefauthors}%
van~der Aalst, W\BPBI M\BPBI P.%
\end{APACrefauthors}%
\unskip\
\newblock
\APACrefYear{2016}.
\newblock
\APACrefbtitle {Process mining: data science in action} {Process mining: data
  science in action}.
\newblock
\APACaddressPublisher{Heidelberg}{Springer}.
\PrintBackRefs{\CurrentBib}

\bibitem [\protect \citeauthoryear {%
Van~Hee%
, Reijers%
, Verbeek%
\BCBL {}\ \BBA {} Zerguini%
}{%
Van~Hee%
\ \protect \BOthers {.}}{%
{\protect \APACyear {2001}}%
}]{%
van2001optimal}
\APACinsertmetastar {%
van2001optimal}%
\begin{APACrefauthors}%
Van~Hee, K.%
, Reijers, H.%
, Verbeek, H.%
\BCBL {}\ \BBA {} Zerguini, L.%
\end{APACrefauthors}%
\unskip\
\newblock
\APACrefYearMonthDay{2001}{}{}.
\newblock
{\BBOQ}\APACrefatitle {On the optimal allocation of resources in stochastic
  workflow nets} {On the optimal allocation of resources in stochastic workflow
  nets}.{\BBCQ}
\newblock
\BIn{} \APACrefbtitle {Proceedings of the Seventeenth UK Performance
  Engineering Workshop} {Proceedings of the seventeenth uk performance
  engineering workshop}\ (\BPGS\ 23--34).
\PrintBackRefs{\CurrentBib}

\bibitem [\protect \citeauthoryear {%
Wei%
\ \BBA {} Blake%
}{%
Wei%
\ \BBA {} Blake%
}{%
{\protect \APACyear {2016}}%
}]{%
wei2016proactive}
\APACinsertmetastar {%
wei2016proactive}%
\begin{APACrefauthors}%
Wei, Y.%
\BCBT {}\ \BBA {} Blake, M\BPBI B.%
\end{APACrefauthors}%
\unskip\
\newblock
\APACrefYearMonthDay{2016}{}{}.
\newblock
{\BBOQ}\APACrefatitle {Proactive virtualized resource management for service
  workflows in the cloud} {Proactive virtualized resource management for
  service workflows in the cloud}.{\BBCQ}
\newblock
\APACjournalVolNumPages{Computing}{98}{5}{523--538}.
\PrintBackRefs{\CurrentBib}

\bibitem [\protect \citeauthoryear {%
Weske%
}{%
Weske%
}{%
{\protect \APACyear {2019}}%
}]{%
DBLP:books/sp/Weske19}
\APACinsertmetastar {%
DBLP:books/sp/Weske19}%
\begin{APACrefauthors}%
Weske, M.%
\end{APACrefauthors}%
\unskip\
\newblock
\APACrefYear{2019}.
\newblock
\APACrefbtitle {Business Process Management - Concepts, Languages,
  Architectures, Third Edition} {Business process management - concepts,
  languages, architectures, third edition}.
\newblock
\APACaddressPublisher{}{Springer}.
\PrintBackRefs{\CurrentBib}

\bibitem [\protect \citeauthoryear {%
Wibisono%
, Nisafani%
, Hyerim~Bae%
\BCBL {}\ \BBA {} You-Jin~Park%
}{%
Wibisono%
\ \protect \BOthers {.}}{%
{\protect \APACyear {2016}}%
}]{%
wibisono2016dynamic}
\APACinsertmetastar {%
wibisono2016dynamic}%
\begin{APACrefauthors}%
Wibisono, A.%
, Nisafani, A\BPBI S.%
, Hyerim~Bae, H\BPBI B.%
\BCBL {}\ \BBA {} You-Jin~Park, Y\BHBI J\BPBI P.%
\end{APACrefauthors}%
\unskip\
\newblock
\APACrefYearMonthDay{2016}{}{}.
\newblock
{\BBOQ}\APACrefatitle {A dynamic and human-centric resource allocation for
  managing business process execution} {A dynamic and human-centric resource
  allocation for managing business process execution}.{\BBCQ}
\newblock
\APACjournalVolNumPages{International Journal of Industrial Engineering:
  Theory, Applications and Practice, 23 (4), 2016}{}{}{}.
\PrintBackRefs{\CurrentBib}

\bibitem [\protect \citeauthoryear {%
Williams%
, Chatterjee%
\BCBL {}\ \BBA {} Rossi%
}{%
Williams%
\ \protect \BOthers {.}}{%
{\protect \APACyear {2008}}%
}]{%
williamsDesignEmergingDigital2008}
\APACinsertmetastar {%
williamsDesignEmergingDigital2008}%
\begin{APACrefauthors}%
Williams, K.%
, Chatterjee, S.%
\BCBL {}\ \BBA {} Rossi, M.%
\end{APACrefauthors}%
\unskip\
\newblock
\APACrefYearMonthDay{2008}{}{}.
\newblock
{\BBOQ}\APACrefatitle {Design of Emerging Digital Services: A Taxonomy} {Design
  of emerging digital services: A taxonomy}.{\BBCQ}
\newblock
\APACjournalVolNumPages{EJIS}{17}{5}{505--517}.
\newblock
\begin{APACrefDOI} \doi{10.1057/ejis.2008.38} \end{APACrefDOI}
\PrintBackRefs{\CurrentBib}

\bibitem [\protect \citeauthoryear {%
Xie%
, Chen%
, Ni%
\BCBL {}\ \BBA {} Wu%
}{%
Xie%
\ \protect \BOthers {.}}{%
{\protect \APACyear {2019}}%
}]{%
xie2019integration}
\APACinsertmetastar {%
xie2019integration}%
\begin{APACrefauthors}%
Xie, Y.%
, Chen, S.%
, Ni, Q.%
\BCBL {}\ \BBA {} Wu, H.%
\end{APACrefauthors}%
\unskip\
\newblock
\APACrefYearMonthDay{2019}{}{}.
\newblock
{\BBOQ}\APACrefatitle {Integration of resource allocation and task assignment
  for optimizing the cost and maximum throughput of business processes}
  {Integration of resource allocation and task assignment for optimizing the
  cost and maximum throughput of business processes}.{\BBCQ}
\newblock
\APACjournalVolNumPages{Journal of Intelligent
  Manufacturing}{30}{3}{1351--1369}.
\PrintBackRefs{\CurrentBib}

\bibitem [\protect \citeauthoryear {%
Xie%
, Chien%
\BCBL {}\ \BBA {} Tang%
}{%
Xie%
\ \protect \BOthers {.}}{%
{\protect \APACyear {2016}}%
}]{%
xie2016dynamic}
\APACinsertmetastar {%
xie2016dynamic}%
\begin{APACrefauthors}%
Xie, Y.%
, Chien, C\BHBI F.%
\BCBL {}\ \BBA {} Tang, R\BHBI Z.%
\end{APACrefauthors}%
\unskip\
\newblock
\APACrefYearMonthDay{2016}{}{}.
\newblock
{\BBOQ}\APACrefatitle {A dynamic task assignment approach based on individual
  worklists for minimizing the cycle time of business processes} {A dynamic
  task assignment approach based on individual worklists for minimizing the
  cycle time of business processes}.{\BBCQ}
\newblock
\APACjournalVolNumPages{Computers \& Industrial Engineering}{99}{}{401--414}.
\PrintBackRefs{\CurrentBib}

\bibitem [\protect \citeauthoryear {%
Xu%
, Liu%
\BCBL {}\ \BBA {} Zhao%
}{%
Xu%
\ \protect \BOthers {.}}{%
{\protect \APACyear {2009}}%
}]{%
xu2009resource}
\APACinsertmetastar {%
xu2009resource}%
\begin{APACrefauthors}%
Xu, J.%
, Liu, C.%
\BCBL {}\ \BBA {} Zhao, X.%
\end{APACrefauthors}%
\unskip\
\newblock
\APACrefYearMonthDay{2009}{}{}.
\newblock
{\BBOQ}\APACrefatitle {Resource planning for massive number of process
  instances} {Resource planning for massive number of process
  instances}.{\BBCQ}
\newblock
\BIn{} \APACrefbtitle {OTM Confederated International Conferences" On the Move
  to Meaningful Internet Systems"} {Otm confederated international conferences"
  on the move to meaningful internet systems"}\ (\BPGS\ 219--236).
\PrintBackRefs{\CurrentBib}

\bibitem [\protect \citeauthoryear {%
Xu%
, Liu%
, Zhao%
\BCBL {}\ \BBA {} Ding%
}{%
Xu%
\ \protect \BOthers {.}}{%
{\protect \APACyear {2013}}%
}]{%
xu2013incorporating}
\APACinsertmetastar {%
xu2013incorporating}%
\begin{APACrefauthors}%
Xu, J.%
, Liu, C.%
, Zhao, X.%
\BCBL {}\ \BBA {} Ding, Z.%
\end{APACrefauthors}%
\unskip\
\newblock
\APACrefYearMonthDay{2013}{}{}.
\newblock
{\BBOQ}\APACrefatitle {Incorporating structural improvement into resource
  allocation for business process execution planning} {Incorporating structural
  improvement into resource allocation for business process execution
  planning}.{\BBCQ}
\newblock
\APACjournalVolNumPages{Concurrency and Computation: Practice and
  Experience}{25}{3}{427--442}.
\PrintBackRefs{\CurrentBib}

\bibitem [\protect \citeauthoryear {%
Xu%
, Liu%
, Zhao%
, Yongchareon%
\BCBL {}\ \BBA {} Ding%
}{%
Xu%
\ \protect \BOthers {.}}{%
{\protect \APACyear {2016}}%
}]{%
xu2016resource}
\APACinsertmetastar {%
xu2016resource}%
\begin{APACrefauthors}%
Xu, J.%
, Liu, C.%
, Zhao, X.%
, Yongchareon, S.%
\BCBL {}\ \BBA {} Ding, Z.%
\end{APACrefauthors}%
\unskip\
\newblock
\APACrefYearMonthDay{2016}{}{}.
\newblock
{\BBOQ}\APACrefatitle {Resource management for business process scheduling in
  the presence of availability constraints} {Resource management for business
  process scheduling in the presence of availability constraints}.{\BBCQ}
\newblock
\APACjournalVolNumPages{ACM Transactions on Management Information Systems
  (TMIS)}{7}{3}{1--26}.
\PrintBackRefs{\CurrentBib}

\bibitem [\protect \citeauthoryear {%
Yaghoibi%
\ \BBA {} Zahedi%
}{%
Yaghoibi%
\ \BBA {} Zahedi%
}{%
{\protect \APACyear {2017}}%
}]{%
yaghoibi2017cycle}
\APACinsertmetastar {%
yaghoibi2017cycle}%
\begin{APACrefauthors}%
Yaghoibi, M.%
\BCBT {}\ \BBA {} Zahedi, M.%
\end{APACrefauthors}%
\unskip\
\newblock
\APACrefYearMonthDay{2017}{}{}.
\newblock
{\BBOQ}\APACrefatitle {Cycle Time Reduction and Runtime Rebalancing by
  Reallocating Dependent Tasks} {Cycle time reduction and runtime rebalancing
  by reallocating dependent tasks}.{\BBCQ}
\newblock
\APACjournalVolNumPages{International Journal of
  Engineering}{30}{12}{1831--1839}.
\PrintBackRefs{\CurrentBib}

\bibitem [\protect \citeauthoryear {%
Yaghoubi%
\ \BBA {} Zahedi%
}{%
Yaghoubi%
\ \BBA {} Zahedi%
}{%
{\protect \APACyear {2016}}%
}]{%
yaghoubi2016resource}
\APACinsertmetastar {%
yaghoubi2016resource}%
\begin{APACrefauthors}%
Yaghoubi, M.%
\BCBT {}\ \BBA {} Zahedi, M.%
\end{APACrefauthors}%
\unskip\
\newblock
\APACrefYearMonthDay{2016}{}{}.
\newblock
{\BBOQ}\APACrefatitle {Resource allocation using task similarity distance in
  business process management systems} {Resource allocation using task
  similarity distance in business process management systems}.{\BBCQ}
\newblock
\BIn{} \APACrefbtitle {2016 2nd International Conference of Signal Processing
  and Intelligent Systems (ICSPIS)} {2016 2nd international conference of
  signal processing and intelligent systems (icspis)}\ (\BPGS\ 1--5).
\PrintBackRefs{\CurrentBib}

\bibitem [\protect \citeauthoryear {%
Yari~Eili%
\ \BBA {} Rezaeenour%
}{%
Yari~Eili%
\ \BBA {} Rezaeenour%
}{%
{\protect \APACyear {2022}}%
}]{%
yari2022survey}
\APACinsertmetastar {%
yari2022survey}%
\begin{APACrefauthors}%
Yari~Eili, M.%
\BCBT {}\ \BBA {} Rezaeenour, J.%
\end{APACrefauthors}%
\unskip\
\newblock
\APACrefYearMonthDay{2022}{}{}.
\newblock
{\BBOQ}\APACrefatitle {A survey on recommendation in process mining} {A survey
  on recommendation in process mining}.{\BBCQ}
\newblock
\APACjournalVolNumPages{Concurrency and Computation: Practice and
  Experience}{34}{26}{e7304}.
\PrintBackRefs{\CurrentBib}

\bibitem [\protect \citeauthoryear {%
Yeon%
, Lee%
, Pham%
\BCBL {}\ \BBA {} Kim%
}{%
Yeon%
\ \protect \BOthers {.}}{%
{\protect \APACyear {2022}}%
}]{%
yeon2022experimental}
\APACinsertmetastar {%
yeon2022experimental}%
\begin{APACrefauthors}%
Yeon, M\BHBI S.%
, Lee, Y\BHBI K.%
, Pham, D\BHBI L.%
\BCBL {}\ \BBA {} Kim, K\BPBI P.%
\end{APACrefauthors}%
\unskip\
\newblock
\APACrefYearMonthDay{2022}{}{}.
\newblock
{\BBOQ}\APACrefatitle {Experimental Verification on Human-Centric Network-Based
  Resource Allocation Approaches for Process-Aware Information Systems}
  {Experimental verification on human-centric network-based resource allocation
  approaches for process-aware information systems}.{\BBCQ}
\newblock
\APACjournalVolNumPages{IEEE Access}{10}{}{23342--23354}.
\PrintBackRefs{\CurrentBib}

\bibitem [\protect \citeauthoryear {%
Yu%
, Jia%
, Liu%
\BCBL {}\ \BBA {} Ma%
}{%
Yu%
\ \protect \BOthers {.}}{%
{\protect \APACyear {2020}}%
}]{%
yu2020task}
\APACinsertmetastar {%
yu2020task}%
\begin{APACrefauthors}%
Yu, W.%
, Jia, M.%
, Liu, C.%
\BCBL {}\ \BBA {} Ma, Z.%
\end{APACrefauthors}%
\unskip\
\newblock
\APACrefYearMonthDay{2020}{}{}.
\newblock
{\BBOQ}\APACrefatitle {Task preemption based on petri nets} {Task preemption
  based on petri nets}.{\BBCQ}
\newblock
\APACjournalVolNumPages{IEEE Access}{8}{}{11512--11519}.
\PrintBackRefs{\CurrentBib}

\bibitem [\protect \citeauthoryear {%
Zelkowitz%
\ \BBA {} Wallace%
}{%
Zelkowitz%
\ \BBA {} Wallace%
}{%
{\protect \APACyear {1998}}%
}]{%
zelkowitz1998experimental}
\APACinsertmetastar {%
zelkowitz1998experimental}%
\begin{APACrefauthors}%
Zelkowitz, M\BPBI V.%
\BCBT {}\ \BBA {} Wallace, D\BPBI R.%
\end{APACrefauthors}%
\unskip\
\newblock
\APACrefYearMonthDay{1998}{}{}.
\newblock
{\BBOQ}\APACrefatitle {Experimental models for validating technology}
  {Experimental models for validating technology}.{\BBCQ}
\newblock
\APACjournalVolNumPages{Computer}{31}{5}{23--31}.
\PrintBackRefs{\CurrentBib}

\bibitem [\protect \citeauthoryear {%
Zhao%
, Liu%
, Dai%
\BCBL {}\ \BBA {} Ma%
}{%
Zhao%
\ \protect \BOthers {.}}{%
{\protect \APACyear {2016}}%
}]{%
zhao2016entropy}
\APACinsertmetastar {%
zhao2016entropy}%
\begin{APACrefauthors}%
Zhao, W.%
, Liu, H.%
, Dai, W.%
\BCBL {}\ \BBA {} Ma, J.%
\end{APACrefauthors}%
\unskip\
\newblock
\APACrefYearMonthDay{2016}{}{}.
\newblock
{\BBOQ}\APACrefatitle {An entropy-based clustering ensemble method to support
  resource allocation in business process management} {An entropy-based
  clustering ensemble method to support resource allocation in business process
  management}.{\BBCQ}
\newblock
\APACjournalVolNumPages{Knowledge and Information Systems}{48}{2}{305--330}.
\PrintBackRefs{\CurrentBib}

\bibitem [\protect \citeauthoryear {%
Zhao%
, Pu%
\BCBL {}\ \BBA {} Jiang%
}{%
Zhao%
\ \protect \BOthers {.}}{%
{\protect \APACyear {2020}}%
}]{%
zhao2020human}
\APACinsertmetastar {%
zhao2020human}%
\begin{APACrefauthors}%
Zhao, W.%
, Pu, S.%
\BCBL {}\ \BBA {} Jiang, D.%
\end{APACrefauthors}%
\unskip\
\newblock
\APACrefYearMonthDay{2020}{}{}.
\newblock
{\BBOQ}\APACrefatitle {A human resource allocation method for business
  processes using team faultlines} {A human resource allocation method for
  business processes using team faultlines}.{\BBCQ}
\newblock
\APACjournalVolNumPages{Applied Intelligence}{50}{9}{2887--2900}.
\PrintBackRefs{\CurrentBib}

\bibitem [\protect \citeauthoryear {%
Zhao%
, Yang%
, Liu%
\BCBL {}\ \BBA {} Wu%
}{%
Zhao%
\ \protect \BOthers {.}}{%
{\protect \APACyear {2015}}%
}]{%
zhao2015optimization}
\APACinsertmetastar {%
zhao2015optimization}%
\begin{APACrefauthors}%
Zhao, W.%
, Yang, L.%
, Liu, H.%
\BCBL {}\ \BBA {} Wu, R.%
\end{APACrefauthors}%
\unskip\
\newblock
\APACrefYearMonthDay{2015}{}{}.
\newblock
{\BBOQ}\APACrefatitle {The optimization of resource allocation based on process
  mining} {The optimization of resource allocation based on process
  mining}.{\BBCQ}
\newblock
\BIn{} \APACrefbtitle {International Conference on Intelligent Computing}
  {International conference on intelligent computing}\ (\BPGS\ 341--353).
\PrintBackRefs{\CurrentBib}

\bibitem [\protect \citeauthoryear {%
Zhao%
, Zeng%
, Zheng%
\BCBL {}\ \BBA {} Yang%
}{%
Zhao%
\ \protect \BOthers {.}}{%
{\protect \APACyear {2017}}%
}]{%
zhao2017resource}
\APACinsertmetastar {%
zhao2017resource}%
\begin{APACrefauthors}%
Zhao, W.%
, Zeng, Q.%
, Zheng, G.%
\BCBL {}\ \BBA {} Yang, L.%
\end{APACrefauthors}%
\unskip\
\newblock
\APACrefYearMonthDay{2017}{}{}.
\newblock
{\BBOQ}\APACrefatitle {The resource allocation model for multi-process
  instances based on particle swarm optimization} {The resource allocation
  model for multi-process instances based on particle swarm
  optimization}.{\BBCQ}
\newblock
\APACjournalVolNumPages{Information Systems Frontiers}{19}{5}{1057--1066}.
\PrintBackRefs{\CurrentBib}

\bibitem [\protect \citeauthoryear {%
Zhou%
\ \BBA {} Chen%
}{%
Zhou%
\ \BBA {} Chen%
}{%
{\protect \APACyear {2008}}%
}]{%
zhou2008project}
\APACinsertmetastar {%
zhou2008project}%
\begin{APACrefauthors}%
Zhou, Y.%
\BCBT {}\ \BBA {} Chen, Y.%
\end{APACrefauthors}%
\unskip\
\newblock
\APACrefYearMonthDay{2008}{}{}.
\newblock
{\BBOQ}\APACrefatitle {Project-oriented resource assignment: from business
  process modelling to business process instantiation with operational
  performance consideration} {Project-oriented resource assignment: from
  business process modelling to business process instantiation with operational
  performance consideration}.{\BBCQ}
\newblock
\APACjournalVolNumPages{International Journal of Computer Integrated
  Manufacturing}{21}{1}{97--110}.
\PrintBackRefs{\CurrentBib}

\end{thebibliography}

\end{document}